\documentclass[12pt]{article}
\usepackage{epsfig,amssymb,amsmath,psfrag}

\catcode`\@=11
\newcommand{\insertfig}[2]{\mbox{\epsfxsize=#1cm \epsfbox{#2.eps}}}

\font\cmss=cmss12 
\def\1{\hbox{{1}\kern-.25em\hbox{l}}}
\def\bfZ{\relax{\hbox{\cmss Z\kern-.4em Z}}}

\def \thesection {\arabic{section}.}

\def \be  {\begin{equation}}
\def \ee  {\end{equation}}
\def \ba  {\begin{eqnarray}}
\def \ea  {\end{eqnarray}}
\def \baa {\begin{eqnarray*}}
\def \eaa {\end{eqnarray*}}
\def \bb  {\begin {thebibliography} }
\def \eb  {\end{thebibliography}}
\def \lab #1 {\label{#1}}

\newcommand\re[1]{(\ref{#1})}

\def \qqqquad {\qquad\qquad}
\def \matrix #1 {\left(\begin{array}{cc} #1 \end{array}\right)}

\def \tr {\mathop{\rm tr}\nolimits}
\def \Im {\mathop{\rm Im}\nolimits}

\def \e  {\mathop{\rm e}\nolimits}
\newcommand\lr[1]{{\left({#1}\right)}}

\newcommand \vev [1] {\langle{#1}\rangle}

\newcommand \ket [1] {|{#1}\rangle}

\newcommand{\as}{\ifmmode\alpha_{\rm s}\else{$\alpha_{\rm s}$}\fi}
\newcommand{\asbar}{\ifmmode\bar{\alpha}_{\rm s}\else{$\bar{\alpha}_{\rm s}$}\fi}

\newcommand{\bit}[1]{\mbox{\boldmath$#1$}}
\newcommand{\ft}[2]{{\textstyle\frac{#1}{#2}}}

\font\cmss=cmss12 
\def\inbar{\,\vrule height1.5ex width.4pt depth0pt}
\def\IC{\relax\hbox{$\inbar\kern-.3em{\rm C}$}}
\def\IZ{\relax{\hbox{\cmss Z\kern-.4em Z}}}
\def\IR{{\hbox{{\rm I}\kern-.2em\hbox{\rm R}}}}

\def\IP{{\hbox{{\rm I}\kern-.2em\hbox{\rm P}}}}
\def\II{\hbox{{1}\kern-.25em\hbox{l}}}

\def\numberbysection{\@addtoreset{equation}{section}
                     \def\theequation{\thesection\arabic{equation}}}
\numberbysection

\newbox\lett\newdimen\lheight\newdimen\lwidth
\def\ontop#1#2{\setbox\lett=\hbox{#2}\lheight\ht\lett
\multiply\lheight by 12 \divide\lheight by 10\relax%
\lwidth\wd\lett \multiply\lwidth by 8 \divide\lwidth by 10\relax #2\kern-\lwidth%
\raise\lheight\hbox{{$\scriptstyle #1$}}\kern.1ex}

\psfrag{dots}[cc][cc]{$...$}
\psfrag{Z1}[cc][cc]{$\scriptstyle Z_1$}
\psfrag{Z2}[cc][cc]{$\scriptstyle Z_2$}
\psfrag{Z3}[cc][cc]{$ $}
\psfrag{ZN-1}[cc][cc]{$ $}
\psfrag{ZN}[cc][cc]{$\scriptstyle Z_L$}

\begin{document}

\begin{titlepage}
\begin{flushright}
\begin{tabular}{l}
LPT--Orsay--04--60 \\
hep-th/0409120
\end{tabular}
\end{flushright}

\vskip1cm

\centerline{\large \bf Dilatation operator in (super-)Yang-Mills theories on the
light-cone}

\vspace{1cm}

\centerline{\sc A.V. Belitsky$^a$, S.\'E. Derkachov$^{b,c}$,
                G.P. Korchemsky$^b$, A.N. Manashov$^{d,c}$}

\vspace{10mm}

\centerline{\it $^a$Department of Physics and Astronomy, Arizona State
University} \centerline{\it Tempe, AZ 85287-1504, USA}

\vspace{3mm}

\centerline{\it $^b$Laboratoire de Physique Th\'eorique\footnote{Unit\'e
                    Mixte de Recherche du CNRS (UMR 8627).},
                    Universit\'e de Paris XI}
\centerline{\it 91405 Orsay C\'edex, France}

\vspace{3mm}

\centerline{\it $^c$Department of Theoretical Physics,  St.-Petersburg State
University}

\centerline{\it 199034, St.-Petersburg, Russia}

\vspace{3mm}

\centerline{\it $^d$Institut f{\"u}r Theoretische Physik, Universit{\"a}t
Regensburg}
\centerline{\it D-93040 Regensburg, Germany}

\def\thefootnote{\fnsymbol{footnote}}%
\vspace{1cm}

\centerline{\bf Abstract}

\vspace{5mm}

The gauge/string correspondence hints that the dilatation operator in gauge
theories with the superconformal $SU(2,2|\mathcal{N})$ symmetry should possess
universal integrability properties for different $\mathcal{N}$. We provide
further support for this conjecture by computing a one-loop dilatation operator
in all (super)symmetric Yang-Mills theories on the light-cone ranging from
gluodynamics all the way to the maximally supersymmetric $\mathcal{N}=4$
theory. We demonstrate that the dilatation operator takes a remarkably simple
form when realized in the space spanned by single-trace products of superfields
separated by light-like distances. The latter operators serve as generating
functions for Wilson operators of the maximal Lorentz spin and the scale
dependence of the two are in the one-to-one correspondence with each other. In
the maximally supersymmetric, $\mathcal{N}=4$ theory all nonlocal light-cone
operators are built from a single CPT self-conjugated superfield while for
$\mathcal{N}=0,1,2$ one has to deal with two distinct superfields and
distinguish three different types of such operators. We find that for the
light-cone operators built from only one species of superfields, the one-loop
dilatation operator takes the same, universal form in all SYM theories and it
can be mapped in the multi-color limit into a Hamiltonian of the
$SL(2|\mathcal{N})$ Heisenberg (super)spin chain of length equal to the number
of superfields involved. For ``mixed'' light-cone operators involving both
superfields the dilatation operator for $\mathcal{N}\le 2$ receives an
additional contribution from the exchange interaction between superfields on
the light-cone which breaks its integrability symmetry and creates a mass gap
in the spectrum of anomalous dimensions.

\end{titlepage}

\setcounter{footnote} 0

\thispagestyle{empty}

\newpage

\pagestyle{empty} {\small\tableofcontents}

\newpage

\pagestyle{plain} \setcounter{page} 1

\section{Introduction}

Four-dimensional gauge theories are expected to admit, at least in the
multi-color limit, a complimentary description via yet to be identified string
theories \cite{PolBook}. The latter operate in terms of collective degrees of
freedom (Faraday lines) which are more appropriate to tackle the strong-coupling
dynamics of Yang-Mills theories. The most prominent and thoroughly verified to
date example of the gauge/string correspondence is the maximally supersymmetric
$\mathcal{N}=4$ Yang-Mills (SYM) theory \cite{GliSchOli77} and its dual description
in terms of a critical string theory with AdS$_5 \times$S$_5$ target space
\cite{Mal97,GunKlePol98,Wit98}. Recently it has been conjectured \cite{Pol04}
that noncritical sigma models, possessing the $\kappa$-symmetry and having
AdS$_5$ geometry as a factor of the target space, are dual to yet unknown
(non)supersymmetric gauge theories which exhibit conformal $SU(2,2|\mathcal{N})$
invariance with $\mathcal{N}=0,1,2$ at finite values of the coupling constant.
Both critical and noncritical sigma models on the anti-de Sitter space turn out
to be completely integrable \cite{ManSurWad02} and it is believed that this
property must manifest itself in hidden symmetries of the corresponding
Yang-Mills theory.

Indeed, it has been known for some time that four-dimensional Yang-Mills theories
exhibit a remarkable phenomenon of integrability. It has been first discovered in
the context of QCD, i.e., $\mathcal{N}=0$ Yang-Mills theory with fundamental
matter, in the studies of the Regge asymptotics of scattering
amplitudes~\cite{Lip94,FadKor94} and anomalous dimensions of high-twist Wilson
operators in multi-color
limit~\cite{BraDerMan98,BraDerKorMan99,Bel99,DerKorMan00}. In the former case,
high-energy asymptotics of the scattering amplitudes is driven by the
contribution of multi-gluonic color-singlet states which can be identified as
eigenstates of the Heisenberg $SL(2,\mathbb{C})$ spin magnet. In the latter case,
the one-loop dilatation operator for a special class of maximal-helicity
high-twist operators can be mapped into a Hamiltonian of a completely integrable
Heisenberg $SL(2,\mathbb{R})$ spin magnet. The number of sites in this spin chain
equals the number of fundamental fields involved and the symmetry group is a
collinear subgroup of the $SO(4,2) \sim SU(2,2)$ conformal group. Although
conformal symmetry of QCD is broken at the quantum level, symmetry breaking
effects arise starting from two loops only \cite{Mul94} (for a review, see
\cite{BraKorMul03}). This implies that to one-loop accuracy, QCD is not distinct
from a conformal field theory. Obviously, the same holds true in supersymmetric
$\mathcal{N} = 1$ and $\mathcal{N} = 2$ SYM theories whereas the $\mathcal{N} =
4$ theory remains conformal to all orders of perturbation theory.

In the present paper, we shall study integrability properties of the one-loop
dilatation operator in $\mathcal{N}-$extended SYM theories. As was already
mentioned, in the $\mathcal{N} = 0$ theory the integrability phenomenon has been
observed in the sector of maximal-helicity operators. The integrability gets
extended to a larger class of operators as one goes over from the
nonsupersymmetric ($\mathcal{N} = 0$) to the maximally supersymmetric
($\mathcal{N}=4$) gauge theory. In particular, in the $\mathcal{N}=4$ model, the
integrability was found in the sector of local scalar operators. In the
multi-color limit, the one-loop mixing matrix for such operators can be mapped
into a Hamiltonian of the $SO(6)$ Heisenberg magnet with the symmetry group
reflecting the $R-$symmetry of the model~\cite{MinZar02}. Eventually, the
$SL(2;\mathbb{R})$ and $SO(6)$ sectors can be unified together into a
$PSU(2,2|4)$ Heisenberg magnet~\cite{BeiKriSta03,BeiSta03}. The gauge/string
correspondence hints that the dilatation operator in gauge theories with the
$SU(2,2|\mathcal{N})$ symmetry should possess universal integrability properties
for different $\mathcal{N}$~\cite{Pol04}. This suggests that integrable
structures found previously in the $\mathcal{N}=0$ and $\mathcal{N}=4$ SYM should
be different facets of the same phenomenon. To address this issue we need an
approach that would allow us to treat simultaneously the operator mixing in
various $\mathcal{N}-$extended SYM theories.

The conventional covariant approach based on calculation of the mixing matrix for
local Wilson operators is not particularly suited for these purposes as it has
the following shortcomings. The form of the mixing matrix depends on the sector
under consideration. For example, it is given by a finite-dimensional matrix for
local composite operators built from fundamental fields without covariant
derivatives and by an infinite-dimensional matrix for operators with an arbitrary
number of derivatives. In addition, due to different particle content of
$\mathcal{N}-$extended SYM theories, the number of possible Wilson operators vary
with ${\cal N}$ and, therefore, one would not expect any connection among the
mixing matrices for different ${\cal N}$. Last but not least, hidden
integrability symmetry is identifiable only in simplest sectors of Wilson
operators but it is not manifest in the most generic case. As was demonstrated in
\cite{BelDerKorMan04}, these drawbacks can be circumvented by studying the mixing
of Wilson operators within a light-cone superspace
formalism~\cite{BriLinNil83,Man83}. Recently this formalism has been applied to
calculating anomalous dimensions of Wilson operators in the $\mathcal{N}=1$ SYM
theory within an effective action approach~\cite{Kir04}. The interaction vertices
in the effective action are manifestly invariant under superconformal
transformations and can be mapped into four-point correlation functions.

The Wilson operators in the $\mathcal{N}-$extended SYM theories are local
composite gauge-invariant operators built as products of an arbitrary number of
fundamental fields and an arbitrary number of covariant derivatives acting on
them. They can be classified according to representations of the superconformal
$SU(2,2|\mathcal{N})$ group. In what follows we shall consider single-trace
Wilson operators possessing the maximal Lorentz spin and minimal twist for a
given number of constituent fields. They belong to the $SL(2|\mathcal{N})$
subgroup of the full superconformal group and are known in QCD as quasipartonic
operators \cite{BukFroKurLip85}. In this paper, we demonstrate that the one-loop
dilatation operator in the $\mathcal{N}-$extended SYM theory acting on the space
spanned by the quasipartonic operators has a universal form in the multi-color
limit and is intrinsically related to a completely integrable $SL(2|\mathcal{N})$
Heisenberg magnet. For $\mathcal{N}=0$ one recovers the $SL(2)$ magnet in the
sector of maximal helicity Wilson
operators~\cite{BraDerMan98,BraDerKorMan99,Bel99,DerKorMan00}, while for
$\mathcal{N}=4$ the $SL(2|4)$ magnet forms an autonomous subsector of a bigger
$PSU(2,2|4)$ magnet~\cite{BeiKriSta03,BeiSta03}.

To study the scale dependence of quasipartonic operators, it is convenient to
switch from local Wilson operators to nonlocal light-cone operators. The latter
are generating functions for the quasipartonic operators and are defined as
\begin{equation}
\label{O}
\mathcal{O}_{i_1\ldots i_L} (z_1,\ldots,z_L) = \tr \{X_{i_1}(n z_1) \ldots
X_{i_L}(n z_L)\} \,,
\end{equation}
where $X_{i}=\{\lambda,\bar\lambda,n^\mu F_{\mu\perp},\phi \}$ is a unified
notation for ``good'' components of fundamental fields in the underlying
$\mathcal{N}-$extended SYM (fermions, field strength tensor, scalars) given by
$N_c \times N_c$ matrices $X_i = X_i^a\, t^a$ with $t^a$ being generators of the
fundamental representation of the $SU(N_c)$ group. The fields in \re{O} are
located along a light-cone direction defined by a light-like vector $n_\mu$ such
that $n_\mu^2=0$ and their position on the light-cone is specified by the
coordinates $z_1,\ldots,z_L$. It is tacitly assumed that the gauge invariance in
\re{O} is restored by inserting Wilson lines $P\exp\big(ig\int_{z_k}^{z_{k+1}}d s
\, n^\mu A_\mu(ns)\big)$ that run along the light-cone between two adjacent
fields. Later, we shall adopt the light-cone axial gauge $n^\mu A_\mu (x) \equiv
A_+ (x) = 0$ in which these Wilson lines are reduced to a unity matrix. Expanding
\re{O} around $z_1=\ldots=z_L=0$ one generates the quasipartonic operators
\be
\mathcal{O}_{i_1\ldots i_L} (z_1,\ldots,z_L)=\sum_{k_1,...,k_L \ge
0}\frac{z_1^{k_1}}{k_1!}\ldots\frac{z_L^{k_L}}{k_L!} \tr \{D_+^{k_1} X_{i_1}(0)
\ldots D_+^{k_L}X_{i_L}(0)\}\,,
\label{OW-W}
\ee
where $D_+ = n^\mu D_\mu$ is a projection of the covariant derivative on the
light-cone. The operators $\tr \{D_+^{k_1} X_{i_1}(0) \ldots
D_+^{k_L}X_{i_L}(0)\}$ have the maximal possible Lorentz spin, $k_1+\ldots+k_L$,
and their twist equals the number of constituents $L$. Among them there are the
operators with no derivatives $\tr \{X_{i_1}(0) \ldots X_{i_L}(0)\}$ as well as
operators involving an arbitrary number of covariant derivatives. These operators
mix under renormalization and the corresponding mixing matrix can be deduced from
the scale dependence of nonlocal light-cone operators \re{O} with a help of
\re{OW-W}.

A very convenient framework for discussing the scale dependence of the
quasipartonic operators in the SYM theories is provided by the light-cone
superspace formalism~\cite{BriLinNil83,Man83}. In this approach, the SYM theory
is quantized on the light-cone and its Lagrangian is built from two distinct
chiral superfields ${\Phi}(x_\mu,\theta^A)$ and ${\Psi}(x_\mu,\theta^A)$ (with
${\scriptstyle A}=1,\ldots,\mathcal{N}$) which comprise all ``good'' components
of the fundamental fields $X_i(x_\mu)$ describing dynamically independent
propagating modes. Both superfields realize a representation of the
superconformal $SU(2,2|\mathcal{N})$ group and carry a definite value of the
conformal spin. While the chiral superfield ${\Psi}$ has the conformal spin
$j_\Psi=(3-\mathcal{N})/2$ which depends on the number of supercharges ${\cal
N}$, the one of ${\Phi}$ equals $j_\Phi=-1/2$. For $\mathcal{N}\le 2$ the two
superfields are independent on each other whereas in the maximally
supersymmetric, $\mathcal{N}=4$ gauge theory they are not independent, ${\Phi}
\sim {\Psi}$.

By definition, the propagating fields $X_i(zn_\mu)$ are the coefficients in the
Taylor expansion of the superfields ${\Phi}(zn_\mu,\theta^A)$ and
${\Psi}(zn_\mu,\theta^A)$ in powers of the odd coordinates $\theta^A$. This
suggests to generalize further \re{O} and consider composite single-trace
operators constructed from an arbitrary number of superfields located on the
light-cone $x_\mu = z n_\mu$. Let us denote ${\Phi}(zn_\mu,\theta^A) \equiv
{\Phi}(Z)$ and ${\Psi} (zn_\mu,\theta^A) \equiv {\Psi}(Z)$ and identify $Z =
(z,\theta^A)$ as a point in the $(\mathcal{N}+1)-$dimensional light-cone
superspace. In general, one can distinguish three types of single-trace
operators:
\begin{enumerate}

\item[(i)] operators built only from ${\Phi}-$superfields:
\begin{equation}
\label{Chiral-sector}
\mathbb{O}_{\Phi\ldots\Phi} (Z_1, \ldots, Z_L) = \tr \{ {\Phi}(Z_1) \ldots
{\Phi}(Z_L)\} \, ,
\end{equation}

\item[(ii)] operators built only from ${\Psi}-$superfields:
\begin{equation}
\mathbb{O}_{\Psi\ldots\Psi} (Z_1,\ldots,Z_L) = \tr \{ {\Psi}(Z_1) \ldots
{\Psi}(Z_L)\} \, ,
\label{Chiral-sector-1}
\end{equation}

\item[(iii)] operators built from  both ${\Phi}-$ and ${\Psi}-$superfields:
\begin{equation}
\label{Mixed-sector}
\mathbb{O}_{\Phi\ldots\Psi} (Z_1,\ldots,Z_L) = \tr \{ {\Phi}(Z_1) \ldots
{\Psi}(Z_L)\} \, ,
\end{equation}

\end{enumerate}
In the ${\cal N} = 4$ SYM theory all three sectors coincide since $\Psi\sim\Phi$.
For ${\cal N} \le 2$ each sector has to be considered separately. Expanding the
operators \re{Chiral-sector} -- \re{Mixed-sector} in
$\theta_1^{A_1}\ldots\theta_L^{A_L}$, one generates all
nonlocal light-cone operators \re{O}, symbolically, \\[0mm]
\be
\mathbb{O}(Z_1,\ldots,Z_L) = \sum_{\{ {i_1},\ldots, {i_L}\}}
\theta_{i_1}^{A_{i_1}} \ldots \theta_{i_L}^{A_{i_L}} \mathcal{O}_{i_1\ldots
i_L}(z_1,\ldots,z_L)\,,
\label{OW-OS}
\ee
where ${\scriptstyle A}_i=1,\ldots,\mathcal{N}$, so that the total number of
$\theta-$variables in this expansion varies between $0$ and $\mathcal{N}^L$.

Combining together \re{O}, \re{OW-W} and \re{OW-OS} one finds that the problem of
finding the scale dependence of (an infinite number of) Wilson, quasipartonic
operators $\tr \{D_+^{k_1} X_{i_1}(0) \ldots D_+^{k_L}X_{i_L}(0)\}$ can be mapped
into the problem of constructing the dilatation operator on the space spanned by
nonlocal (super-)light-cone operators \re{Chiral-sector} -- \re{Mixed-sector}
(see Eq.~\re{EQ} below). As we will show below, to one-loop accuracy in the
multi-color limit, the operators \re{OW-OS} mix under renormalization with
single-trace light-cone operators built from the same number of ${\Phi}-$ and
${\Psi}-$superfields but ordered differently inside the trace. This allows one to
realize the one-loop dilatation operator for the operators \re{OW-OS} as a
quantum-mechanical Hamiltonian $\mathbb{H}$ acting on $L$ superfields. The
resulting one-loop Callan-Symanzik equation for the nonlocal operators \re{OW-OS}
reads
\be
\left\{ \mu\frac{\partial}{\partial\mu} +
\beta_\mathcal{N}(g)\frac{\partial}{\partial g}
+ L \gamma_\mathcal{N}(g) \right\} \mathbb{O}(Z_1,Z_2,\ldots,Z_L) = - \frac{g^2
N_c}{8\pi^2} \left[ \mathbb{H}\cdot \mathbb{O}\right](Z_1,Z_2,\ldots,Z_L)\, ,
\label{EQ}
\ee
where $\beta_\mathcal{N}(g)$ is the beta-function in the SYM theory and
$\gamma_\mathcal{N}(g)=\beta_\mathcal{N}(g)/g$ is the anomalous dimension of the
superfields in the light-like axial gauge $A_+(x)=0$. The superconformal
invariance of the SYM theory imposes restrictions on the possible form of the
one-loop dilatation operator and allows one to fix $\mathbb{H}$ up to a scalar
function. We will determine this function performing an explicit calculation of
Feynman supergraphs in an $\mathcal{N}-$extended SYM theory.

To one-loop order the dilatation operator $\mathbb{H}$ has a two-particle
structure. In addition, in the multi-color limit the interaction can happen only
between two neighboring superfields and, therefore, $\mathbb{H}$ is given by the
sum over the nearest neighbors
\be
\mathbb{H} = \mathbb{H}_{12} + \ldots + \mathbb{H}_{L,L-1} + \mathbb{H}_{L,1}\,.
\label{H-multi-color}
\ee
Here the two-particle kernel $\mathbb{H}_{k,k+1}$ acts locally on the superfields
with the coordinates $Z_k$ and $Z_{k+1}$ and leaves the remaining superfields
intact. The explicit form of $\mathbb{H}_{k,k+1}$ depends on the superfields
involved. For $\mathcal{N}=4$ the light-cone operators \re{OW-OS} are built from
the superfields $\Phi$ only, Eq.~\re{Chiral-sector}, and, therefore,
$\mathbb{H}_{k,k+1}$ coincides with the dilatation operator in the
$\Phi\Phi-$sector, $\mathbb{H}_{k,k+1}=\mathbb{H}_{\Phi\Phi}$. For
$\mathcal{N}\le 2$ one has to distinguish four different operators
$\mathbb{H}_{\Phi\Phi}$, $\mathbb{H}_{\Psi\Psi}$, $\mathbb{H}_{\Phi\Psi}$ and
$\mathbb{H}_{\Psi\Phi}$. They define the two-particle evolution kernel
$\mathbb{H}_{k,k+1}$ in the $\Phi\Phi-$, $\Psi\Psi-$, $\Phi\Psi-$ and
$\Psi\Phi-$sectors, respectively. Later we shall often use a unifying notation
for the superfields, ${\Phi}_{j_\Phi} ={\Phi}$ and ${\Phi}_{j_\Psi} ={\Psi}$, and
combine these operators into a $2\times 2$ matrix $\mathbb{H}_{ab}$ (with
$a,b={\scriptstyle \Phi,\Psi}$).

The outcome of our consideration can be summarized in a few equations for the
two-particle dilatation operators $\mathbb{H}_{ab}$. These operators admit the
following representation
\begin{equation}
\mathbb{H}_{ab} = \left[ \mathbb{V}^{(j_a, j_b)} - \mathbb{V}^{(j_a, j_b)}_{\rm
ex} \right] (1 - \Pi_{ab}) \qquad (a,b={\scriptstyle \Phi,\Psi})\, ,
\label{ansatz-gen-H12}
\end{equation}
where $j_a$ is the conformal spin of the corresponding superfield ($j_\Phi=-1/2$
and $j_\Psi=(3 - {\cal N})/2$) and the operators $\mathbb{V}^{(j_a, j_b)}$ ,
$\mathbb{V}^{(j_a, j_b)}_{\rm ex}$ and $\Pi_{ab}$ are defined as follows. The
kernel $\mathbb{V}^{(j_a, j_b)}$ describes the ``diagonal'' transition
$\Phi_{j_a}\Phi_{j_b}\to \Phi_{j_a}\Phi_{j_b}$. It is given by the following
integral operator
\begin{eqnarray}
\mathbb{V}^{(j_a, j_b)} \, {\Phi}_{j_a} (Z_1) {\Phi}_{j_b} (Z_2) \!\!\!&=&\!\!\!
\int_0^1 \frac{d\alpha}{\alpha} \, \bigg\{ 2 {\Phi}_{j_a} (Z_1) {\Phi}_{j_b}
(Z_2)
\label{V-super}
\\
&& \hspace*{-40mm} - (1-\alpha)^{2 j_a - 1} {\Phi}_{j_a} ((1-\alpha) Z_1 + \alpha
Z_2) {\Phi}_{j_b} (Z_2) - (1-\alpha)^{2 j_b - 1} {\Phi}_{j_a} (Z_1) {\Phi}_{j_b}
((1-\alpha)Z_2 + \alpha Z_1) \bigg\} \, , \nonumber
\end{eqnarray}
which displaces the superfields in the direction of each other,
$$
\Phi_j((1-\alpha) Z_1 + \alpha Z_2) \equiv \Phi_j((1-\alpha) z_1 + \alpha
z_2,(1-\alpha)\theta_1^A + \alpha \theta_2^A) \, .
$$
The term with $\mathbb{V}^{(j_a, j_b)}_{\rm ex}$ arises in \re{ansatz-gen-H12}
only for $j_a\neq j_b$, that is, $\mathbb{V}^{(j_\Psi, j_\Psi)}_{\rm ex} =
\mathbb{V}^{(j_\Phi, j_\Phi)}_{\rm ex}=0$. The kernel $\mathbb{V}^{(j_\Phi,
j_\Psi)}_{\rm ex}$ describes the exchange transition $\Phi\Psi \to \Psi\Phi$
\begin{eqnarray} \mathbb{V}^{(j_\Phi, j_\Psi)}_{\rm ex}
\, {\Phi} (Z_1) {\Psi} (Z_2) = \int_0^1 d\alpha \frac{\alpha^{3 - {\cal
N}}}{(1-\alpha)^2} \, {\Psi} ((1-\alpha) Z_1 + \alpha Z_2) {\Phi} (Z_2) \, .
\label{V-ex-super}
\label{ansatz-V-ex}
\end{eqnarray}
The evolution kernel $\mathbb{V}^{(j_\Psi, j_\Phi)}_{\rm ex}$ describes the
transition $\Psi\Phi \to \Phi\Psi$ and is given by the same expression with the
superfields ${\Phi}$ and ${\Psi}$ interchanged in both sides of \re{ansatz-V-ex}.

The integral in \re{ansatz-V-ex} is divergent for $\alpha\to 1$. The same problem
arises in \re{V-super} if at least one of the superfields carries a negative
conformal spin, $j_\Phi=-1/2$. In Eq.~\re{ansatz-gen-H12}, divergences are
removed by the operator $(1-\Pi_{ab})$, which is a projector. For $\mathcal{N}\le
2$, in the $\Psi\Psi-$sector, the projector is not required, $\Pi_{\Psi\Psi}=0$,
since the superfield ${\Psi}$ has a positive conformal spin $j_\Psi =(3 - {\cal
N})/2> 0$. The expressions for the projectors $\Pi_{\Phi\Phi}$, $\Pi_{\Phi\Psi}$
and $\Pi_{\Psi\Phi}$ will be given below (see Eqs.~\re{Pi12-main},
\re{projector-phi-psi} and \re{projector-psi-phi}).

There exist nontrivial relations between the two-particle dilatations operators,
Eqs.~\re{ansatz-gen-H12} -- \re{V-ex-super}, for different $\mathcal{N}$. Namely,
the one-loop dilatation operator in SYM theories with ${\cal N} \le 2$
supersymmetries can be obtained from the dilatation operator in the maximally
supersymmetric, $\mathcal{N}=4$ theory through a ``method of
truncation''~\cite{BriTol83}. It amounts to reducing the number of ``odd''
dimensions in the light-cone superspace from $\mathcal{N}=4$ down to
$\mathcal{N}=0$. In this way one finds that two seemingly different expressions
for the evolution kernels \re{V-super} and \re{ansatz-V-ex} for $\mathcal{N}\le
2$ follow from the kernel $\mathbb{V}^{(-1/2,-1/2)}$ in the $\mathcal{N}=4$
theory. Similar relation between the evolution kernels also at work in the
opposite direction. Namely, the expressions for the kernel $\mathbb{V}^{(j_a,
j_b)}$ and $\mathbb{V}_{\rm ex}^{(j_a, j_b)}$ in the $\mathcal{N}=0$ theory can
be generalized to arbitrary $\mathcal{N}$ by simply extending the one-dimensional
light-cone direction to the $(\mathcal{N}+1)-$dimensional superspace, $z\to
Z=(z,\theta^A)$.

The two-particle evolution kernels, Eqs.~\re{ansatz-gen-H12} -- \re{ansatz-V-ex},
allow us to construct a one-loop dilatation operator \re{H-multi-color}. Its
eigenvalues determine the spectrum of anomalous dimensions of \textsl{all}
quasipartonic operators in SYM theories with $\mathcal{N}=0,1,2,4$ supercharges.
Notice that the two-particle kernel in the $\Phi\Phi-$sector,
$\mathbb{H}_{\Phi\Phi}$, does not depend on the number of supercharges and,
therefore, the one-loop dilatation operator \re{H-multi-color} acting on the
light-cone operators \re{Chiral-sector} has a universal form in the SYM theories
with $0\le \mathcal{N}\le 4$. For the light-cone operators \re{Chiral-sector-1}
and \re{Mixed-sector} the dilatation operator depends on $\mathcal{N}$ through
the dependence of two-particle kernels $\mathbb{H}_{\Phi\Psi}$,
$\mathbb{H}_{\Psi\Phi}$ and $\mathbb{H}_{\Psi\Psi}$, Eq.~\re{ansatz-gen-H12}, on
the conformal spin of the superfield ${\Psi}$, $j_\Psi=(3-\mathcal{N})/2$.

It turns out, the one-loop dilatation operator defined in Eqs.~\re{H-multi-color}
-- \re{V-ex-super} has a hidden integrability symmetry: the two-particle kernel
$\mathbb{V}^{(j_1, j_2)}$ can be identified as a Hamiltonian of the Heisenberg
$SL(2|\mathcal{N})$ spin chain consisting of two sites~\cite{Kulish,Frahm,DKK}.
As a consequence, for the light-cone operators \re{Chiral-sector} and
\re{Chiral-sector-1} the one-loop dilatation operator coincides in the
multi-color limit with a Hamiltonian of a completely integrable
$SL(2|\mathcal{N})$ spin chain of the length equal to the number of superfields
$L$. For $\mathcal{N}\le 2$, the dilatation operator acting on the ``mixed''
light-cone operators \re{Mixed-sector} receives an additional contribution from
the exchange interaction $\mathbb{V}_{\rm ex}^{(j_a, j_b)}$. This interaction
breaks integrability symmetry of the dilatation operator and leads to appearance
of a \textsl{mass gap} in the spectrum of the anomalous dimensions of the
operators \re{Mixed-sector}~\cite{BraDerKorMan99,Bel99}.

Some of the results were reported in an earlier Letter \cite{BelDerKorMan04}. In
this paper we provide a detailed account on our approach and present new results.
The paper is organized as follows. In Sect.~2, we review the
Brink-Lindgren-Nilsson and Mandelstam approaches to light-cone SYM theories. In
Sect.~3 we discuss the superconformal symmetry of the SYM theories on the
light-cone and conjecture the form of the one-loop dilatation operator on the
basis of symmetry consideration alone. To verify the conjecture, we perform in
Sect.~4 the one-loop calculation of renormalization group kernels in the
$\mathcal{N}-$extended SYM theories and establish the relation between the
one-loop dilatation operators for different $\mathcal{N}$. In Sect.~5, we apply
the obtained expressions for the dilatation operator to evaluate the one-loop
anomalous dimensions of Wilson operators and demonstrate their agreement with the
known results. In Sect.~6, we reveal a hidden symmetry of the one-loop dilatation
operator in the SYM theory on the light-cone and discuss its relation to
Heisenberg (super)spin chains. Our conclusions are summarized in Sect.~6. Four
Appendices contain a detailed derivation of the results formulated in the body of
the paper.

\section{Super-Yang-Mills theories on the light-cone}

To calculate a one-loop dilatation operator in (super) Yang-Mills theories we
shall apply the ``light-cone formalism'' \cite{KogSop,BriLinNil83,Man83}. In this
formalism one integrates out non-propagating components of fields and formulates
the (super) Yang-Mills action in terms of ``physical'' degrees of freedom.
Although the resulting action is not manifestly covariant under the Poincar\'e
transformations, the main advantage of the light-cone formalism for SYM theories
is that the $\mathcal{N}-$extended supersymmetric algebra is closed off-shell for
the propagating fields and there is no need to introduce auxiliary fields. This
allows one to design a unifying light-cone superspace formulation of various
$\mathcal{N}-$extended SYM, including the ${\cal N} = 4$ theory for which a
covariant superspace formulation does not exist.

Following \cite{KogSop,BriLinNil83,Man83}, we split four components of the gauge
field $A_\mu(x)=A_\mu^a(x)\, t^a$, with $t^a$ being generators of the fundamental
representation of the $SU(N_c)$, into two longitudinal, $A_\pm(x)$, and two
transverse holomorphic and antiholomorphic components, $A(x)$ and $\bar A(x)$,
respectively,
\begin{equation}
A_\pm \equiv \ft1{\sqrt{2}}(A_0 \pm A_3 )\,,\qquad A \equiv \ft1{\sqrt{2}}(A_1 +
i A_2) \, , \qquad
\bar A \equiv A^* = \ft1{\sqrt{2}}(A_1 - i A_2)
\, .
\label{gauge-perp}
\end{equation}
In the light-cone formalism, one quantizes the SYM theory in a noncovariant,
light-cone gauge $A^a_+(x) = 0$. Making a similar decomposition of (Majorana)
fermion fields ${\psi}(x)={\psi}^a(x) \, t^a$ into the so-called ``bad'' and
``good'' components with a help of projectors $ \Pi_\pm = \ft12 \gamma_\pm
\gamma_\mp$ ($\Pi_\pm^2=\Pi_\pm$ and $\Pi_\pm\Pi_\mp=0$)
\be
{\psi} = {\Pi}_+ {\psi} + {\Pi}_- {\psi} \equiv {\psi}_+ + {\psi}_-\,,
\label{good-bad}
\ee
one finds that the fields ${\psi}_-(x)$ and $A_-(x)$ can be integrated out in
this gauge. The resulting action of the SYM theory is expressed in terms of
``physical'' fields: complex gauge field, $A(x)$, ``good'' components of fermion
fields ${\psi}_+(x)$ and, in general, complex scalar fields $\phi(x)$. Finally,
one combines these fields into superfields and rewrites the SYM action on the
light-cone as an integral over the superspace.

At present there exist two different superspace formulations of the SYM theory on
the light-cone. In the Brink-Lidgren-Nilsson formulation~\cite{BriLinNil83}, the
superspace has $2\mathcal{N}$ odd directions, $\theta^A$ and $\bar \theta_A$ with
$A=1,\ldots, \mathcal{N}$, and the light-cone action is build from chiral and
antichiral superfields. In the Mandelstam formulation~\cite{Man83}, the
superspace has only $\mathcal{N}$ odd directions, $\theta^A$ with $A=1,\ldots,
\mathcal{N}$, and the light-cone action involves two distinct chiral superfields.
In this Section, we shall review both formulations and demonstrate their
equivalence.

\subsection{Brink-Lindgren-Nilsson formalism}

Let us start from the ${\cal N} = 4$ SYM and reduce step-by-step the number of
supersymmetries descending down to ${\cal N} = 0$ SYM (pure gluodynamics).

\subsubsection{$\mathcal{N}=4$ theory}

In the $\mathcal{N}=4$ model, the propagating modes consist of the complex field
$A(x)$ describing transverse components of the gauge field, complex Grassmann
fields $\lambda^A(x)$ defining ``good'' components of four Majorana fermions (see
Eqs.~\re{psi-Weyl} and \re{LCprojection}) and a matrix of complex scalar fields
$\phi^{AB}(x)$ (with ${\scriptstyle A,B} =1, \ldots,4$) satisfying $
\phi^{AB}=-\phi^{BA}$. Fields conjugated to them are $\bar A(x)$,
$\bar\lambda_A(x)$ and $\bar \phi_{AB}= \left( \phi^{AB} \right)^* =
\ft12\varepsilon_{ABCD} \phi^{CD}$, respectively.

In the light-cone formalism, all propagating modes can be combined into a single
complex scalar $\mathcal{N}=4$ superfield
\begin{eqnarray}
{\Phi} (x, \theta^A, \bar\theta_A) = {\rm e}^{\frac12 \bar\theta \cdot \theta \,
\partial_+} \bigg\{ \partial_+^{-1}A(x) \!\!\!&+&\!\!\! \theta^A
\partial_+^{-1}\bar\lambda_A (x) + \frac{i}{2!} \theta^A \theta^B \bar \phi_{AB}
(x)
\nonumber\\
&+&\!\!\! \frac{1}{3!} \varepsilon_{ABCD} \theta^A \theta^B \theta^C \lambda^D
(x) - \frac{1}{4!} \varepsilon_{ABCD} \theta^A \theta^B \theta^C \theta^D
\partial_+ \bar{A} (x)
\bigg\} . \quad \ \label{N=4-field}
\end{eqnarray}
Here $\bar\theta\cdot \theta \equiv \bar\theta_A \theta^A$ and the nonlocal
operator $\partial_+^{-1}$ is defined using the Mandelstam-Leibbrandt
prescription \cite{Man83} (see Eq.~\re{ML-pre}). It is tacitly assumed that
${\Phi}={\Phi}^a \, t^a$ with $t^a$ being the generators of the fundamental
representation of the $SU(N_c)$ group.

The light-cone action of the $\mathcal{N} = 4$ SYM reads~\cite{BriLinNil83}
\ba
S_{\mathcal{N}=4} \!\!\!&=&\!\!\! \int d^4 x\, d^4\theta\, d^4 \bar\theta\,
\Bigg\{ \frac12 \bar{\Phi}^a\frac{\Box}{\partial_+^2}{\Phi}^a - \frac23 g f^{abc}
\lr{ \frac{1}{\partial_+}\bar{\Phi}^a{\Phi}^b\bar\partial{\Phi}^c +
\frac{1}{\partial_+}{\Phi}^a\bar{\Phi}^b\partial\bar{\Phi}^c } \nonumber
\\
&&\qquad\qquad\qquad - \frac12 g^2 f^{abc}f^{ade} \left( \frac{1}{\partial_+}
({\Phi}^b\partial_+{\Phi}^c) \frac{1}{\partial_+}
(\bar{\Phi}^d\partial_+\bar{\Phi}^e) +
\frac12{\Phi}^b\bar{\Phi}^c{\Phi}^d\bar{\Phi}^e \right) \Bigg\} \, ,
\label{N=4-S}
\ea
where $\bar{\Phi}= ({\Phi} (x, \theta^A, \bar\theta_A))^*$ is a
conjugated superfield%
\footnote{Complex conjugation for Grassmann variables is specified in
\re{cc-rule1} and \re{cc-rule2}. }, $f^{abc}$ are the structure constants of the
$SU(N_c)$ group, the light-cone derivatives $\partial_+$, $\partial$ and
$\bar\partial$ are defined in \re{Derivatives} and the integration measure over
Grassmann variables is normalized as in \re{NormalizationSuperspace}. The Green's
functions computed from \re{N=4-S} do not contain ultraviolet divergences to all
orders of perturbation theory and, therefore, the $\mathcal{N}=4$ light-cone
action \re{N=4-S} defines an ultraviolet (UV) finite quantum field
theory~\cite{SohWes81,BriLinNil83,HowSteow84,Man83}.

The ${\mathcal{N}=4}$ light-cone superfield \re{N=4-field} has the following
unique properties. It comprises {\sl all} propagating fields of the model, and
expansion in $\theta^A$ can be viewed as an expansion in different helicity
components: $+1$ for $A(x)$, $1/2$ for $\bar\lambda_A (x)$, $0$ for
$\bar\phi_{AB}$, $-1/2$ for $\lambda^A (x)$ and $-1$ for $\bar A(x)$. As a
consequence, the conjugated superfield is {\sl not} independent and is related to
${\Phi} (x, \theta^A,
\bar\theta_A)$ as
\be
\bar{\Phi} (x, \theta^A , \bar\theta_A)
= - \frac1{4!} \partial_+^{-2} \varepsilon^{ABCD} {D_A D_B D_C D_D} {\Phi} (x,
\theta^A, \bar\theta_A) \, .
\label{N=4-bar-field}
\ee
Here the notation was introduced for the covariant derivatives in the superspace
\be
\label{Cov-der}
D_A = \partial_{\theta^A} - \ft{1}{2}\bar\theta_A
\partial_+
\, , \qquad
\bar D^A = \partial_{\bar\theta_A}-\ft{1}{2}\theta^A
\partial_+
\, ,
\ee
satisfying $\{ D_A,D_B\} =\{ \bar D^A, \bar D^B\}= 0$ and $\{ D_A, \bar D^B\} = -
\delta_A^B\partial_+$. The superfields \re{N=4-field} and \re{N=4-bar-field} obey
the chirality conditions
\be
\bar{D}^B {\Phi} (x , \theta^A , \bar\theta_A)
= D_B \bar{\Phi} (x, \theta^A , \bar\theta_A) = 0 \, .
\label{chirality}
\ee
As usual, they imply that the dependence of the chiral superfield ${\Phi}(x ,
\theta^A ,\bar\theta_A)$ on $\bar \theta_A$ and antichiral superfield
$\bar{\Phi}(x , \theta^A , \bar\theta_A)$ on $\theta_A$ can be absorbed into a
redefinition of the space-time coordinate $x_\mu$.

Notice that the lowest two components of the superfield \re{N=4-field} are
nonlocal fields. As a consequence, the expansion of the light-cone operators
\re{Chiral-sector} around the origin in the superspace yields both Wilson
operators and ``spurious'' operators involving the fields $\partial_+^{-1} A(0)$,
$A(0)$ and $\partial_+^{-1}\bar\lambda_A(0)$. The latter operators do not have a
clear physical meaning and their appearance is an artefact of the light-cone
superspace formalism. We shall return to this issue in Sect.~3.4.

\subsubsection{$\mathcal{N}=2$ theory}

The light-cone formulation of Yang-Mills theories with less supersymmetry can be
obtained from $\mathcal{N}=4$ SYM through a ``method of truncation''
\cite{BriTol83}. It is based on the following identity:
\be
\int d^4 x \, d^\mathcal{N} \theta \, d^\mathcal{N} \bar\theta \, \mathcal{L}
({\Phi},\bar{\Phi}) = (-1)^\mathcal{N}\int d^4 x \, d^{\mathcal{N}-1} \theta \,
d^{\mathcal{N}-1} \bar\theta \, \left[
\bar D^\mathcal{N} D_\mathcal{N} \mathcal{L}({\Phi},\bar{\Phi})
\right] \bigg|_{\theta^{\cal N} = \bar\theta_{\cal N} = 0} \, ,
\label{truncation}
\ee
with the covariant derivatives $D_\mathcal{N}$ and $\bar D^\mathcal{N}$ defined
in \re{Cov-der}. Subsequently applying \re{truncation}, one can rewrite the
action of the $\mathcal{N}=4$ model in terms of the $\mathcal{N}=2$ light-cone
Yang-Mills chiral superfield ${\Phi}^{(2)} (x, \theta^A,
\bar\theta_A)$ coupled to the $\mathcal{N}=2$ Wess-Zumino chiral superfield
${\Psi}^{(2)}_{\scriptscriptstyle \rm WZ} (x, \theta^A,
\bar\theta_A)$
\be
{\Phi}^{(2)} = {\Phi}^{(4)} (x, \theta^A , \bar\theta_A) \bigg|_{\theta^3 =
\bar\theta_3 = 0 \atop \theta^4 = \bar\theta_4 = 0} \, , \qquad
{\Psi}^{(2)}_{\scriptscriptstyle \rm WZ} = D_3\, {\Phi}^{(4)}(x, \theta^A ,
\bar\theta_A) \bigg|_{\theta^3 = \bar\theta_3 = 0 \atop \theta^4 =
\bar\theta_4 = 0} \, .
\label{N=4->2}
\ee
Here the superscript refers to the underlying $\mathcal{N}-$extended SYM and
${\Phi}^{(4)}$ is given by \re{N=4-field}. The conjugated (antichiral)
superfields are
\be
\bar{\Phi}^{(2)} = \bar{\Phi}^{(4)} (x, \theta^A , \bar\theta_A)
\bigg|_{\theta^3 = \bar\theta_3 = 0 \atop \theta^4 = \bar\theta_4 = 0} \, ,
\qquad \bar{\Psi}^{(2)}_{\scriptscriptstyle \rm WZ} = \bar D^3\,
\bar{\Phi}^{(4)}(x, \theta^A ,
\bar\theta_A) \bigg|_{\theta^3 = \bar\theta_3 = 0 \atop \theta^4 = \bar\theta_4 =
0} \, .
\ee
Expansion of the $\mathcal{N}=4$ chiral superfield \re{N=4-field} over the
$\mathcal{N}=2$ chiral (${\Phi}^{(2)}$, ${\Psi}^{(2)}_{\scriptscriptstyle \rm
WZ}$) and antichiral ($\bar{\Phi}^{(2)}$, $\bar{\Psi}^{(2)}_{\scriptscriptstyle
\rm WZ}$) superfields looks as follows
\be
{\Phi}^{(4)} (x, \theta^A , \bar\theta_A)={\rm e}^{\frac12 (\bar\theta_3
\theta^3+\bar\theta_4 \theta^4) \, \partial_+} \bigg\{ {\Phi}^{(2)}+\theta^3
{\Psi}^{(2)}_{\scriptscriptstyle \rm WZ}-\theta^4\,
\partial_+^{-1} \bar D^1\bar D^2 \bar{\Psi}^{(2)}_{\scriptscriptstyle \rm WZ}
-\theta^3\theta^4\bar D^1\bar D^2\bar{\Phi}^{(2)}\bigg\}\, .
\label{Phi2-from-Phi4}
\ee
Substitution of this relation into \re{truncation} yields the action \re{N=4-S}
rewritten as the $\mathcal{N}=2$ SYM theory coupled to the $\mathcal{N}=2$
Wess-Zumino multiplet. To obtain the light-cone formulation of the
$\mathcal{N}=2$ SYM theory it suffices to put ${\Psi}^{(2)}_{\scriptscriptstyle
\rm WZ}=\bar{\Psi}^{(2)}_{\scriptscriptstyle \rm WZ}=0$. In this way, one finds
the $\mathcal{N}=2$ action as \cite{BriTol83}
\ba
S_{\mathcal{N}=2} = \int d^4 x \, d^2 \theta \, d^2\bar\theta \bigg\{
\!\!\!&-&\!\!\!
\bar{\Phi}^a{\Box}{\Phi}^a
+ 2 g f^{abc} \lr{
\partial_+ {\Phi}^a \bar{\Phi}^b \bar\partial {\Phi}^c
+
\partial_+ \bar{\Phi}^a {\Phi}^b \partial \bar{\Phi}^c
}
\nonumber \\
&-&\!\!\! 2 g^2 f^{abc}f^{ade} \frac{1}{\partial_+}\lr{\partial_+{\Phi}^b
\bar D^1 \bar D^2 \bar{\Phi}^c}\frac{1}{\partial_+}
\lr{ \partial_+\bar{\Phi}^d D_1 D_2 {\Phi}^e } \bigg\} \, ,
\label{N=2-S}
\ea
where ${\Phi}\equiv {\Phi}^{(2)}(x, \theta^A , \bar\theta_A)$ is a complex chiral
$\mathcal{N}=2$ superfield and $\bar{\Phi}$ is a conjugated antichiral
superfield. Substituting \re{N=4-field} into \re{N=4->2} one gets
\be
{\Phi} (x , \theta^A , \bar\theta_A) = {\rm e}^{\frac12 \bar\theta \cdot \theta
\, \partial_+} \bigg\{
\partial_+^{-1} A(x)
+ \theta^A \partial_+^{-1}\bar\lambda_A (x) + \frac{i}{2!} \varepsilon_{AB}
\theta^A \theta^B \bar \phi (x) \bigg\} \, ,
\label{N=2-field}
\ee
with $\bar \phi \equiv \bar \phi_{12}(x)$ and ${\scriptstyle A,B} = 1,2$. The
antichiral superfield $\bar{\Phi} (x, \theta^A,
\bar\theta_A)$ involves the fields $\bar A(x)$, $\lambda^A$ and $\phi$,
and in distinction with the $\mathcal{N=}4$ model, it is independent on the
chiral superfield ${\Phi}^{(2)}(x, \theta^A , \bar\theta_A)$.

The propagating fields in the $\mathcal{N}=2$ theory \re{N=2-S} are the complex
gauge field $A(x)$, one complex scalar field $\phi(x)$ and two complex Grassmann
fields $\lambda^A(x)$ (${\scriptstyle A}=1,2$) describing ``good'' components of
two Majorana fermions. By construction, the $\mathcal{N}=2$ SYM action differs
from the $\mathcal{N}=4$ SYM action by the contribution of the Wess-Zumino
superfield ${\Psi}^{(2)}_{\scriptscriptstyle \rm WZ}(x, \theta^A ,
\bar\theta_A)$. Had we retained this superfield, the two theories would be
equivalent. For ${\Psi}^{(2)}_{\scriptscriptstyle \rm WZ}=0$, the properties of
the theory are changed drastically: the $\mathcal{N}=2$ SYM acquires a
nonvanishing $\beta-$function and its conformal symmetry is broken on the quantum
level.

\subsubsection{$\mathcal{N}=1$ theory}

As a next step, one applies \re{truncation} to truncate the $\mathcal{N}=2$ down
to $\mathcal{N}=1$ SYM. Similar to the previous case, one defines two chiral
superfields
\be
{\Phi}^{(1)} = {\Phi}^{(2)} (x, \theta^A ,
\bar\theta_A) |_{\theta^2=\bar\theta_2=0}\,,\qquad
{\Psi}^{(1)}_{\scriptscriptstyle \rm WZ} = D_2 {\Phi}^{(2)} (x, \theta^A ,
\bar\theta_A)|_{\theta^2 =
\bar\theta_2 = 0}
\ee
and puts ${\Psi}^{(1)}_{\scriptscriptstyle \rm WZ} = 0$ to retain  the
contribution of the $\mathcal{N}=1$ SYM superfield only. This leads to
\ba
S_{\mathcal{N}=1} = \int d^4 x \, d \theta \, d \bar\theta \bigg\{
\bar{\Phi}^a{\Box}\, \partial_+ {\Phi}^a
\!\!\!&+&\!\!\! 2 g f^{abc} \lr{
\partial_+{\Phi}^a \partial_+\bar{\Phi}^b\bar\partial{\Phi}^c
-
\partial_+\bar{\Phi}^a \partial_+{\Phi}^b\partial\bar{\Phi}^c
}
\nonumber\\
&+&\!\!\! 2 g^2 f^{abc}f^{ade} \frac{1}{\partial_+} \lr{
\partial_+{\Phi}^b
\bar D^1\partial_+\bar{\Phi}^c}\frac{1}{\partial_+}
\lr{\partial_+\bar{\Phi}^d D_1\partial_+{\Phi}^e} \bigg\} \, , \ \
\label{N=1-S}
\ea
where the $\mathcal{N}=1$ light-cone chiral superfield ${\Phi} \equiv
{\Phi}^{(1)}(x, \theta, \bar\theta)$ is given by
\be
{\Phi} (x ,  \theta , \bar\theta ) = {\rm e}^{\frac12  \bar\theta \theta
\partial_+} \bigg\{
\partial_+^{-1}A(x) + \theta\, \partial_+^{-1} \bar\lambda  (x)
\bigg\} \, .
\label{N=1-field}
\ee
Here $\bar\lambda = \bar\lambda_1(x)$, and $\bar{\Phi}=({\Phi} (x , \theta ,
\bar\theta ))^*$ is a conjugated, antichiral $\mathcal{N}=1$ superfield. In the
$\mathcal{N}=1$ light-cone action \re{N=1-S}, the propagating fields are the
complex gauge field $A(x)$ and one complex Grassmann field $\lambda(x)$
describing the ``good'' component of Majorana fermion.

\subsubsection{$\mathcal{N}=0$ theory}

Finally, we use \re{truncation} to truncate the $\mathcal{N}=1$ theory down to
$\mathcal{N}=0$ Yang-Mills theory. The resulting light-cone action takes the form
\ba
S_{\mathcal{N}=0} = \int d^4 x \bigg\{
\bar{\Phi}^a{\Box} \, \partial_+^2 {\Phi}^a
\!\!\!&-&\!\!\! 2 g f^{abc} \lr{
\partial_+{\Phi}^a \partial_+^2 \bar{\Phi}^b \bar\partial {\Phi}^c
+
\partial_+\bar{\Phi}^a \partial_+^2{\Phi}^b\partial\bar{\Phi}^c
}
\nonumber\\
&-&\!\!\! 2 g^2 f^{abc} f^{ade} \frac{1}{\partial_+} \lr{
\partial_+ {\Phi}^b
\partial_+^2 \bar{\Phi}^c
} \frac{1}{\partial_+} \lr{
\partial_+\bar{\Phi}^d \partial_+^2{\Phi}^e
} \bigg\} \, ,
\label{N=0-S}
\ea
where the $\mathcal{N}=0$ field is given by
\be
{\Phi} (x) = {\Phi}^{(1)}(x, \theta , \bar\theta )|_{\theta =
\bar\theta = 0} =
\partial_+^{-1} A(x)
\, ,
\label{N=0-field}
\ee
and $\bar{\Phi} (x)=\partial_+^{-1}\bar A(x)$.  Eq.~\re{N=0-S} coincides with the
well-known expression for the light-cone action of $SU(N_c)$
gluodynamics~\cite{DerKir01}.

\subsection{Mandelstam formalism}

In the Brink-Lindgren-Nilsson formalism, the light-cone action $S_\mathcal{N} =
\int d^4 x \, d^\mathcal{N} \theta \, d^\mathcal{N} \bar\theta \,
\mathcal{L}_{\scriptscriptstyle\rm BLN} ({\Phi},\bar{\Phi})$ involves both
chiral and antichiral superfields. The same action can be rewritten in terms of
chiral superfields only, without any reference to the conjugated
$\bar\theta_A-$variables. To this end, one trades the antichiral superfield
$\bar{\Phi}$ for yet another chiral superfield
\be
\label{new-field}
\bar{\Phi}(x,\theta^A,\bar\theta_A)
= (-1)^{\mathcal{N}-1} \partial_+^{-2} {D_{ 1}} \ldots D_{\mathcal{N}} \, {\Psi}
( x, \theta^A ,\bar\theta_A) \, .
\ee
The inverse relation looks as
\be
{\Psi} (x, \theta^A,\bar\theta_A) = - \partial_+^{2 - \mathcal{N}}
\bar D^\mathcal{N}
\ldots
\bar D^1
\bar{\Phi}(x,\theta^A,\bar\theta_A)\,,
\label{Psi-from-Phi}
\ee
so that $\bar D^B\, {\Psi} (x, \theta^A,\bar\theta_A)=0$. Comparing
\re{new-field} with \re{N=4-bar-field} one finds that for $\mathcal{N}=4$,
${\Psi} (x, \theta^A,\bar\theta_A) = {\Phi}(x,\theta^A,\bar\theta_A)$. For
$\mathcal{N}\le 2$ the chiral superfields ${\Phi}(x,\theta^A,\bar\theta_A)$ and
${\Psi} (x,\theta^A,\bar\theta_A)$ are independent on each other. Their explicit
expressions are given below (see Eqs.~\re{M=0-field} -- \re{M=4-field}).

Making use of \re{new-field} one can rewrite the light-cone SYM actions defined
in the previous section in terms of chiral superfields ${\Phi}$ and ${\Psi}$.
In general, a chiral field satisfies the relation ${\Phi}
(x_\mu,\theta^A,\bar\theta_A) = {\Phi}(x_\mu + \ft12\bar\theta \cdot\theta\,
n_\mu,\theta^A,0)$, and, as a consequence, its $\bar\theta_A-$dependence can be
eliminated by substituting $x_\mu \to x_\mu - \ft12\bar\theta\cdot\theta
n_\mu$. Under this transformation the chiral superfields entering
$S_\mathcal{N}$ are replaced by the following expressions
\ba
\label{MandelstamPhi}
{\Phi}_{\scriptscriptstyle\rm M} (x, \theta^A) &=& {\rm e}^{- \ft12
\bar\theta\cdot\theta\partial_+} {\Phi}(x, \theta^A, \bar\theta_A) =
{\Phi}(x, \theta^A, 0)
\\
{\Psi}_{\scriptscriptstyle\rm M} (x, \theta^A) &=& {\rm e}^{- \ft12
\bar\theta\cdot\theta\partial_+} {\Psi}(x, \theta^A, \bar\theta_A) =
{\Psi}(x, \theta^A, 0) \, . \nonumber
\ea
In a similar manner, one redefines the covariant derivatives acting on new
superfields
\ba
D_{{\scriptscriptstyle\rm M},A} \!\!\!&=&\!\!\! {\rm e}^{-\ft12
\bar\theta\cdot\theta\partial_+} D_{A} {\rm e}^{\ft12
\bar\theta\cdot\theta\partial_+} =
\partial_{\theta^A} - \bar\theta_A \partial_+
\, , \nonumber \\
\bar D_{\scriptscriptstyle\rm M}{}^A\!\!\!&=&\!\!\! {\rm e}^{-\ft12
\bar\theta\cdot\theta\partial_+}
\bar D^{A}{\rm e}^{\ft12 \bar\theta\cdot\theta\partial_+}
=
\partial_{\bar\theta_A}
\, , \label{derivatives-M}
\ea
so that $\bar D_{\scriptscriptstyle\rm M}{}^B{\Phi}_{\scriptscriptstyle\rm M}
(x, \theta^A) =\bar D_{\scriptscriptstyle\rm
M}{}^B{\Psi}_{\scriptscriptstyle\rm M} (x, \theta^A)=0$. Then, one performs
integration over the $\bar\theta_A-$vari\-ables inside $S_\mathcal{N}$ and
obtains the light-cone SYM action in the Mandelstam formulation
\be
S_\mathcal{N} = \int d^4 x\, d^\mathcal{N}\theta \,
\mathcal{L}_{\scriptscriptstyle\rm M} ({\Phi}_{\scriptscriptstyle\rm
M},{\Psi}_{\scriptscriptstyle\rm M}) \, , \qquad
\mathcal{L}_{\scriptscriptstyle\rm M} = \int d^\mathcal{N} \bar\theta\,
\mathcal{L}_{\scriptscriptstyle\rm BLN} =
\partial_{\bar\theta_\mathcal{N}}
\ldots
\partial_{\bar\theta_1}
\mathcal{L}_{\scriptscriptstyle\rm BLN} \, .
\label{S-B2M}
\ee
It depends on the chiral superfields ${\Phi}_{\scriptscriptstyle\rm M}$ and
${\Psi}_{\scriptscriptstyle\rm M}$ and involves only ``half'' of odd variables.
From now on, we will suppress the subscript ``${\scriptscriptstyle\rm M}$'' on
the superfields and use only the Mandelstam fields throughout our subsequent
presentation. This will not lead to a confusion anyway, since the Lagrangian
$\mathcal{L}_{\scriptscriptstyle\rm M}$ is evaluated for $\bar\theta_A = 0$, so
that Eq.\ (\ref{MandelstamPhi}) is at work.

Combining together \re{new-field}, \re{MandelstamPhi} and \re{S-B2M} we find from
\re{N=4-S}, \re{N=2-S}, \re{N=1-S} and \re{N=0-S} that the resulting expression
for the light-cone action in the Mandelstam formalism can be written in the
following form
\begin{eqnarray}
S_{\mathcal{N}=0, 1, 2} \!\!\!&=&\!\!\! -\sigma_\mathcal{N} \int d^4 x \,
d^\mathcal{N}\theta\bigg( {\Psi}^a \Box {\Phi}^a + 2 g f^{abc}
\partial_+ {\Phi}^a
\bar\partial {\Phi}^b {\Psi}^c + 2g f^{abc} \partial_+^{2-\mathcal{N}} {\Phi}^a
 {}[
\partial_+^{-2}\partial {\Psi}^b,\, \partial_+^{-1} {\Psi}^c ]
\nonumber
\\
&&\qquad\qquad - 2(-1)^\mathcal{N} g^2 f^{abc} f^{ade} \partial_+^{-2} \left(
\partial_+ {\Phi}^b {\Psi}^c \right)  \left[\partial_+^{-1} {\Psi}^d
, \,
\partial_+^{2-\mathcal{N}}{\Phi}^e \right] \bigg)
\, ,
\label{N=0-M}
\end{eqnarray}
and
\begin{eqnarray}
S_{\mathcal{N}=4} \!\!\!&=&\!\!\! - \int d^4x\, d^4\theta \bigg( \frac12 {\Phi}^a
\Box {\Phi}^a + \frac23 g f^{abc} \partial_+ {\Phi}^a
\bar\partial {\Phi}^b {\Phi}^c + \frac23 g f^{abc}
\partial_+^{-1}{\Phi}^a
{}\left[\partial_+^{-2}{\Phi}^b, \,\partial\partial_+^{-2}{\Phi}^c\right]
\label{N=4-M}
\\
&&\qquad - \frac12 g^2 f^{abc} f^{ade} \left\{ \partial_+^{-2} \left(
{\Phi}^b\partial_+ {\Phi}^c \right)   \left[
\partial_+^{-2}{\Phi}^d ,\,\partial_+^{-1}{\Phi}^e \right] -
\frac12{\Phi}^b{\Phi}^d  \left[
\partial_+^{-2}{\Phi}^c,\,
\partial_+^{-2}{\Phi}^e \right] \right\} \bigg) \, . \nonumber
\end{eqnarray}
Here $\sigma_\mathcal{N}=(-1)^{\mathcal{N}(\mathcal{N}+1)/2}$ is the signature
factor and the notation was introduced for a ``square bracket''. For two
arbitrary superfields ${\Phi}_1(x,\theta^A)$ and ${\Phi}_2(x,\theta^A)$, it is
defined as (for $\mathcal{N}\ge 1$)
\be
 {}[ {\Phi}_1  , \, {\Phi}_2 ] =
\prod_{A = 1}^{\mathcal{N}}  \left( {\partial_{\theta^A}^{(1)}}
{\partial_+^{(2)}} - {\partial_{\theta^A}^{(2)}}{\partial_+^{(1)}} \right)
{\Phi}_1 \, {\Phi}_2 \, ,
\label{Q-op}
\ee
where the ordering of fermion derivatives is from the left to right, i.e.,
$\prod_{A=1}^{\mathcal{N}}\partial_{\theta^A} =\partial_{\theta^1}\ldots
\partial_{\theta^\mathcal{N}}$, and the superscript indicates the field to
which the derivative is applied. For $\mathcal{N}=0$ one has $[ {\Phi}_1 , \,
{\Phi}_2 ] = {\Phi}_1\, {\Phi}_2$.

Substituting \re{N=4-field}, \re{N=2-field}, \re{N=1-field} and \re{N=0-field}
into \re{Psi-from-Phi} and \re{MandelstamPhi}, one finds the explicit expressions
for the chiral superfields ${\Phi}(x,\theta^A)$ and ${\Psi}(x,\theta^A)$ in the
Mandelstam formulation
\be
{\Phi} (x) = \partial_+^{-1} A(x)\,,\qquad {\Psi} (x) = - \partial_+
\bar A(x)
\label{M=0-field}
\ee
for $\mathcal{N}=0$,
\ba
{\Phi}  (x ,  \theta ) &=&
\partial_+^{-1}A(x) + \theta\, \partial_+^{-1} \bar\lambda  (x)
\nonumber\\[2mm]
{\Psi}  (x, \theta) &=& - \lambda(x) + \theta\partial_+ \bar A(x)
\ea
for $\mathcal{N}=1$,
\ba
{\Phi} (x , \theta^A ) &=&
\partial_+^{-1} A(x)
+ \theta^A \partial_+^{-1}\bar\lambda_A (x) + \frac{i}{2!} \varepsilon_{AB}
\theta^A \theta^B \bar \phi (x)  \, ,
\nonumber\\
{\Psi} (x,\theta^A) &=& i \phi(x) - \varepsilon_{AB} \theta^A \lambda^B (x) +
\frac12 \varepsilon_{AB} \theta^A\theta^B \partial_+
\bar A (x) \, ,
\label{M=2-field}
\ea
for $\mathcal{N}=2$, and
\begin{eqnarray}
{\Phi} (x, \theta^A) &=& {\Psi} (x, \theta^A) =
\partial_+^{-1}A(x) +\theta^A
\partial_+^{-1}\bar\lambda_A (x) + \frac{i}{2!} \theta^A \theta^B \bar \phi_{AB}
(x)
\nonumber\\
&+&\!\!\! \frac{1}{3!} \varepsilon_{ABCD} \theta^A \theta^B \theta^C \lambda^D
(x) - \frac{1}{4!} \varepsilon_{ABCD} \theta^A \theta^B \theta^C \theta^D
\partial_+ \bar{A} (x)
\label{M=4-field}
\end{eqnarray}
for $\mathcal{N}=4$. The following comments are in order.

As we demonstrated in this section, the Brink-Lidgren-Nilsson and Mandelstam
formulations of the SYM theory on the light-cone are equivalent.%
\footnote{ Although the expressions for the light-cone action, Eqs.~\re{N=0-M}
and \re{N=4-M}, differ from those proposed by Mandelstam in Ref.~\cite{Man83}, we
demonstrate their equivalence in Appendix~A4.} In what follows we shall rely on
Eqs.~\re{N=0-M} and \re{N=4-M} since they are more suitable for our purposes.

In the Mandelstam formalism, for $\mathcal{N}\le 2$ chiral superfields
${\Phi}(x,\theta^A)$ and ${\Psi}(x,\theta^A)$ describe a half of the propagating
fields each.  Notice that the superfield ${\Psi} (x, \theta^A)$ is bosonic for
$\mathcal{N}=0, 2, 4$ and fermionic for $\mathcal{N}=1$. The important difference
between the superfields ${\Phi}(x,\theta^A)$ and ${\Psi}(x,\theta^A)$ is that the
former involves nonlocal fields, $\partial_+^{-1}A$ and
$\partial_+^{-1}\bar\lambda$, whereas the latter contains only local primary
fields: scalars, $\phi$, fermions, $\lambda^A$, and gauge strength tensor
projected onto the light-cone, $n^\mu F_{\mu\perp}=(\partial_+ A,\partial_+
\bar A)$ in the axial gauge $(n\cdot A)\equiv A_+(x)=0$. For $\mathcal{N}\le 2$,
one could have avoided nonlocal operators from the very beginning if the SYM
theory were reformulated in terms of two superfields, chiral ${\Psi}(x,\theta^A)$
and antichiral $\bar{\Psi}(x,\bar\theta_A)=({\Psi}(x,\theta^A))^\dagger$, by
making use of the relation
\be
{\Phi}(x,\theta^A)=-(-i)^{\mathcal{N}}\partial_+^{-2}\partial_{\bar\theta_\mathcal{N}}
\ldots\partial_{\bar\theta_1}\bar{\Psi}(x-\bar\theta\cdot\theta,\bar\theta_A)\,,
\label{bad}
\ee
which follows from \re{new-field}, \re{MandelstamPhi} and \re{derivatives-M}. The
reason why we prefer to deal with the superfields ${\Phi}(x,\theta^A)$ and
${\Psi}(x,\theta^A)$ is that substitution of \re{bad} into \re{N=0-M} will break
invariance of the light-cone action under translations in the superspace and, as
a consequence, the resulting expression for the dilatation operator acting on the
light-cone operators involving antichiral superfield $\bar{\Psi}(x,\bar\theta_A)$
is more complicated.

\section{Superconformal invariance on the light-cone}

The $\mathcal{N}-$extended SYM theory is invariant on the classical level under
superconformal $SU(2,2|\mathcal{N})$ transformations. They include
\begin{itemize}
\item Conformal $SO(4,2)$ symmetry generated by translations $\mathbf{P}_\mu$,
Lorentz transformations $\mathbf{M}_{\mu\nu}$, dilatations $\mathbf{D}$ and
special conformal transformations $\mathbf{K}_\mu$;

\item Poincar\'e supersymmetry generated by the supercharges $\mathbf{Q}_{\alpha
A}$ and their conjugates $\mathbf{\bar Q}^{\dot\alpha A}$;

\item Conformal supersymmetry generated by the supercharges $\mathbf{S}^A_\alpha$
and their conjugates $\mathbf{\bar S}_{A}^{\dot\alpha}$;

\item $R-$symmetry generated by the bosonic chiral charge $\mathbf{R}$, and, in
case of extended $N\ge 2$ supersymmetry, isotopic $SU(\mathcal{N})$ symmetry
generated by charges ${\mathbf{T}_A}^B$ satisfying the $SU(\mathcal{N})$
commutation relations.
\end{itemize}
The odd charges $\mathbf{Q}_{\alpha A}$, $\mathbf{\bar Q}^{\dot\alpha A}$,
$\mathbf{S}^A_\alpha$ and $\mathbf{\bar S}_{A}^{\dot\alpha}$ are two-dimensional
Weyl spinors ($\alpha=1,2$ and ${\scriptstyle A}=1,\ldots,\mathcal{N}$). On the
quantum level, the superconformal symmetry is broken in $\mathcal{N}=0$,
$\mathcal{N}=1$ and $\mathcal{N}=2$ SYM. In the $\mathcal{N}=4$ SYM theory, it
survives to all loops but is reduced due to a $U_R(1)-$anomaly down to the
$PSU(2,2|4)$ group. The symmetry breaking effects manifest themselves starting
from two-loops and, therefore, the one-loop dilatation operator in the
$\mathcal{N}-$extended SYM enjoys the full $SU(2,2|\mathcal{N})$ symmetry.

The superfield ${\Phi}(x,\theta^A)$ (and ${\Psi}(x,\theta^A)$) realizes a
representation of the superconformal algebra. Its infinitesimal variations under
the $SU(2,2|\mathcal{N})$ transformations look as
\be
\delta_{G}\, {\Phi}(x,\theta^A) = i [ {\Phi}(x,\theta^A), \mathbf{G} ] = -  G\,
{\Phi}(x,\theta^A)\,,
\label{G-rep}
\ee
where $\mathbf{G}= \varepsilon^\mu \mathbf{P}_\mu,\ \varepsilon^{\mu\nu}
\mathbf{M}_{\mu\nu},\ldots $ and $\mathbf{G}= \xi^{\alpha A} \mathbf{Q}_{\alpha
A}, \ \chi^{\alpha}_A \,\mathbf{S}_{\alpha}^A,\ldots $ for odd generators with
$\xi^{\alpha A}$, $\chi^{\alpha}_A$ being constant Grassman-valued Weyl spinors.
In \re{G-rep}, the quantum-field operator $\mathbf{G}$ is represented by an
operator $G$ acting on the superfield. In the light-cone formalism, the
$SU(2,2|\mathcal{N})$ charges can be split into ``kinematical'' and ``dynamical''
charges. For the former, the operator $G$ is given by linear differential
operators acting on even and odd coordinates of the superfield, while for the
latter it is realized nonlinearly and, in general, does not preserve the
number of superfields~\cite{BriTol83}.%
\footnote{Since nonlinear terms are accompanied by powers of the coupling
constant, they do not intervene to the lowest order.}

\subsection{Collinear supergroup}

In this paper, we shall calculate the one-loop dilatation operator in the
$\mathcal{N}-$extended SYM theory, acting on single-trace operators built from
chiral superfields ${\Phi}(z n^\mu,\theta^A)$ and ${\Psi}(z n^\mu,\theta^A)$ both
located on the light-cone along the $n-$direction ($n_\mu^2=0$)
\be
\mathbb{O}(Z_1,Z_2,\ldots,Z_L) = \tr\{{\Phi}(Z_1){\Psi}(Z_2)\ldots {\Phi}(Z_L)
\}\,.
\label{O-def}
\ee
Hereafter, we shall use a short-hand notation for the arguments of superfields on
the light-cone, ${\Phi}(z_k n^\mu,\theta_k^A)\equiv {\Phi}(Z_k)$ where
$Z_k=(z_k,\theta_k^A)$ specifies the position of the $k$th superfield in the
superspace. We recall that in the $\mathcal{N}=4$ SYM theory we have only one
operator \re{Chiral-sector}, while for $\mathcal{N}\le 2$ one has to distinguish
three different sets, Eqs.~\re{Chiral-sector} -- \re{Mixed-sector}. The operators
\re{O-def} are generating functions of local composite operators in the
underlying $\mathcal{N}-$extended SYM theory. The latter operators can be
obtained from \re{O-def} by substituting the superfields $\Phi_{j}=
\{\Phi,\Psi\}$ by their expansion around the origin in the superspace
\be
{\Phi}_{j}(Z) = {\Phi}_{j}(0) + Z\cdot \partial_Z {\Phi}_{j}(0)+\frac12(Z\cdot
\partial_Z)^2 {\Phi}_{j}(0) + \ldots\,,
\ee
where $Z\cdot \partial_Z = z\partial_z + \theta^A
\partial_{\theta^A}$.

The superfields in \re{EQ} are located on the light-cone along the
`$+$'-direction defined by the light-like vector $n_\mu$. To work out the
restrictions on $\mathbb{H}$ due to superconformal invariance, we have to
restrict ourselves to the superconformal transformations \re{G-rep} that map the
light-cone operators \re{EQ} into itself. It is well-known that in
nonsupersymmetric Yang-Mills theories such transformations correspond to the
so-called collinear $SL(2)$ subgroup of the conformal $SO(4,2)$ group~(see the
review~\cite{BraKorMul03}). They are generated by the charges ${\mathbf P}_+$,
${\mathbf M}_{-+}$, ${\mathbf D}$ and ${\mathbf K}_-$ which form the $SL(2)$
algebra.

Supersymmetry enlarges the $SL(2)$ subgroup. Examining the $SU(2,2|\mathcal{N})$
commutation relations one finds that the resulting collinear superalgebra
involves the additional charges: the $U(1)$ chiral charge ${\mathbf R}$, the
$SU(\mathcal{N})$ charges ${\mathbf T}_A{}^B$, helicity operator ${\mathbf
M}_{12}$ and the ``odd'' charges $\mathbf{Q}_{+  A}$ , $\mathbf{\bar Q}_{+}^{
A}$, ${\mathbf S}_{-}^A$ and $\mathbf{\bar S}_{- A}$.%
\footnote{Here the $+/-$ subscript indicates ``good''/``bad'' components of the
corresponding Weyl spinors, $\mathbf{Q}_{\alpha A}$, $\mathbf{\bar Q}^{\dot\alpha
A}$, $\mathbf{S}^A_\alpha$ and $\mathbf{\bar S}_{A}^{\dot\alpha}$ (see
Appendix~A2 for the definition).} In the light-cone formalism, such one-component
spinors are described by a complex Grassmann field without any Lorentz index.
Introducing
linear combinations of the charges
\ba
&
\begin{array}{llll}
  \mathbf{L}^-=-i \mathbf{P}_+\,,\quad
  & \mathbf{L}^+=\frac{i}2 \mathbf{K}_-\,,\quad
  & \mathbf{L}^0=\frac{i}2 ( \mathbf{D} + \mathbf{M}_{-+} )\,,\quad
  & \mathbf{E}=i ( \mathbf{D} - \mathbf{M}_{-+} )\,,
  \\[5mm]
  \mathbf{V}_{A}^-= \frac{i\varrho}{2} \mathbf{Q}_{+ A}\,,
  & {\mathbf{W}}^{A,-} =-\frac{\varrho}2 \mathbf{\bar Q}{}_{+}^A\,,
  & \mathbf{W}^{A,+}=-\frac{i}{2\varrho}   \mathbf{S}_-^{A}\,,
  & {\mathbf{V}}_{A}^+=\frac1{2\varrho} \mathbf{{\bar S}}{}_{-A}\,, \\
\end{array}
& \nonumber
\\[4mm]
& \mathbf{B}=\ft14 (1 - \ft{4}{{\cal N}}) \mathbf{R} + \ft12   \mathbf{M}_{12}\,,
&
\label{collinear}
\ea
one finds that together with ${\mathbf T}_A{}^B$ they satisfy the
$SL(2|\mathcal{N})$ (graded) commutation relations. In Eq.~\re{collinear} the
normalization factor $\varrho=2^{1/4}$ was introduced to bring these relations to
their canonical form~\cite{FraSorSci}. To save space we do not display them here.

Using the technique of induced representations~\cite{GatGriRocSie83,Sohnius85},
one can obtain representation of the generators of the collinear superalgebra
\re{collinear} for a general chiral superfield ${\Phi} (z n_\mu, \theta^A )$. The
relevant center elements of the superalgebra are
\begin{eqnarray}
\begin{array}{ll}
{}[{\mathbf M}_{-+} , {\Phi} (0,0)] = -is\, {\Phi} (0,0) \, , & \quad {}[{\mathbf
D} , {\Phi} (0,0)] = - i \ell\, {\Phi} (0,0)
\, , \\[3mm]
{}[{\mathbf M}_{12} , {\Phi} (0,0)] =  h\, {\Phi} (0,0)
 \,
, & \quad {}[{\mathbf R} , {\Phi} (0,0)] = r\, {\Phi} (0,0) \, ,
\end{array}
\end{eqnarray}
where $\ell$, $s$, $h$ and $n$ are correspondingly the canonical dimension  of
the superfield, projection of its spin on the `$+$'- direction, its helicity and
its $R-$charge. This leads to
\be
{}[{\mathbf L}^0 , {\Phi} (0,0)] = j\, {\Phi} (0,0)\,,\qquad {}[{\mathbf E} ,
{\Phi} (0,0)] = t\, {\Phi} (0,0)\,,\qquad {}[{\mathbf B} , {\Phi} (0,0)] = b\,
{\Phi} (0,0)
\label{parameters}
\ee
where $j = \ft12 (s + \ell)$ is the {\sl conformal spin},  $t = \ell - s$ is the
{\sl twist} and $b=\ft14 (1 - \ft{4}{{\cal N}}) r + \ft12 h$ is the $B-$charge of
the superfield~\cite{BraKorMul03,Sohnius85}. For the chiral fields the charges
$b$ and $j$ are related as~\cite{Sohnius85}
\be
b=-j\,.
\ee
The parameters $j$ and $t$ define the so-called ``atypical'' representation of
the collinear $SL(2|\mathcal{N})$ supergroup that we shall denote as
$\mathcal{V}_{j}$. In this representation, the charges \re{collinear} are
realized as differential operators acting on the light-cone coordinates of the
chiral superfield ${\Phi} (z n_\mu, \theta^A )$
\begin{equation}
\label{sl2}
\begin{array}{llll}
{L}^- = -\partial_z \, , \ & {L}^+ = 2 j\, z + z^2\partial_z + z \left(
\theta\cdot \partial_\theta \right) \, , \ & {L}^0 = j + z
\partial_z + \ft12\left( \theta\cdot \partial_\theta \right)
\, , \ & {E}= t \, , \\ [4mm] {W}{}^{A,-} = \theta^A \, \partial_z \, , \ &
{W}{}^{A,+} = \theta^A [ 2j  +  z \partial_z +  \left( \theta\cdot
\partial_\theta \right) ] \, , \ & {V}^-_{A} = \partial_{\theta^A} \, , \ &
{V}^+_{A} = z\partial_{\theta^A} , \!\!\!\\ [4mm] \, \ & {T}_B{}^A = \theta^A
\partial_{\theta^B} - \ft1{\mathcal{N}} \, \delta_B^A \left( \theta\cdot
\partial_\theta \right) \, , \ & {B} = - j - \ft12 \left( 1 - \ft{2}{{\cal N}}
\right) \left( \theta\cdot \partial_\theta \right) \, , \ &
\end{array}
\end{equation}
where $\partial_z \equiv \partial/\partial z$ and $\theta \cdot \partial_\theta
\equiv \theta^A \partial/\partial\theta^A$.

Let us identify the values of the conformal spin, $j$, and twist, $t$, for the
chiral superfields ${\Phi}(x,\theta^A)$ and ${\Psi}(x,\theta^A)$ in Mandelstam
formulation, Eqs.~\re{M=0-field} -- \re{M=4-field}. According to \re{parameters},
they are determined by the properties of the lowest component of the superfields,
${\Phi}(0,0)=\partial_+^{-1} A(0)$ and ${\Psi}(0,0)=-
\partial_+ \bar A(0)\,, - \lambda(0)\,,i\phi(0)$ for $\mathcal{N}=0,1,2$, respectively.
Therefore, for the scalar chiral superfield ${\Phi}(x,\theta^A)$ one has
$\ell=r=0$, $s=-1$ and $h=1$ leading to
\be
j_\Phi = - \frac12 \, , \qquad t_\Phi = 1
\label{j=-1/2}
\ee
independently on $\mathcal{N}$. Similarly, for the chiral superfield
${\Psi}(x,\theta^A)$ one gets $\ell=2-\mathcal{N}/2$, $s=-h=1-\mathcal{N}/2$ and
$r=\mathcal{N}$ leading to
\be
j_\Psi =  \frac{3-\mathcal{N}}2 \, , \qquad t_\Psi = 1\,.
\label{Psi-j}
\ee
We see that the chiral superfields ${\Phi}(x,\theta^A)$ and ${\Psi}(x,\theta^A)$
have the same twist $t=1$ but different conformal spins. Notice that $j_\Psi \ge
1/2$ for $\mathcal{N}\le 2$ while $j_\Phi$ is negative for all $\mathcal{N}$. As
we will show below, this difference has important consequences for the properties
of the dilatation operator.

For the nonlocal light-cone operators \re{O-def}, the generators of the
superconformal $SL(2|\mathcal{N})$ transformations act on the tensor product of
the atypical representations $\mathcal{V}_{j_\Psi}$ and  $\mathcal{V}_{j_\Phi}$
corresponding to constituent superfields
\be
\mathcal{V}_L =\mathcal{V}_{j_\Phi}\otimes \mathcal{V}_{j_\Psi}\otimes\cdots
\otimes\mathcal{V}_{j_\Phi}\,.
\label{V-quantum}
\ee
They are given by the sum of differential operators \re{sl2} acting on the
coordinates of the superfields, $Z_k=(z_k,\theta^A_k)$ with $j=j_\Psi$ or
$j=j_\Phi$ depending on the superfield. Since the twist generator $E$ in \re{sl2}
is a c-number, the twist of the nonlocal operator \re{O-def} is equal to the sum
of twists of the superfields leading to $t_\mathbb{O}=L$. Obviously, the local
composite operators generated by $\mathbb{O}(Z_1,\ldots,Z_L)$ have the same
twist. Such operators are known as quasipartonic operators. In a general
classification of local operators, they carry a maximal Lorentz spin and have a
minimal possible twist.

The superconformal invariance implies that the evolution equation \re{EQ} has to
be invariant under the $SL(2|\mathcal{N})$ transformations of the superfields.
For a general light-cone superfield ${\Phi}_j(Z)\equiv {\Phi}_j(z,\theta^A)$ with
the conformal spin $j$, these transformations are generated by the operators
\re{sl2}: The operators ${L}^-$, ${L}^+$ and ${L}^0$ generate projective, $SL(2)$
transformations on the light-cone
\ba
&& \e^{\epsilon L^-} {\Phi}_j(Z) = {\Phi}_j(z-\epsilon,\theta^A)\,,
\nonumber \\[3mm]
&& \e^{\epsilon L^0} {\Phi}_j(Z) = \e^{j \epsilon}{\Phi}_j(\e^{j \epsilon }z
,\e^{j \epsilon/2}\theta^A)\,,
\nonumber \\
&& \e^{\epsilon L^+} {\Phi}_j(Z) = (1-\epsilon
z)^{-2j}{\Phi}_j\left(\frac{z}{1-\epsilon z},\frac{\theta^A}{1-\epsilon
z}\right)\,.
\label{SL2-subgroup}
\ea
The operators ${W}{}^{A,-}$ and ${V}^-_{A}$ generate translations in the
superspace and correspond to supersymmetric transformations of the components of
the superfield
\ba
 \e^{ \xi\cdot {V}^- } {\Phi}_j(Z)  &=& {\Phi}_j(z ,\theta^A+\xi^A)\,,\qquad
\nonumber
\\
 \e^{\xi\cdot {W}^- } {\Phi}_j(Z)  &=& {\Phi}_j(z+\xi\cdot \theta ,\theta^A)\,.
\label{V-minus}
\ea
The operators ${W}{}^{A,+}$ and ${V}^+_{A}$ generate conformal transformations in
the superspace
\ba
\e^{ \xi\cdot {V}^+ } {\Phi}_j(Z) &=& {\Phi}_j(z,\theta^A+z\xi^A)\,,\qquad
\nonumber \\
\e^{ \xi\cdot {W}^+ } {\Phi}_j(Z) &=&
(1-\xi\cdot\theta)^{-2j}{\Phi}_j\left(\frac{z
}{1-\xi\cdot\theta},\frac{\theta^A}{1-\xi\cdot\theta}\right)\,.
\label{V-plus}
\ea
Then, the evolution equation \re{EQ} is invariant under supersymmetric
transformations of the superfields ${\Phi}(Z)$ and ${\Psi}(Z)$ provided that the
Hamiltonian $\mathbb{H}$ commutes with the $SL(2|\mathcal{N})$ generators
\be
 \mathbb{H} \cdot   G\, \mathbb{O}(Z_1,\ldots,Z_L)
 =   G \cdot \mathbb{H} \, \mathbb{O}(Z_1,\ldots,Z_L)
\label{Superconform}
\ee
where $ G=\{L^0,L^\pm,V_A^\pm, W^{A,\pm}, T_B{}^A, B\}$ are the
$SL(2|\mathcal{N})$ generators acting on the tensor product \re{V-quantum}, that
is $ G=\sum_{k=1}^L  G_k$ with $ G_k$ given by the differential operators
\re{sl2} acting on the coordinates of the $k$th superfield. Substituting
\re{H-multi-color} into \re{Superconform} one finds that the two-particle kernel
$\mathbb{H}_{k,k+1}$ has to be an $SL(2|\mathcal{N})$ invariant operator
\be
 [ \mathbb{H}_{k,k+1}, G_k + G_{k+1}] = 0\,.
\label{2PI-constraint}
\ee
In the next section, we present a general expression for the operator
$\mathbb{H}_{k,k+1}$ satisfying \re{2PI-constraint}.

\subsection{The $SL(2|\mathcal{N})$ invariant operators}

By definition, the two-particle kernel $\mathbb{H}_{12}$ governs the scale
dependence of the product of two chiral superfields ${\Phi}_{j_1}(Z_1)
{\Phi}_{j_2}(Z_2)$ carrying the conformal spins $j_1$ and $j_2$. As before,
${\Phi}_{j}(Z)$ stands for the superfields in an $\mathcal{N}-$extended SYM
theory, ${\Phi}(Z)$ and ${\Psi}(Z)$, with the conformal spins $j_\Phi=-1/2$ and
$j_\Psi=(3-\mathcal{N})/2$, respectively.

As we will demonstrate in Sect.~4, to one-loop order the operator
$\mathbb{H}_{12}$ does not change the number of superfields and can be realized
as a quantum mechanical Hamiltonian acting on the coordinates of the superfields,
$Z_1$ and $Z_2$. In addition, if the superfields are not identical,
$\mathbb{H}_{12}$ may exchange the superfields inside the single trace,
Eq.~\re{O-def}. Therefore, defining the two-particle kernel $\mathbb{H}_{12}$ one
has to distinguish two different channels ${\Phi}_{j_1}{\Phi}_{j_2} \to
{\Phi}_{j_1}{\Phi}_{j_2}$ and ${\Phi}_{j_1}{\Phi}_{j_2} \to
{\Phi}_{j_2}{\Phi}_{j_1}$. Let us denote the corresponding evolution kernels as
$\mathbb{V}^{(j_1j_2)}$ and $\mathbb{V}^{(j_1j_2)}_{\rm ex}$, respectively. By
definition, they act on the tensor product of two $SL(2|\mathcal{N})$ chiral (or
atypical) representations as
\be
\mathbb{V}^{(j_1j_2)}:\quad \mathcal{V}_{j_1} \otimes \mathcal{V}_{j_2} \to
\mathcal{V}_{j_1} \otimes \mathcal{V}_{j_2} \,,\qquad \mathbb{V}^{(j_1j_2)}_{\rm
ex}:\quad  \mathcal{V}_{j_1} \otimes \mathcal{V}_{j_2} \to \mathcal{V}_{j_2}
\otimes \mathcal{V}_{j_1}\,.
\label{V-times-V}
\ee
We shall argue below that the $SL(2|\mathcal{N})$ invariance fixes these
operators up to a scalar function. To this end, we will make use of the $SL(2)$
subgroup of the full superconformal group generated by the operators $L^0$, $L^+$
and $L^-$, Eqs.~\re{SL2-subgroup} to construct the $SL(2)$ invariant operators
$\mathbb{V}^{(j_1j_2)}$ and $\mathbb{V}^{(j_1j_2)}_{\rm ex}$ and, then,
generalize them to ensure invariance under the $SL(2|\mathcal{N})$
transformations.

Additional complication arises due to the fact that the $SL(2|\mathcal{N})$
representation $\mathcal{V}_{j}$ is reducible for $j=-1/2$, that is, for the
superfield ${\Phi}(Z)$, Eq.~\re{j=-1/2}. We shall assume for the moment that the
representations $\mathcal{V}_{j_1}$ and $\mathcal{V}_{j_2}$ are irreducible in
\re{V-times-V} and extend analysis to the spin $j=-1/2$ representations in
Sect.~3.3.

\subsubsection{The $SL(2)$ invariant operators}

Let us consider a nonlocal light-cone operator built from two chiral superfields
${\Phi}_j(z,\theta^A=0)$ ``living'' along the $z-$axis in the superspace
\be
\mathbb{O}_{j_1j_2}(z_1,z_2) = {\Phi}_{j_1}(z_1,0){\Phi}_{j_2}(z_2,0)\,.
\label{O-SL2}
\ee
According to \re{SL2-subgroup}, the superfield ${\Phi}_j(z,0)$ is transformed
under the $SL(2;\mathbb{R})$ transformations as
\be
z\to \frac{az+b}{cz+d}\,,\qquad {\Phi}_j(z,0) \to (cz+d)^{-2j}
{\Phi}_j\left(\frac{az+b}{cz+d},0\right)
\label{SL2-transformations}
\ee
with $ad-bc=1$. The generators of these transformations are
\be
{l}_j^- = -\partial_z \, , \qquad {l}_j^+ = 2 j\, z + z^2\partial_z \, , \qquad
{l}_j^0 = j + z
\partial_z\,.
\label{sl2-generators}
\ee
They are obtained from the generators $L^0$, $L^+$ and $L^-$, Eq.~\re{sl2}, by
neglecting terms involving $\theta-$variables. The $SL(2|\mathcal{N})$ invariant
kernels  $\mathbb{V}^{(j_1j_2)}$ and $\mathbb{V}^{(j_1j_2)}_{\rm ex}$ acting on
the product of superfields \re{O-SL2} should be invariant under the $SL(2)$
transformations \re{SL2-transformations}.

According to \re{SL2-transformations}, $\Phi_j(z,0)$ realizes the spin$-j$
representation of the $SL(2,\mathbb{R})$ group,
$\mathcal{V}_{j}^{\scriptscriptstyle \rm SL(2)}$. Indeed, as follows from
Eqs.~\re{M=0-field} and \re{M=4-field}, the superfield $\Phi_j(z,0)$ is given by
its lowest component which in its turn is a primary of the $SL(2,\mathbb{R})$
group with the conformal spin $j$. The light-cone operators \re{O-SL2} belong to
the tensor product of two $SL(2;\mathbb{R})$ representations
$\mathcal{V}_{j_1}^{\scriptscriptstyle \rm SL(2)}\otimes
\mathcal{V}_{j_2}^{\scriptscriptstyle \rm SL(2)}$ labelled by the spins $j_1$ and
$j_2$. The operators  $\mathbb{V}^{(j_1j_2)}$ and $\mathbb{V}^{(j_1j_2)}_{\rm
ex}$, Eq.~\re{V-times-V}, act on this product as
\ba
&& \mathbb{V}^{(j_1j_2)}: \mathcal{V}_{j_1}^{\scriptscriptstyle \rm SL(2)}
\otimes \mathcal{V}^{\scriptscriptstyle \rm SL(2)}_{j_2} \to
\mathcal{V}^{\scriptscriptstyle \rm SL(2)}_{j_1} \otimes
\mathcal{V}^{\scriptscriptstyle \rm SL(2)}_{j_2}\,,\qquad
\nonumber \\[2mm]
&& \mathbb{V}^{(j_1j_2)}_{\rm ex}: \mathcal{V}_{j_1}^{\scriptscriptstyle \rm
SL(2)} \otimes \mathcal{V}^{\scriptscriptstyle \rm SL(2)}_{j_2} \to
\mathcal{V}^{\scriptscriptstyle \rm SL(2)}_{j_2} \otimes
\mathcal{V}^{\scriptscriptstyle \rm SL(2)}_{j_1}\,.
\ea
Such operators have been studied thoroughly in the context of QCD conformal
operators. As was shown in Ref.~\cite{DKM02}, the $SL(2)$ invariant operators
$\mathbb{V}^{(j_1j_2)}$ and $\mathbb{V}^{(j_1j_2)}_{\rm ex}$ defined in this way
have the following general form
\ba
\mathbb{V}^{(j_1j_2)}\cdot \mathbb{O}_{j_1j_2}(z_1,z_2) \!\!\! &=&\!\!\!
\e^{i\pi(j_1+j_2)}\int [\mathcal{D} w_1]_{j_1} \int [\mathcal{D}
w_2]_{j_2}\,\mathbb{O}_{j_1j_2}(w_1,w_2)
\label{general-SL2} \\
&& \qquad\qquad \times(z_1-\bar w_1)^{-2j_1} (z_2-\bar w_2)^{-2j_2} f(\xi)\,,
\nonumber
\\
\mathbb{V}^{(j_1j_2)}_{\rm ex}\cdot \mathbb{O}_{j_1j_2}(z_1,z_2)\!\!\! &=&\!\!\!
\e^{i\pi(j_1+j_2)}\int [\mathcal{D}w_1]_{j_1} \int [\mathcal{D}
w_2]_{j_2}\,\mathbb{O}_{j_2j_1}(w_1,w_2)
\label{general-ex-SL2} \\
&& \qquad\qquad\times(z_1-\bar w_2)^{-2j_1}  (z_2-\bar w_1)^{-2j_2} f_{\rm
ex}(\xi)\,. \nonumber
\ea
Here a notation was introduced for the $SL(2)$ invariant measure
\be
[\mathcal{D} w]_j =  \frac{2j-1}{\pi} d^2 w \,(\Im w)^{2j-2}\,\theta(\Im w)\,,
\ee
with the integration region extended over the upper half-plane in the complex
$w-$plane, $\bar w_k = w_k^*$. Also, $f(\xi)$ and $f_{\rm ex}(\xi)$ are arbitrary
functions of the harmonic ratio
\be
\xi=\frac{(z_1-\bar w_2)(z_2-\bar w_1)}{(z_1-\bar w_1)(z_2-\bar w_2)}\,.
\label{xi}
\ee
It is straightforward to verify that the operators $\mathbb{V}^{(j_1j_2)}$ and
$\mathbb{V}^{(j_1j_2)}_{\rm ex}$ are invariant under the $SL(2)$ transformations
\re{SL2-transformations}.

We would like to stress that the explicit form of the functions $f(\xi)$ and
$f_{\rm ex}(\xi)$ is not fixed by the $SL(2)$ invariance. These functions
determine the dilatation operator in the $\mathcal{N}-$extended SYM theory and
one might expect that they should depend on $\mathcal{N}$. Nevertheless, as we
will show in Sect.~4 by an explicit calculation of the one-loop dilatation
operator, the functions $f(\xi)$ and $f_{\rm ex}(\xi)$ have the \textsl{same},
universal form in all SYM theories
\ba
\label{FandFex}
& f(\xi)&\!\!\! =\ln \xi+\psi(2j_1)+\psi(2j_2) - 2\psi(1)\,,
\\[2mm]
& f_{\rm ex}(\xi)&\!\!\! =\xi^{2j_1}\theta(j_2-j_1) + \xi^{2j_2}\theta(j_1-j_2)
\,. \nonumber
\ea
Substituting this ansatz into \re{general-SL2} and \re{general-ex-SL2}, one
performs the integration with a help of the identity \re{SL2-identity} and
obtains the following expression for the $SL(2)$ invariant operators acting on
the product of two superfields \re{O-SL2}
\ba
\mathbb{V}^{(j_1j_2)}\mathbb{O}_{j_1j_2}(z_1,z_2) &=& \int_0^1
\frac{d\alpha}{\alpha} \, \bigg\{ - \bar\alpha^{2 j_1 - 1}
\mathbb{O}_{j_1j_2}(\bar\alpha z_1 + \alpha z_2,z_2)
\label{V-SL2-ansatz}\\
&& \hspace*{14mm} - \bar\alpha^{2 j_2 - 1} \mathbb{O}_{j_1j_2}(z_1,\bar\alpha z_2
+ \alpha z_1) + 2 \mathbb{O}_{j_1j_2}(z_1,z_2) \bigg\}\,,
\nonumber \\
\mathbb{V}^{(j_1j_2)}_{\rm ex}\mathbb{O}_{j_1j_2}(z_1,z_2) &=&
\theta(j_2-j_1)\int_0^1 d\alpha\,
\bar\alpha^{2j_1-1}\alpha^{2(j_2-j_1)-1} \mathbb{O}_{j_2j_1}(
\bar\alpha z_1+\alpha z_2,z_2)
\label{V-ex-SL2-ansatz}
\\
&+&\theta(j_1-j_2)\int_0^1 d\alpha\,
\bar\alpha^{2j_2-1}\alpha^{2(j_1-j_2)-1} \mathbb{O}_{j_2j_1}(z_1,
\bar\alpha z_2+\alpha z_1)\,, \nonumber
\ea
where $\bar{\alpha} \equiv 1 - \alpha$. These operators have a simple
interpretation: they displace the superfields along the light-cone in the
direction of each other with the weight functions depending on their conformal
spins.

The operators \re{V-SL2-ansatz} and \re{V-ex-SL2-ansatz} commute with the
$SL(2;\mathbb{R})$ generators \re{sl2-generators} acting on the tensor product
$\mathcal{V}_{j_1}^{\scriptscriptstyle \rm SL(2)}\otimes
\mathcal{V}_{j_2}^{\scriptscriptstyle \rm SL(2)}$ and, therefore, they are
functions of the two-particle conformal spin $\mathbb{J}_{12}$ defined through
the $SL(2)$ Casimir operator
\be
l^2 = (l^0)^2-l^0+l^+l^- = \mathbb{J}_{12}(\mathbb{J}_{12}-1)\,,
\label{Sl2-Casimir}
\ee
with $l^\alpha = l^\alpha_{j_1} + l^\alpha_{j_2}$ for $\alpha=0,\pm$ and the
$SL(2)$ generators $l^\alpha_{j}$ given by \re{sl2-generators}. To establish the
explicit form of the dependence of $\mathbb{V}^{(j_1j_2)}$ and $\mathbb{V}_{\rm
ex}^{(j_1j_2)}$ on the conformal spin $\mathbb{J}_{12}$, it suffices to compare
their action on the space of test functions which belong to the tensor product
$\mathcal{V}_{j_1}^{\scriptscriptstyle \rm SL(2)}\otimes
\mathcal{V}_{j_2}^{\scriptscriptstyle \rm SL(2)}$. This space is spanned by
homogeneous polynomials of two variables $z_1$ and $z_2$ and it possesses the
highest weights ${O}_{j_1j_2}^{(n)}(z_1,z_2)=(z_1-z_2)^n$. These states satisfy
the relations
\be
l^- {O}_{j_1j_2}^{(n)}(z_1,z_2)=0\,,\qquad
\mathbb{J}_{12}\,{O}_{j_1j_2}^{(n)}(z_1,z_2)=(n+j_1+j_2){O}_{j_1j_2}^{(n)}(z_1,z_2)
\label{Casimir-SL2}
\ee
and, most importantly, they diagonalize the kernels \re{V-SL2-ansatz} and
\re{V-ex-SL2-ansatz}. One replaces $\mathbb{O}_{j_1j_2}(z_1,z_2)$ in
\re{V-SL2-ansatz} and \re{V-ex-SL2-ansatz} by the highest weights
\be
\mathbb{O}_{j_1j_2}(z_1,z_2) \quad\to\quad
{O}_{j_1j_2}^{(n)}(z_1,z_2)=(z_1-z_2)^n\,,
\ee
calculates the corresponding eigenvalues of the operators
$\mathbb{V}^{(j_1j_2)}$ and $\mathbb{V}_{\rm ex}^{(j_1j_2)}$ and casts them
into an operator form with a help of \re{Casimir-SL2} to get
\ba
\mathbb{V}^{(j_1j_2)} &=&
\psi\left(\mathbb{J}_{12}+j_1-j_2\right)+\psi\left(\mathbb{J}_{12}-j_1+j_2\right)-2\psi(1)\,,
\label{SL-invariant-form}
\\[3mm]
\mathbb{V}_{\rm ex}^{(j_1j_2)}
&=&\mathbb{P}_{12}\frac{\Gamma(\mathbb{J}_{12}-|j_1-j_2|)}{\Gamma(\mathbb{J}_{12}+|j_1-j_2|)}
{\Gamma(2|j_1-j_2|)}\,, \nonumber
\ea
where $\mathbb{P}_{12}$ is a permutation operator, $\mathbb{P}_{12}
\mathbb{O}_{j_1j_2}(z_1,z_2)=\mathbb{O}_{j_2j_1}(z_2,z_1)$, and
$\psi(x)=d\ln\Gamma(x)/dx$ is the Euler $\psi-$function. Since $|j_1-j_2|$ takes
(half)integer values, the operator $\mathbb{V}_{\rm ex}^{(j_1j_2)}$ is a rational
function of the conformal spin $\mathbb{J}_{12}$. The operator
$\mathbb{V}^{(j_1j_2)}$ is well-known in the theory of lattice integrable models.
It can be identified as a two-particle Hamiltonian of a completely integrable
Heisenberg $SL(2;\mathbb{R})$ spin chain. As was mentioned in the Introduction,
it is this property that is responsible for remarkable integrability symmetry of
the one-loop dilatation operator in the SYM theory on the
light-cone~\cite{BelDerKorMan04}.

\subsubsection{From the light-cone to the superspace}

As a next step, we have to restore the dependence of the superfields in
\re{O-SL2} on the odd coordinates $\theta_1^A$ and $\theta_2^A$ and ``lift'' the
relations \re{V-SL2-ansatz} and \re{V-ex-SL2-ansatz} from the light-cone to the
superspace, $z\to Z=(z,\theta)$. One possibility could be to generalize the
relations \re{general-SL2} and \re{general-ex-SL2} and write down expressions for
an $SL(2|\mathcal{N})$ invariant operators as integrals over the representation
space of the $SL(2|\mathcal{N})$ group. We shall choose however another route
which is much simpler and leads immediately to the same final expressions.

Let us apply the superconformal transformations generated by the
$SL(2|\mathcal{N})$ charges $V_A^+$ and $V_A^-$ to Eq.\ \re{O-SL2}. Taking into
account \re{V-minus} and \re{V-plus}, we obtain that these generators displace
the chiral superfields along odd directions in the superspace and do not alter
their positions on the light-cone
\be
\e^{\xi\cdot V^-+\epsilon\cdot V^+} \mathbb{O}_{j_1j_2}(z_1,z_2)=
{\Phi}_{j_1}(z_1,\xi^A+ z_1 \epsilon^A){\Phi}_{j_2}(z_2,\xi^A+ z_2
\epsilon^A)\equiv \mathbb{O}_{j_1j_2}(Z_1,Z_2)\,.
\label{lift}
\ee
Denoting $\theta_1^A=\xi^A+ z_1 \epsilon^A$ and $\theta_2^A=\xi^A+ z_2
\epsilon^A$, one finds that for $z_1\neq z_2$ and $z_1,z_2\neq 0$ the superfields
in \re{lift} are located in two different points of the superspace,
$Z_1=(z_1,\theta_1^A)$ and $Z_2=(z_2,\theta_2^A)$. Since the $SL(2|\mathcal{N})$
invariant operators $\mathbb{V}_{12}$ and $\mathbb{V}_{12}^{\rm (ex)}$ have to
commute with the generators $V_A^\pm$, one gets
\be
\mathbb{V}^{(j_1j_2)}\mathbb{O}_{j_1j_2}(Z_1,Z_2)=\e^{\xi\cdot V^-+\epsilon\cdot
V^+} \mathbb{V}^{(j_1j_2)}\mathbb{O}_{j_1j_2}(z_1,z_2)
\label{V-lift}
\ee
and similar for $\mathbb{V}^{(j_1j_2)}_{\rm ex}$. This relation allows one to
reconstruct the operators $\mathbb{V}^{(j_1j_2)}$ and $\mathbb{V}^{(j_1j_2)}_{\rm
ex}$ acting on the superfields
$\mathbb{O}_{j_1j_2}(Z_1,Z_2)={\Phi}_{j_1}(Z_1){\Phi}_{j_2}(Z_2)$ from their
expressions on the light-cone, Eqs.~\re{V-SL2-ansatz} and \re{V-ex-SL2-ansatz}.

The transformations \re{lift} and \re{V-lift} amount to replacing the arguments
of the superfields
\be
{\Phi}_j(\alpha z_1 + \bar\alpha z_2,0) \to {\Phi}_j(\alpha Z_1 +
\bar\alpha Z_2)\,,
\label{lift1}
\ee
with $Z_k=(z_k,\theta_k^A)$. As a consequence, the $SL(2)$ invariant operators
\re{V-SL2-ansatz} and \re{V-ex-SL2-ansatz} are transformed into
\begin{eqnarray}
\mathbb{V}^{(j_1, j_2)} \, \mathbb{O}_{j_1j_2}(Z_1,Z_2) \!\!\!&=&\!\!\! \int_0^1
\frac{d\alpha}{\alpha} \, \bigg\{ 2 \mathbb{O}_{j_1j_2}(Z_1,Z_2)
\label{V-gen-SL2N}
\\
\!\!\!&-&\!\!\! \bar\alpha^{2 j_1 - 1} \mathbb{O}_{j_1j_2} (\bar\alpha Z_1 +
\alpha Z_2,Z_2) - \bar\alpha^{2 j_2 - 1} \mathbb{O}_{j_1j_2} (Z_1,\bar\alpha Z_2
+ \alpha Z_1) \bigg\} \, . \nonumber
\\
\mathbb{V}^{(j_1, j_2)}_{\rm ex} \, \mathbb{O}_{j_1j_2}(Z_1,Z_2) \!\!\!&=&\!\!\!
\theta(j_2-j_1)\int_0^1 d\alpha \,\bar\alpha^{2j_1-1} \alpha^{2j_2-2j_1-1}
\mathbb{O}_{j_2j_1}(\bar\alpha Z_1+\alpha Z_2,Z_2)
\label{V-ex-gen-SL2N} \\
\!\!\!&+&\!\!\! \theta(j_1-j_2)\int_0^1 d\alpha \,{\bar\alpha}^{2j_2-1}
\alpha^{2j_1-2j_2-1} \mathbb{O}_{j_2j_1}(Z_1,\bar\alpha Z_2+\alpha Z_1) \,.
\nonumber
\end{eqnarray}
The only difference with the previous case, Eqs.~\re{V-SL2-ansatz} and
\re{V-ex-SL2-ansatz}, is that $Z=(z,\theta^A)$ has nonvanishing odd coordinates
and displacement of the superfields takes place along the line in the superspace
connecting two points $Z_1$ and $Z_2$. Obviously, for $\theta_1^A=\theta_2^A=0$
one recovers the $SL(2)$ expressions. One can verify that \re{V-gen-SL2N} and
\re{V-ex-gen-SL2N} are invariant under the $SL(2|\mathcal{N})$ transformations,
Eqs.~\re{SL2-subgroup} -- \re{V-plus}.

Eqs.~\re{V-gen-SL2N} and \re{V-ex-gen-SL2N} define the $SL(2|\mathcal{N})$
invariant operators $\mathbb{V}^{(j_1j_2)}$ and $\mathbb{V}^{(j_1j_2)}_{\rm ex}$
acting on the product of two {\sl chiral} superfields, ${\Phi}_{j_1}(Z_1)
{\Phi}_{j_2}(Z_2)$. In the next section, we shall apply \re{V-gen-SL2N} and
\re{V-ex-gen-SL2N} to construct an ansatz for the one-loop dilatation operator
acting on the space spanned by the light-cone operators \re{Chiral-sector} --
\re{Mixed-sector} built from the chiral superfields ${\Phi}(Z)$ and ${\Psi}(Z)$.

\subsection{Ansatz for the dilatation operator}

In the $\mathcal{N}-$extended SYM theory, the conformal spins of the chiral
superfields, ${\Phi}$ and ${\Psi}$, take the values $j_\Phi=-1/2$ and
$j_\Psi=(3-\mathcal{N})/2$, respectively. Substituting $j_1=j_\Phi=-1/2$ into
\re{V-super} one encounters a problem: due to the presence of the factor
$\bar\alpha^{2j_1-1}$ the integral over $\alpha$ is divergent for $\alpha\to 1$
and, therefore, the corresponding operator $\mathbb{V}^{(j_1j_2)}$ is not
well-defined. The problem arises every time the operators $\mathbb{V}^{(j_1j_2)}$
and $\mathbb{V}^{(j_1j_2)}_{\rm ex}$ are applied to the product of superfields
with at least one of them carrying the conformal spin $j_1=- 1/2$, that is, in
the $\Phi\Phi-$, $\Phi\Psi-$ and $\Psi\Phi-$sectors. As we will see later in this
section, this divergence is ultimately related to the fact that the
$SL(2|\mathcal{N})$ representation defined by the superfield $\Phi_j(Z)$ is
reducible for $j =- 1/2$, that is, the corresponding representation space
$\mathcal{V}_j$ contains an invariant subspace. The above mentioned divergences
originate from the states belonging to this subspace.

For $j_1=-1/2$ the divergences in \re{V-super} originate from the first two terms
of the expansion of nonlocal operator $\mathbb{O}_{-1/2,j_2} (\bar\alpha
Z_1+\alpha Z_2,Z_2)$ around $\alpha=1$
\be
\mathbb{O}_{-1/2,j_2}(\bar\alpha Z_1+\alpha
Z_2,Z_2)={\Phi}_{-1/2}(Z_2){\Phi}_{j_2}(Z_2) +\bar\alpha
Z_{12}\cdot\partial_{Z_2}{\Phi}_{-1/2}(Z_2){\Phi}_{j_2}(Z_2)+O(\bar\alpha^2)\,.
\ee
Here the expansion coefficients are local operators defined at the point $Z_2$.
These operators belong to the same $SL(2|\mathcal{N})$ module as the operators
${\Phi}_{-1/2}(0){\Phi}_{j_2}(0)$ and
$\partial_Z{\Phi}_{-1/2}(0){\Phi}_{j_2}(0)$, since they are obtained from the
latter through translations in the superspace \re{V-minus}. According to
\re{M=0-field} -- \re{M=4-field}
\be
{\Phi}_{-1/2}(0) = \partial_+^{-1} A(0)\,,\qquad
\partial_z{\Phi}_{-1/2}(0) = A(0)\,,\qquad
\partial_{\theta^A}{\Phi}_{-1/2}(0) = \partial_+^{-1}\bar \lambda_A(0)\,,
\label{spurious}
\ee
with $\partial_+ A(0) = n^\mu F_{\mu z}(0)$ in the light-like gauge $A_+=0$.
Notice that all fields in \re{spurious} are \textsl{nonlocal}, spurious
operators. Their definition involves the inverse derivative $\partial_+^{-1}$,
which is not a well-defined integral operator. In the momentum representation, it
induces a spurious pole at $k_+=0$ and the properties of the fields \re{spurious}
depend on the prescription adopted to regularize the pole. Throughout this paper
we shall define the operator $\partial_+^{-1}$ using the Mandelstam-Leibbrandt
prescription (see Eq.~\re{ML-pre}).

One of the advantages of this prescription is that $1/k_+-$factors do not induce
additional singularities inside Feynman integrals and calculating superficial
divergence index of diagrams one can treat them on equal footing with the
conventional Feynman propagators. This property plays a crucial role in
establishing a UV finiteness of the $\mathcal{N}=4$ SYM
theory~\cite{SohWes81,BriLinNil83,Man83,HowSteow84}. In the present case, it also
has important consequences for renormalization properties of composite operators
involving spurious fields \re{spurious}. As we will show in Sect.~4, the
additional $1/k_+-$factors improve convergence properties of Feynman integrals in
a SYM theory, and, as a consequence, the one-loop corrections to certain
operators involving the nonlocal fields \re{spurious} are ultraviolet finite. The
UV finite spurious operators include $\Phi(0)\Phi(0)$, $\Phi(0)\partial_{Z_a}
\Phi(0)$, $\partial_{Z_a}\Phi(0)\partial_{Z_b}\Phi(0)$, $\partial_{Z_a}
\Phi(0)\Psi(0)$, ... with
$\partial_{Z_a}\Phi=(\partial_z\Phi,\partial_{\theta^A}\Phi)$. Notice that this
set does not comprise \textsl{all} operators involving the fields \re{spurious}.
For instance, the operators like $\Phi(0)\partial_z^n \Phi(0)$ (with $n\ge 3$)
mix under renormalization with ``physical'' operators
$\partial_z^m\Phi(0)\partial_z^{n-m}\Phi(0)$ and acquire a nontrivial anomalous
dimension.

Let us consider separately UV finite spurious operators in the $\Phi\Phi-$,
$\Phi\Psi-$ and $\Psi\Phi-$sectors. In the $\Phi\Phi-$sector, they are given by a
bilinear product of the fields \re{spurious}, like ${\Phi}(0){\Phi}(0)$,
${\Phi}(0)\partial_z{\Phi}(0)$,
$\partial_{\theta^A}{\Phi}(0)\partial_z{\Phi}(0)$, and their $SL(2|\mathcal{N})$
descendants. For our purposes it is convenient to introduce a ``spurious''
superfield
\be
{\Phi}_{\rm sp}(Z)={\Phi}(0)+Z\cdot\partial_Z {\Phi}(0)\,.
\label{aux-field}
\ee
and treat the product ${\Phi}_{\rm sp}(Z_1){\Phi}_{\rm sp}(Z_2)$ as a generating
function for such operators. As was just mentioned, ${\Phi}_{\rm
sp}(Z_1){\Phi}_{\rm sp}(Z_2)$ does not acquire anomalous dimension and,
therefore, it has to be annihilated by the one-loop dilatation operator
\be
\mathbb{H}_{\Phi\Phi}\,{\Phi}_{\rm sp}(Z_1){\Phi}_{\rm sp}(Z_2) = 0\,.
\label{zero-sector}
\ee
Let us confront \re{zero-sector} with  properties of the $SL(2|\mathcal{N})$
invariant operator, Eq.~\re{V-super}. One applies $\mathbb{V}^{(-1/2,-1/2)}$ to
the product of two superfields ${\Phi}_{\rm sp}(Z_1){\Phi}_{\rm sp}(Z_2)$ and,
instead of getting zero, arrives at a divergent integral over the
$\alpha-$parameter. To remove divergencies and, at the same time, to reproduce
\re{zero-sector}, it suffices to introduce a projection operator,
$\Pi_{\Phi\Phi}^2=\Pi_{\Phi\Phi}$, such that
\be
(1-\Pi_{\Phi\Phi})\,{\Phi}_{\rm sp}(Z_1){\Phi}_{\rm sp}(Z_2)=0\,.
\label{condition-phi}
\ee
Making use of $\Pi_{\Phi\Phi}$ one can construct an integral operator
\be
\mathbb{H}_{\Phi\Phi}^{\rm (ansatz)} = \mathbb{V}^{(-1/2,-1/2)}
(1-\Pi_{\Phi\Phi})\,.
\label{ansatz-PhiPhi}
\ee
It verifies \re{zero-sector} and coincides with \re{V-super} on the subspace of
light-cone operators annihilated by $\Pi_{\Phi\Phi}$. To preserve the
superconformal symmetry one requires that the projector $\Pi_{\Phi\Phi}$ has to
be an $SL(2|\mathcal{N})$ invariant operator. It acts on the tensor product
$\mathcal{V}_{-1/2}\otimes \mathcal{V}_{-1/2}$, Eq.~\re{V-times-V}, and has a
general form \re{V-gen-SL2N}. The corresponding scalar function $\varphi$ is
uniquely fixed by the condition \re{condition-phi}. Going over through the
calculation (see Appendix~D) one finds that $\varphi(\xi)=c_1\,
\delta(1-\xi)+c_2\,\delta'(1-\xi)$ with $c_1$ and $c_2$ being some coefficients.
In this way, we obtain the following expression for the projector
\be
\Pi_{\Phi\Phi}\, \mathbb{O} (Z_1,Z_2) = \ft12\left(1+ Z_{12}\cdot
\partial_{Z} \right)\mathbb{O}(Z,Z_2)\bigg|_{Z=Z_2}
+ \ft12\left(1+ Z_{21}\cdot
\partial_{Z}\right)\mathbb{O}(Z_1,Z)\bigg|_{Z=Z_1},
\label{Pi12-main}
\ee
where $(Z_{12}\cdot \partial_{Z})\equiv (z_1-z_2)\partial_{z}
+(\theta_{1}^A-\theta_{2}^A) \partial_{\theta^A}$. One verifies that the operator
$\Pi_{\Phi\Phi}$, defined in this way, indeed satisfies \re{condition-phi}.

Let us now examine composite operators in the $\Phi\Psi-$sector. We remind that
the superfield ${\Psi}(Z)$ involves only physical fields and the UV finite
spurious operators in this sector are ${\Phi}(0){\Psi}(0)$,
$\partial_{Z_a}{\Phi}(0){\Psi}(0)$ and their $SL(2|\mathcal{N})$ descendants like
$\partial_+^n(\partial_{Z_a}{\Phi}(0){\Psi}(0))$ with $n$ positive. Similar to
the previous case, these operators have to be annihilated by the one-loop
dilatation operator in the $\Phi\Psi-$sector
\be
\mathbb{H}_{\Phi\Psi}\, {\Phi}_{\rm sp}(Z_1){\Psi}(0) = 0\,,
\label{zero-anom-dim}
\ee
with the auxiliary superfield ${\Phi}_{\rm sp}$ defined in \re{aux-field}. In
general, $\mathbb{H}_{\Phi\Psi}$ is given by a linear combination of the
operators $\mathbb{V}^{(-1/2,j_\Psi)}$ and $\mathbb{V}_{\rm ex}^{(-1/2,j_\Psi)}$
defined in \re{V-super} and \re{ansatz-V-ex}, respectively. As before, one can
fulfill \re{zero-anom-dim} at an expense of introducing yet another projection
operator
\be
\mathbb{H}_{\Phi\Psi}^{\rm (ansatz)} = \left[\mathbb{V}^{(-1/2,j_\Psi)}+c
\mathbb{V}_{\rm ex}^{(-1/2,j_\Psi)} \right] (1-\Pi_{\Phi\Psi})\,,
\label{ansatz-PhiPsi}
\ee
with a constant $c$. Its value $c=-1$ will be fixed in Sect.~4. The projector
$\Pi_{\Phi\Psi}$ acts on the tensor product $\mathcal{V}_{-1/2}\otimes
\mathcal{V}_{j_\Psi}$ and satisfies
\be
(1-\Pi_{\Phi\Psi})\, {\Phi}_{\rm sp}(Z_1){\Psi}(0)=0\,.
\label{pro-mixed-sector}
\ee
Looking for $\Pi_{\Phi\Psi}$ in the form of a general $SL(2|\mathcal{N})$
invariant operator, Eq.~\re{V-gen-SL2N}, one uses \re{pro-mixed-sector} to fix
the corresponding scalar function $\varphi$ and obtains (see Appendix~B for
details)
\be
\Pi_{\Phi\Psi} \mathbb{O}(Z_1,Z_2)=\mathbb{O}(Z_2,Z_2)+Z_{12}\cdot \partial_Z
\mathbb{O}(Z,Z_2)\bigg|_{Z=Z_2}\,.
\label{projector-phi-psi}
\ee
One verifies that $(1-\Pi_{\Phi\Psi})$ annihilates the operators
$\mathbb{O}(Z_1,Z_2)$ linear in $Z_1$ and, therefore, \re{pro-mixed-sector} is
automatically satisfied.

Finally, one examines UV finite spurious operators in the $\Psi\Phi-$sector. The
only difference with the previous case is that the ${\Psi}-$ and
${\Phi}-$superfields have to be interchanged inside the trace, so that a
generalization of \re{ansatz-PhiPsi} is straightforward
\be
\mathbb{H}_{\Psi\Phi}^{\rm (ansatz)} = \left[\mathbb{V}^{(j_\Psi,-1/2)} -
\mathbb{V}_{\rm ex}^{(j_\Psi,-1/2)} \right] (1-\Pi_{\Psi\Phi})\,,
\label{ansatz-PsiPhi}
\ee
where the projector is defined as
\be
\Pi_{\Psi\Phi} \mathbb{O}(Z_1,Z_2)=\mathbb{O}(Z_1,Z_1)-Z_{12}\cdot \partial_Z
\mathbb{O}(Z_1,Z)\bigg|_{Z=Z_1}\,.
\label{projector-psi-phi}
\ee
After having inserted the projectors into the expression for the dilatation
operator, Eq.~\re{ansatz-PhiPhi}, \re{ansatz-PsiPhi} and \re{ansatz-PhiPsi}, we
achieved two goals simultaneously. Firstly, the dilatation operator annihilates
UV finite spurious operators built from the fields \re{spurious}. Secondly, the
resulting integrals over the $\alpha-$parameter are convergent and the
corresponding integral operators are well-defined. The projectors are not
necessary in the $\Psi\Psi-$sector since the ${\Psi}-$superfield only involves
physical fields and spurious operator do not appear
\be
\mathbb{H}_{\Psi\Psi}^{\rm (ansatz)} = \mathbb{V}^{(j_\Psi,j_\Psi)} \, .
\label{ansatz-PsiPsi}
\ee

To summarize, the one-loop dilatation operator in the $\mathcal{N}-$extended SYM
theory is given in the multi-color limit by \re{H-multi-color} with the
$SL(2|\mathcal{N})$ invariant two-particle kernel $\mathbb{H}_{k,k+1}$ having a
different form for $\mathcal{N}=4$ and $\mathcal{N}\le 2$:
\begin{itemize}
\item For $\mathcal{N}=4$ one finds
\be
\mathbb{H}_{k,k+1}\bigg|_{\mathcal{N}=4} =\mathbb{H}_{\Phi\Phi}=
\mathbb{V}^{(-1/2,-1/2)} (1-\Pi_{\Phi\Phi})\,,
\label{H-4-matrix}
\ee
where the operators  $\mathbb{V}^{(-1/2,-1/2)}$ and $\Pi_{\Phi\Phi}$,
Eqs.~\re{V-super} and \re{Pi12-main}, act on the superfields with the coordinates
$Z_k$ and $Z_{k+1}$.

\item For $\mathcal{N}\le 2$ the two-particle kernel has a different form in the
$\Phi\Phi-$, $\Phi\Psi-$, $\Psi\Phi-$ and $\Psi\Psi-$sectors and can be
represented as a $2\times 2$ matrix
\be
[\mathbb{H}_{k,k+1}]_{ab}\bigg|_{\mathcal{N}\le 2} = \mathbb{H}_{ab} =
\left[\mathbb{V}^{(j_a,j_b)} - \mathbb{V}_{\rm ex}^{(j_a,j_b)} \right]
(1-\Pi_{ab})\,, \label{H-matrix}
\ee
where $a,b={\scriptstyle \Phi,\Psi}$. Here $\mathbb{V}_{\rm
ex}^{(j_\Psi,j_\Psi)}=\mathbb{V}_{\rm ex}^{(j_\Phi,j_\Phi)}=0$ and the projectors
$\Pi_{ab}$ were defined in \re{Pi12-main}, \re{projector-phi-psi},
\re{projector-psi-phi} with $\Pi_{\Psi\Psi}=0$.

\end{itemize}
The eigenvalues of the dilatation operator defined in this way determine the
anomalous dimensions of {\sl all} quasipartonic operators in the SYM theories
with $0\le \mathcal{N} \le 4$. We will demonstrate in Sect.~5 that
Eqs.~\re{H-4-matrix} and \re{H-matrix} lead to the expressions for the anomalous
dimensions which are in agreement with the known results in the
$\mathcal{N}=0$~\cite{GroWil73,BukFroKurLip85},
$\mathcal{N}=1$~\cite{KouRos82,BelMul99,Kir04} and
$\mathcal{N}=4$~\cite{MinZar02,KotLip02,BeiSta03} theories.

\subsection{$SL(2|\mathcal{N})$ invariant form of the dilatation operator}

By construction, the two-particle evolution kernels $\mathbb{H}_{ab}^{\rm
(ansatz)}$ (with $a,b={\scriptstyle \Phi,\Psi}$), Eqs.~\re{ansatz-PhiPhi},
\re{ansatz-PsiPhi} and \re{ansatz-PsiPsi}, commute with the $SL(2|\mathcal{N})$
generators \re{sl2} acting on the tensor product $\mathcal{V}_{j_a}\otimes
\mathcal{V}_{j_b}$. As in the $SL(2)$ case, Eq.~\re{SL-invariant-form}, one can
express the kernels $\mathbb{H}_{ab}^{\rm (ansatz)}$ as functions of the
two-particle \textsl{super}conformal spin $\mathbb{J}_{ab}$. It is defined
through the two-particle $SL(2|\mathcal{N})$ Casimir operator
\be
\mathbb{L}^2_{ab} = (L^0)^2 + L^+ L^- + (\mathcal{N}-1) L^0
+\frac{\mathcal{N}}{\mathcal{N}-2} B^2 - V^+_A W^{A,-} - W^+_A V^{A,-}-\ft12\,
T^B{}_A T^A{}_B\,.
\label{Casimir}
\ee
where $G=\{L^0,L^\pm,B,V^\pm_A,W^{A,\pm},T^A{}_B\}$ are the $SL(2|\mathcal{N})$
generators acting on the tensor product $\mathcal{V}_{j_a}\otimes
\mathcal{V}_{j_b}$, that is, $G=G_{j_a}+G_{j_b}$ with $G_j$ given by \re{sl2}.
Then, the two-particle {super}conformal spin $\mathbb{J}_{12}$ is defined as
\be
\mathbb{L}^2_{ab} =\mathbb{J}_{ab}(\mathbb{J}_{ab}-1)+C_{ab}
\label{J-ab}
\ee
where $C_{ab}=\mathcal{N}(j_a+j_b)[1+(j_a+j_b)/(\mathcal{N}-2)]$ is a c-valued
constant introduced for the latter convenience (see Eq.~\re{conv}). For
$\mathcal{N}=0$, the relation \re{Casimir} coincides with the $SL(2)$ Casimir
\re{Sl2-Casimir}. The contribution of the $B-$charge to \re{Casimir} is divergent
for $\mathcal{N}=2$. This singularity is spurious since the $B-$charge,
Eq.~\re{sl2}, is reduced for $\mathcal{N}=2$ to a c-number, $B=-j$, and,
therefore, it can be removed by subtracting constant $C_{ab}$ from the right-hand
of \re{Casimir} and \re{J-ab}.

As before, to find the explicit form of the dependence of $\mathbb{H}_{ab}^{\rm
(ansatz)}$ on $\mathbb{J}_{ab}$, we shall examine the action of both operators on
the highest weights in $\mathcal{V}_{j_a}\otimes \mathcal{V}_{j_b}$ that we
denote as ${O}_{j_aj_b}^{(n)}(Z_1,Z_2)$. By definition, these states are
annihilated by ``lowering'' operators $L^-$, $W^{A,-}$ and $V^-_A$ defined in
\re{sl2}
\be
L^- {O}_{j_1j_2}^{(n)}(Z_1,Z_2) = W^{A,-} {O}_{j_1j_2}^{(n)}(Z_1,Z_2) = V^-_A
{O}_{j_1j_2}^{(n)}(Z_1,Z_2)=0\,.\vspace*{2mm}
\ee
Solving these relations one finds the highest weights as ($1\le k \le
\mathcal{N}-1$ and $0 \le n < \infty$)
\be
{O}_{j_1j_2}^{(0)}=1\,,\qquad {O}_{j_1j_2}^{(k)}=\theta_{12}^{A_1}\ldots
\theta_{12}^{A_k} \,,\qquad {O}_{j_1j_2}^{(n+\mathcal{N})}=\varepsilon_{A_1\ldots
A_{\mathcal{N}}}\theta_{12}^{A_1}\ldots \theta_{12}^{A_\mathcal{N}}z_{12}^n\,,
\ee
where $\theta_{12}^A= \theta_1^A - \theta_2^A$ and $z_{12}=z_1-z_2$. These states
diagonalize the two-particle Casimir operator \re{Casimir} and carry a definite
value of the superconformal spin
\be
\left(\mathbb{J}_{12}^2-C_{12}\right){O}_{j_1j_2}^{(l)} =
(l+j_1+j_2)(l+j_1+j_2-1) {O}_{j_1j_2}^{(l)} = J_{12}
(J_{12}-1){O}_{j_1j_2}^{(l)}\,,
\label{conv}
\ee
where $J_{12}=l+j_1+j_2$ is the eigenvalue of the two-particle spin
$\mathbb{J}_{12}$, Eq.~\re{J-ab}.

Let us now substitute $\mathbb{O}_{j_1j_2}\to {O}_{j_1j_2}^{(l)}$ in
\re{V-gen-SL2N} and \re{V-ex-gen-SL2N}. One verifies that both operators become
diagonal and the corresponding eigenvalues look
as\\[-4mm]
\ba
\mathbb{V}^{(j_1, j_2)}\,{O}_{j_1j_2}^{(l)} &=& \left[\psi\left(
{J}_{12}+j_1-j_2\right)+\psi\left( {J}_{12}
-j_1+j_2\right)-2\psi(1)\right]{O}_{j_1j_2}^{(l)}
\label{SL-superinvatiant-form}
\\[3mm]
\mathbb{V}_{\rm ex}^{(j_1j_2)}\,{O}_{j_1j_2}^{(l)}
&=&\frac{\Gamma({J}_{12}-|j_1-j_2|)}{\Gamma({J}_{12}+|j_1-j_2|)}
{\Gamma(2|j_1-j_2|)}\,{O}_{j_2j_1}^{(l)} \nonumber
\ea
where $J_{12}=l+j_1+j_2$ with $l\ge 0$ and $j_1\neq j_2$ in the second relation.
Eq.~\re{SL-superinvatiant-form} generalizes the $SL(2)$ expressions
\re{SL-invariant-form} to the case of the $SL(2|\mathcal{N})$ invariant
operators.

Let us set in \re{SL-superinvatiant-form} $j_1=j_2=j_\Psi=(3-\mathcal{N})/2$.
According to \re{ansatz-PsiPsi}, the resulting expression for
$\mathbb{V}^{(j_\Psi, j_\Psi)}$ gives the two-particle kernel in the
$\Psi\Psi-$sector (for $\mathcal{N}=0,1,2$)
\be
\mathbb{H}_{\Psi\Psi}^{\rm
(ansatz)}=2\left[\psi(\mathbb{J}_{\Psi\Psi})-\psi(1)\right]\,,
\label{H-PsiPsi-J}
\ee
with $\mathbb{J}_{\Psi\Psi}$ having the eigenvalues
$\mathbb{J}_{\Psi\Psi}=3-\mathcal{N}+l$. Then, one puts $j_1=j_2=-1/2$ in
\re{SL-superinvatiant-form} that corresponds to going over to the
$\Phi\Phi-$sector. We find that $J_{\Phi\Phi}=-1+l$ and, as a consequence, the
eigenvalues of $\mathbb{V}^{(-1/2, -1/2)}$ take infinite values for $l=0,1$. It
is this divergence that we encountered at the beginning of Sect.~3.3. The
expression for the one-loop dilatation operator in the $\Phi\Phi-$sector,
Eq.~\re{ansatz-PhiPhi}, involves the projector $\Pi_{\Phi\Phi}$. As follows from
its definition \re{Pi12-main}, the operator $(1-\Pi_{\Phi\Phi})$ annihilates two
highest weights with $l=0,1$ leading to
\be
(1-\Pi_{\Phi\Phi}){O}_{\Phi\Phi}^{(l)}={O}_{\Phi\Phi}^{(l)}\,\theta(l-1)\,,
\ee
where the $\theta-$function is defined in such a way that $\theta(n)=0$ for $n\le
0$ and $\theta(n)=1$ for $n>0$. Combining this relation together with
\re{SL-superinvatiant-form}, we get from \re{ansatz-PhiPhi} the following
expression for the two-particle evolution kernel in the
$\Phi\Phi-$sector\\[0mm]
\be
\mathbb{H}_{\Phi\Phi}^{\rm
(ansatz)}=2\left[\psi(\mathbb{J}_{\Phi\Phi})-\psi(1)\right]\theta(\mathbb{J}_{\Phi\Phi})\,.
\label{H-PhiPhi-J}
\ee\\[0mm]
Similarly, the projector $(1-\Pi_{\Phi\Psi})$ entering the expression for the
kernel in the $\Phi\Psi-$sector, Eq.~\re{ansatz-PhiPsi}, annihilates the highest
weights with $l=0,1$, so that $
(1-\Pi_{\Phi\Psi})\mathbb{O}_{\Phi\Psi}^{(l)}=\theta(l-1)\mathbb{O}_{\Phi\Psi}^{(l)}\,.
$ As a result, substituting $j_1=-1/2$ and $j_2=(3-\mathcal{N})/2$ into
\re{SL-superinvatiant-form} we find from \re{ansatz-PhiPsi}
\ba
\mathbb{H}_{\Phi\Psi}^{\rm (ansatz)}\!\!\!&=&\!\!\!\bigg[\psi\left(
\mathbb{J}_{\Phi\Psi}+c_\mathcal{N}\right)+\psi\left( \mathbb{J}_{\Phi\Psi}
-c_\mathcal{N}\right)-2\psi(1)
\label{H-PhiPsi-J}\\
&& \qqqquad\qquad -\mathbb{P}_{\Phi\Psi}
\frac{\Gamma(\mathbb{J}_{\Phi\Psi}-c_\mathcal{N})}{\Gamma(\mathbb{J}_{\Phi\Psi}+c_\mathcal{N})}
{\Gamma(2c_\mathcal{N})} \bigg]\theta(\mathbb{J}_{\Phi\Psi}-c_\mathcal{N})\,,
\nonumber
\ea
where $c_\mathcal{N}=2-\mathcal{N}/2$ and $\mathbb{P}_{\Phi\Psi}$ is a
permutation operator, $\mathbb{P}_{\Phi\Psi}
\mathbb{O}_{\Phi\Psi}(Z_1,Z_2)=\mathbb{O}_{\Psi\Phi}(Z_2,Z_1)$. In this case, the
two-particle spin takes the eigenvalues
$\mathbb{J}_{\Phi\Psi}=c_\mathcal{N}-1+l$, that is integer for $\mathcal{N}=0,2$
and half-integer for $\mathcal{N}=1$. Finally, the two-particle kernel in the
$\Psi\Phi-$sector, Eq.~\re{ansatz-PsiPhi}, is given by the same expression
\re{H-PhiPsi-J} modulo substitution $\mathbb{J}_{\Phi\Psi}\to
\mathbb{J}_{\Psi\Phi}$ and $\mathbb{P}_{\Phi\Psi}\to \mathbb{P}_{\Psi\Phi}$
\be
\mathbb{H}_{\Psi\Phi}^{\rm (ansatz)}=
\mathbb{P}_{\Psi\Phi}\,\mathbb{H}_{\Phi\Psi}^{\rm (ansatz)}\,
\mathbb{P}_{\Psi\Phi}\,,
\label{H-PsiPhi-J}
\ee
where the permutation operator acts as $\mathbb{P}_{\Psi\Phi}
\mathbb{O}_{\Psi\Phi}(Z_1,Z_2)=\mathbb{O}_{\Phi\Psi}(Z_2,Z_1)$. In
Eq.~\re{H-PhiPsi-J}, the term involving the permutation operator describes the
exchange interaction between the superfields. The $\theta-$functions in
\re{H-PhiPhi-J} and \re{H-PhiPsi-J} are induced by the projectors $(1-\Pi_{ab})$
in Eq.~\re{H-matrix}. They assign zero anomalous dimensions to the spurious
operators involving nonlocal fields \re{spurious}.

\subsection{Wilson operators}

The two-particle evolution kernels, Eqs.~\re{H-4-matrix} and \re{H-matrix},
involve the additional projection operators due to the presence of nonlocal
fields $\partial_+^{-1} A(0)$, $A(0)$ and $\partial_+^{-1}\bar\lambda^A(0)$ in
the expansion of the superfield $\Phi(Z)$ around $Z=0$. One can avoid the
spurious operators from the start by subtracting from ${\Phi}(Z)$ the first two
terms of its expansion around $Z=0$
\be
{\Phi}_{\scriptscriptstyle \rm W}(Z) ={\Phi}(Z)-{\Phi}(0)-Z\cdot\partial_Z
{\Phi}(0)={\Phi}(Z)-{\Phi}_{\rm sp}(Z)\,.
\label{Phi-Wilson}
\ee
and introducing a nonlocal light-cone operator $\mathbb{O}_{\scriptscriptstyle
\rm W}$ built from the superfields ${\Psi}(Z)$ and ${\Phi}_{\scriptscriptstyle
\rm W}(Z)$
\be
\mathbb{O}_{\scriptscriptstyle \rm
W}(Z_1,Z_2,\ldots,Z_L)=\tr\{{\Phi}_{\scriptscriptstyle \rm
W}(Z_1){\Psi}(Z_2)\ldots {\Phi}_{\scriptscriptstyle \rm W}(Z_L) \} =
\Pi_{\scriptscriptstyle \rm W}\cdot \mathbb{O}(Z_1,Z_2,\ldots,Z_L)\,.
\label{Wilson-gen-func}
\ee
By construction, the expansion of $\mathbb{O}_{\scriptscriptstyle \rm
W}(Z_1,\ldots,Z_L)$ around $Z_k=0$ generates only ``physical'', Wilson operators.
Here a notation was introduced for the operator ${\Pi}{\scriptscriptstyle \rm W}$
which removes ``spurious'' operators from the light-cone operator. It is easy to
verify that $\Pi_{\scriptscriptstyle\rm W}$ is a projector,
$(\Pi_{\scriptscriptstyle\rm W})^2=\Pi_{\scriptscriptstyle\rm W}$. The chiral
superfields ${\Phi}_{\scriptscriptstyle\rm W}(Z)$ and ${\Psi}(Z)$ span all
propagating fields in the SYM theory, Eqs.~\re{M=0-field} -- \re{M=4-field}. For
$Z=(z,\theta^A)=0$, the derivatives of these superfields along the ``odd''
directions in the superspace generate all field components, while the derivatives
along the ``even'' direction induce light-cone derivatives. In this way,
Eq.~\re{Wilson-gen-func} generates an infinite set of quasipartonic operators.

The Wilson operators mix under renormalization among themselves and form a closed
sector with respect to the action of the dilatation operator $\mathbb{H}$.
Applying the projector $\Pi_{\scriptscriptstyle\rm W}$ to both sides of the
evolution equation \re{EQ} one finds that in order to ensure this property one
has to require that $\Pi_{\scriptscriptstyle\rm
W}\mathbb{H}\,\mathbb{O}(Z_1,\ldots,Z_L)= \Pi_{\scriptscriptstyle\rm
W}\mathbb{H}\,\mathbb{O}_{\scriptscriptstyle \rm W}(Z_1,\ldots,Z_L)$, or
equivalently
\be
\Pi_{\scriptscriptstyle\rm W}\mathbb{H}(1-\Pi_{\scriptscriptstyle\rm W})= 0\,.
\label{W2W}
\ee
Let us examine the difference between two light-cone operators
\be
\mathbb{O}_{\rm sp}=\mathbb{O}(Z_1,\ldots,Z_L)- \mathbb{O}_{\scriptscriptstyle
\rm W}(Z_1,\ldots,Z_L)=(1-\Pi_{\scriptscriptstyle\rm
W})\mathbb{O}(Z_1,\ldots,Z_L)\,.
\ee
According to \re{Phi-Wilson}, it involves at least one spurious superfield
\re{aux-field}. The operators $\mathbb{O}_{\rm sp}$ mix under renormalization
among themselves and with the Wilson operators $\mathbb{O}_{\scriptscriptstyle
\rm W}$. The corresponding evolution kernels are given by
$\mathbb{H}_{\scriptscriptstyle \rm sp}= (1-\Pi_{\scriptscriptstyle\rm
W})\mathbb{H}(1-\Pi_{\scriptscriptstyle\rm W})$ and
$\mathbb{H}_{{\scriptscriptstyle \rm sp}\to{\scriptscriptstyle \rm
W}}=(1-\Pi_{\scriptscriptstyle\rm W})\mathbb{H}\Pi_{\scriptscriptstyle\rm W}$,
respectively. It is convenient to treat $\mathbb{O}_{\scriptscriptstyle \rm W}$
and $\mathbb{O}_{\scriptscriptstyle \rm sp}$ as two components of the same vector
and represent the dilatation operator $\mathbb{H}$ as a triangular $2\times 2$
matrix
\be
\label{SpWoperatorMixing}
\mathbb{O}(Z_1,\ldots,Z_L)=\left({\mathbb{O}_{\scriptscriptstyle \rm W} \atop
\mathbb{O}_{\scriptscriptstyle \rm sp} } \right)\,,\qquad \mathbb{H}=\left(
\begin{array}{lc} \mathbb{H}_{\scriptscriptstyle \rm
W} & 0 \\ \mathbb{H}_{{\scriptscriptstyle \rm sp}\to{\scriptscriptstyle \rm W}} &
\mathbb{H}_{\scriptscriptstyle \rm sp} \end{array} \right),
\ee
where the integral operator $\mathbb{H}_{\scriptscriptstyle \rm W}$ maps the
Wilson operators into themselves
\be
\mathbb{H}_{\scriptscriptstyle \rm W}\equiv\Pi_{\scriptscriptstyle\rm W}
\mathbb{H}=\Pi_{\scriptscriptstyle\rm W} \mathbb{H}\,\Pi_{\scriptscriptstyle\rm
W}\,.
\label{H-projected}
\ee
The zero entry in \re{SpWoperatorMixing} reflects the fact that the Wilson
operators can not mix with the spurious operators whereas the opposite is
possible.

The dilatation operator $\mathbb{H}_{\scriptscriptstyle \rm W}$ governs the scale
dependence of the operators $\mathbb{O}_{\scriptscriptstyle \rm
W}(Z_1,\ldots,Z_L)$. As was shown in Sect.~3.4, the two-particle kernels are
functions of the two-particle superconformal spin,
$\mathbb{H}_{k,k+1}=h(\mathbb{J}_{k,k+1})$, Eqs.~\re{H-PsiPsi-J} --
\re{H-PsiPhi-J}. It follows from \re{H-projected} and \re{H-multi-color} that the
two-particle kernel $ \mathbb{H}^{\scriptscriptstyle \rm
W}_{k,k+1}=\Pi_{\scriptscriptstyle\rm W} h (\mathbb{J}_{k,k+1})\,
\Pi_{\scriptscriptstyle\rm W}$ is given by the same function with
$\mathbb{J}_{k,k+1}$ replaced by a ``projected'' superconformal spin
$\mathbb{J}^{\scriptscriptstyle \rm W}_{k,k+1}=\Pi_{\scriptscriptstyle\rm
W}\mathbb{J}_{k,k+1}\Pi_{\scriptscriptstyle\rm W}$
\be
\mathbb{H}_{k,k+1}=h(\mathbb{J}_{k,k+1})\quad \longrightarrow \quad
 \mathbb{H}^{\scriptscriptstyle \rm
W}_{k,k+1}  =h(\mathbb{J}^{\scriptscriptstyle \rm W}_{k,k+1})\,.
\ee
Thus, the two-particle kernels $\mathbb{H}^{\scriptscriptstyle \rm W}_{k,k+1}$
have the same eigenvalues as the operators \re{H-PsiPsi-J} -- \re{H-PsiPhi-J}.

\section{One-loop dilatation operator}

The one-loop dilatation operator acting on single-trace nonlocal light-cone
operators, Eqs.~\re{Chiral-sector}--\re{Mixed-sector}, is given in the
multi-color limit by the sum over the two-particle evolution kernels
\re{H-multi-color}. The superconformal invariance of the SYM theory on the
light-cone allows one to determine a general form of the two-particle kernels in
various sectors, Eqs.~\re{ansatz-PhiPhi}, \re{ansatz-PhiPsi} and
\re{ansatz-PsiPhi}, but the obtained expressions \re{V-gen-SL2N} and
\re{V-ex-gen-SL2N} involve some unknown scalar functions $f$ and $f_{\rm ex}$.
Based on previous QCD calculations, we conjectured that these functions
should be given by \re{FandFex} leading to the expressions for the one-loop
dilatation operator summarized in the Introduction, Eqs.~\re{ansatz-gen-H12} --
\re{ansatz-V-ex}. In this section, we shall confirm these assertions by
calculating the one-loop corrections to the nonlocal light-cone operators,
Eqs.~\re{Chiral-sector}--\re{Mixed-sector}, and matching their divergent part
into a general expression for the one-loop dilatation operator.

We remind that the $\mathcal{N}=4$ SYM theory involves only one chiral light-cone
superfield and, in order to identify the two-particle kernel $\mathbb{H}_{12}$
entering \re{H-multi-color}, one has to calculate one-loop corrections to the
operator $\Phi(Z_1) \Phi(Z_2)$. For $\mathcal{N}\le 2$, the SYM theories are
formulated in terms of two independent chiral superfields and, therefore, there
are three additional sectors $\Psi(Z_1)\Psi(Z_2)$, $\Phi(Z_1)\Psi(Z_2)$ and
$\Psi(Z_1)\Phi(Z_2)$. In what follows we shall denote the corresponding
two-particle kernels as $\mathbb{H}_{\Phi\Phi}$, $\mathbb{H}_{\Psi\Psi}$,
$\mathbb{H}_{\Phi\Psi}$ and $\mathbb{H}_{\Psi\Phi}$. The first two kernels will
be calculated in Sect.~4.1 and the remaining two in Sect.~4.2.

To calculate the anomalous dimension of the light-cone operators
$\mathbb{O}(Z_1,\ldots,Z_L)$ we apply an approach well-known in perturbative QCD
\cite{BraGeyRob87,BalBra89}. Let us consider the matrix element of this operator
between the vacuum and a reference state, $\vev{0|\mathbb{O}(Z_1,\ldots,Z_L)|P}$.
Since the anomalous dimension of the operator does not depend on the choice of
the state $\ket{P}$, one can choose it at will, from convenience considerations
alone. To this end, we apply the Fourier transformation and
expand the superfield over the plane waves in the superspace
\be
{\Phi} (x, \theta^A) = \int \frac{d^4 p}{(2\pi)^4} \int d^{\mathcal{N}} \pi
\, \textrm{e}^{i \, p \cdot x \, + \pi_A \theta^A} \widetilde {\Phi} (p,
\pi_A) \, ,
\label{Super-Fourier}
\ee
where $\pi_A$ is the Grassmann valued momentum conjugated to the odd coordinates
$\theta^A$ and $p_\mu$ defines the momentum of the field components entering into
expansion of the superfield. Similar expansion holds for the superfield ${\Psi}
(x, \theta^A)$. Let us define $\ket{P}$ to be a state describing $L$ particles
with (super)momenta $P_k=(p_{k,\mu},\pi_{k,A})$
\be
\ket{P}= \left(\prod_{k=1}^L
\frac{i\sigma_{\mathcal{N}}N_c^2}{p_k^2}\right)^{-1}\tr \{\widetilde {\Psi}
(P_1)\ldots \widetilde {\Phi} (P_L) \}\ket{0}
\label{P-state}
\ee
The total (super)momentum of the state is $P=\sum_{k=1}^L P_k$. In addition, we
choose four-dimensional momenta of all particles, $p_{k,\mu}$, to be aligned
along the same direction in Minkowski space-time, close to the ``$-$'' direction
on the light-cone
\be
\bit{p}_{k,\perp}=0\,,\qquad p_{k+}\ll p_{k-}\,,\qquad p_k^2=2p_{k+}p_{k-}\,.
\label{kinematics}
\ee
Then, in the Born approximation, the matrix element
$\vev{0|\mathbb{O}(Z_1,\ldots,Z_L)|P}$ is given by the product of plane waves
accompanied by the propagators (see Eq.~\re{propagator}, Appendix~C). The
latter are cancelled against the prefactor in the right-hand side\ of
\re{P-state} leading in the multi-color limit to%
\footnote{If all particles entering $\ket{P}$ are identical, the right-hand
side\ of \re{Born} is given in the multi-color limit by a sum over cyclic
permutations of their momenta.}
\begin{equation}
\vev{0|\mathbb{O}^{(0)}(Z_1,\ldots,Z_L)|P}
=
\prod_{k=1}^L \textrm{e}^{- i \, P_k \cdot Z_k}
=
\parbox[c][19mm]{50mm}{\insertfig{4.3}{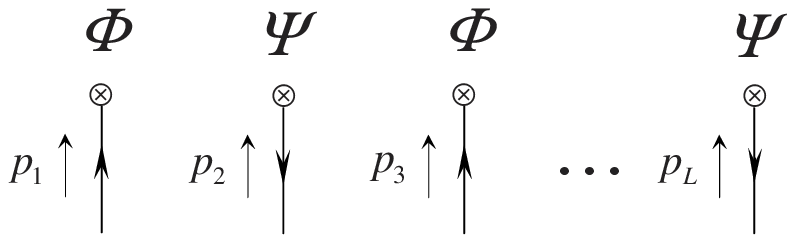}}
\label{Born}
\end{equation}
Here $Z_k=(z_k,\theta_k^A)$ defines the position of the $k-$th superfield in the
superspace and we used the notation for a scalar product in the superspace $i P
\cdot Z = i p_+ z + \pi_A \theta^A$ with $p_{+} = (p\cdot n)$. The superscript
$\scriptstyle (0)$ indicates that the matrix element is evaluated at the Born
level. For $\mathcal{N}\le 2$, to distinguish the superfields ${\Phi}(Z)$ and
${\Psi}(Z)$, we shall denote them by lines with the incoming and outgoing arrows,
respectively. In particular, in our notations the right-hand side of \re{Born}
corresponds to the following operator $\mathbb{O} (Z_1,\ldots,Z_L) = \tr\{{\Phi}
(Z_1){\Psi}(Z_2){\Phi}(Z_3) \ldots {\Psi}(Z_L) \}$. For $\mathcal{N}=4$ we shall
denote the superfield ${\Phi}(Z)$ by a line without an arrow.

Substituting \re{Super-Fourier} into the light-cone SYM actions \re{N=0-M} and
\re{N=4-M}, it is straightforward to work out the Feynman diagram technique for
calculating perturbative corrections to \re{Born}. The Feynman rules involve
three elements: propagators of the superfields, triple  and quartic interaction
vertices. For $\mathcal{N}\le 2$ , the interaction vertices are $\Phi\Psi\Psi$,
$\Psi\Psi\Phi$ and $\Phi\Phi\Psi\Psi$, whereas for $\mathcal{N}=4$ they are
$\Phi\Phi\Phi$ and $\Phi\Phi\Phi\Phi$. Their explicit expressions are given in
Appendix~C. For $\mathcal{N}=0$ similar technique has been worked out in
Ref.~\cite{DerKir01}. As was demonstrated there, the use of the light-cone action
simplifies significantly the calculation of evolution kernels as compared to a
conventional ``covariant'' approach based on the full Yang-Mills action.

Calculating one-loop corrections to the matrix element
$\vev{0|\mathbb{O}^{(1)}(Z_1,\ldots,Z_L)|P}$, we shall apply the dimensional
regularization and evaluate the momentum integrals in $D=4-2\varepsilon$
dimensions
\be
\int \frac{d^4 p}{(2\pi)^4} \to \mu^{4-D} \int \frac{d^D p}{(2\pi)^D}
\ee
with the scale $\mu$ playing the role of a UV cut-off. According to the evolution
equation \re{EQ}, the one-loop dilatation operator $\mathbb{H}$ is related to the
coefficient in front of a pole $1/\varepsilon$ in the expression for the matrix
element of the nonlocal light-cone operator $\mathbb{O}(Z_1,\ldots,Z_L)$
\be
\vev{0|\mathbb{O}^{(1)}|P}= -\frac{g^2
N_c}{(4\pi)^2}\frac{\mu^{2\varepsilon}}{\varepsilon} \vev{0|\left[
\mathbb{H}\cdot \mathbb{O}^{(0)} +L \gamma_\mathcal{N}^{(0)} \mathbb{O}^{(0)}
 \right]|P}+\ldots\,,
\label{pole-H}
\ee
where ellipses denote terms regular for $\varepsilon\to 0$ and
$\gamma_\mathcal{N}^{\scriptscriptstyle (1)}$ defines the one-loop correction to
the anomalous dimension of the superfield, $\gamma_\mathcal{N} =\frac{g^2
N_c}{(4\pi)^2}\gamma_\mathcal{N}^{\scriptscriptstyle (0)} + \mathcal{O}(g^4)$.
Note that in the SYM theory on the light-cone cone, the anomalous dimensions of
the superfields $\Phi(Z)$ and $\Psi(Z)$ are equal to each other and are
proportional to the $\beta-$function, $\gamma_\mathcal{N}=\beta_\mathcal{N}(g)/g$
(see Appendix~D1). The reason why we split the right-hand side\ of \re{pole-H}
into the sum of two terms is that the second term containing
$\gamma_\mathcal{N}^{(1)}$ comes entirely from diagrams containing
self-energy corrections and can be separated from the very beginning. In what
follows, we will not display this term and tacitly imply that it should be added
to the final expression for $\vev{0|\mathbb{O}^{(1)}|P}$.

\subsection{Diagonal sector}

Let us calculate one-loop corrections to the matrix elements of single-trace
operators involving the products ${\Phi}(Z_1) {\Phi}(Z_2)$ and
${\Psi}(Z_1) {\Psi}(Z_2)$ and use them to determine the two-particle
evolution kernels $\mathbb{H}_{\Phi\Phi}$ and $\mathbb{H}_{\Psi\Psi}$,
respectively.

\subsubsection{$\mathcal{N}\le 2$}

\begin{figure}[t]
\begin{center}
\mbox{
\begin{picture}(0,135)(240,0)
\put(20,0){\insertfig{15}{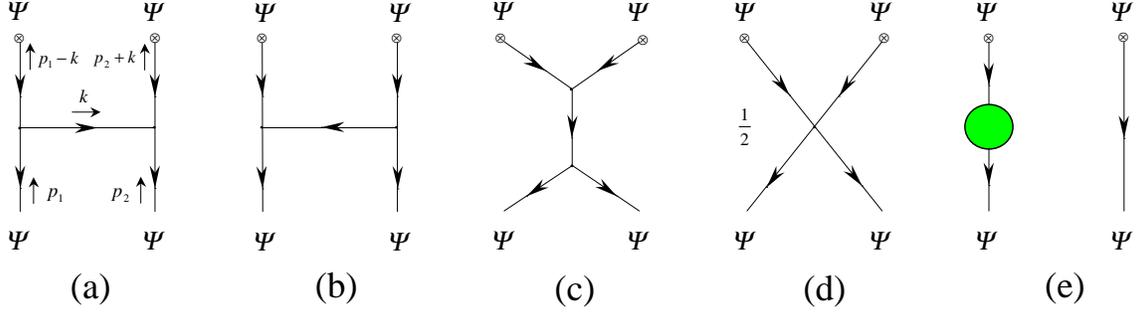}}
\end{picture}
}
\end{center}
\vspace*{-5mm} \caption{\label{HolomorphicKernel} Feynman diagrams contributing
to the one-loop dilatation operator in the $\Psi\Psi-$sector.}
\end{figure}

We start with the $\Psi\Psi-$sector. For $\mathcal{N}\le 2$, the one-loop Feynman
diagrams contributing to $\vev{0|\tr \{{\Psi}(Z_1) {\Psi}(Z_2)...\}|P}$
are shown in Figure~\ref{HolomorphicKernel}. Let us examine the diagrams one
after another.

The diagram Fig.~\ref{HolomorphicKernel}(e) describes the self-energy correction
to the superfield and contributes to the one-loop anomalous dimension
$\gamma_\mathcal{N}(g)$, Eqs.~\re{EQ} and \re{pole-H}. Its calculation can be
found in Appendix~D1. For the annihilation diagram,
Fig.~\ref{HolomorphicKernel}(c), one applies the Feynman rules (see Appendix~C)
and finds that it gives rise to an integral proportional to the holomorphic
component of the loop momenta, $k=(k_1+ik_2)/\sqrt{2}$ (see Eqs.~\re{brackets}
and \re{kinematics})
\be
(p_1-k,p_2+k) = - (p_1+p_2)_+k\,.
\label{annihilation}
\ee
As a result, it equals zero upon integration over the transverse momenta $\int
d^2\bit{k}_\perp\equiv \int dk_1dk_2$. For the sum of the remaining three
diagrams, Figs.~\ref{HolomorphicKernel}(a), (b) and (d), one gets the following
Feynman integral%
\footnote{Hereafter, to simplicity formulae, we do not display the factors
$\e^{-iz_n p_n-\pi_{nA}\theta_n^A}$ corresponding to noninteracting superfields
with the coordinates $Z_n=(z_n,\theta_n^A)$ with $n=3,\ldots,L$.}
(the details can be found in Appendix~D2)
\ba
\vev{0|\mathbb{O}_{\Psi\Psi}^{(1)}(Z_1,Z_2,...)|P} \!\!\!&=&\!\!\! -ig^2 N_c
\mu^{4-D}\!\!\int \frac{d^D k}{(2\pi)^D}\textrm{e}^{- i z_1(p_1 - k) - iz_2
(p_2+k)} \int d^\mathcal{N} \pi \,\textrm{e}^{- (\pi_1-\pi)_A\theta_1^A -
(\pi_2+\pi)_A\theta_2^A} \nonumber
\\
&&\hspace*{-30mm}
\times\left[\frac{\delta^{(\mathcal{N})}\left(\pi-\pi_1\frac{k_+}{p_{1+}}\right)}
{k^2(p_1-k)^2}\frac{p_{1+}}{k_+}\left(\frac{p_{1+}\!\!-k_+}{p_{1+}}\right)^{2-\mathcal{N}}
-\frac{\delta^{(\mathcal{N})}\left(\pi-\pi_2\frac{k_+}{p_{2+}}\right)}{k^2(p_2+k)^2}
\frac{p_{2+}}{k_+}\left(\frac{p_{2+}\!\!+k_+}{p_{2+}}\right)^{2-\mathcal{N}}
\right]\!\!,
 \nonumber\\{}
\label{Psi-Psi-int}
\ea
where the poles at $k_+=0$ are regularized using the Mandelstam-Leibbrandt
prescription, Eq.~\re{ML-def}. Here $\delta(\pi)$ is the Dirac $\delta-$function
for odd coordinates defined in \re{delta-function}
\be
\int d^\mathcal{N} \pi \, \delta^{(\mathcal{N})}(\pi - \pi')\,
\textrm{e}^{\pi_A\theta^A} = \textrm{e}^{\pi_A'\theta^A}\,.
\label{delta-pi}
\ee

For $\mathcal{N}=0$ in \re{Psi-Psi-int}, the integral  over the odd momenta $\pi$
is absent and the odd $\delta-$functions are replaced by $1$. For $\mathcal{N}\ge
1$, the $\pi-$integral in \re{Psi-Psi-int} is trivial due to \re{delta-pi}, while
integration over the loop momentum $k_\mu$ can be easily performed with a help of
the identity \re{PPmomentumIntegral} (for $n=1$). In this way, one can express the
divergent (for $\varepsilon\to 0$) part of \re{Psi-Psi-int} as a sum of plane
waves integrated over a scalar variable $\alpha$ which has the meaning of the
momentum fraction $k_+=\alpha p_+$
\begin{eqnarray}
\vev{0|\mathbb{O}_{\Psi\Psi}^{(1)}(Z_1,Z_2,...)|P}\!\!\!&=&\!\!\! - \frac{g^2
N_c}{(4 \pi)^2 }\frac{\mu^{2\varepsilon}}{\varepsilon} \int_0^1 \frac{d
\alpha}{\alpha} \bigg\{ 2\, {\rm e}^{ -i P_1\cdot Z_1
  - i P_2\cdot Z_2}
\label{Psi-Psi-div}
\\
&&\hspace*{-20mm} - (1 - \alpha)^{2-\mathcal{N}} \left[{\rm e}^{- i P_1
\cdot((1-\alpha) Z_1 + \alpha Z_2) - i P_2\cdot Z_2} + {\rm e}^{- i P_1\cdot Z_1 -
i P_2\cdot ( (1-\alpha) Z_2 + \alpha Z_1)}\right] \bigg\} \, . \nonumber
\end{eqnarray}
Here the notation was introduced for the scalar product in the superspace between
the vectors $Z_k=(z_k,\theta_k^A)$ and $P=(p_{k+},\pi_{k,A})$ (with $k=1,2$)
\be
i(P\cdot Z) \equiv ip_+ z + \pi_A \theta^A
\ee
Making use of \re{Born}, one can rewrite the right-hand side\ of \re{Psi-Psi-div} in terms
of the Born level matrix element leading to
\begin{eqnarray}
\mathbb{O}_{\Psi\Psi}^{(1)}(Z_1,Z_2,...)\!\!\!&=&\!\!\! - \frac{g^2 N_c}{(4
\pi)^2 }\frac{\mu^{2\varepsilon}}{\varepsilon} \int_0^1 \frac{d \alpha}{\alpha}
\bigg\{ 2\, \mathbb{O}_{\Psi\Psi}^{(0)}(Z_1,Z_2,...)
\label{Psi-Psi-vev}
\\
&&\hspace*{-20mm} - (1 - \alpha)^{2-\mathcal{N}} \left[
\mathbb{O}_{\Psi\Psi}^{(0)}((1-\alpha) Z_1 + \alpha Z_2,Z_2,...) +
\mathbb{O}_{\Psi\Psi}^{(0)}(Z_1,(1-\alpha) Z_2 + \alpha Z_1,...)\right] \bigg\}
\, . \nonumber
\end{eqnarray}
Matching this expression into \re{pole-H} and keeping in mind that the term
involving $\gamma_\mathcal{N}^{(1)}$ in \re{pole-H} comes from the self-energy
diagram, one identifies the two-particle evolution kernel $\mathbb{H}_{\Psi\Psi}$
governing the scale dependence of $\tr\{\Psi(Z_1)\Psi(Z_2)\ldots \}$ in the
$\mathcal{N}-$extended SYM theory
\begin{eqnarray}
\mathbb{H}_{\Psi\Psi}{\Psi} (Z_1) {\Psi}
(Z_2)\bigg|_{\mathcal{N}=0,1,2} \!\!\!&=&\!\!\! \int_0^1 \frac{d \alpha}{\alpha}
\bigg\{ 2 {\Psi} (Z_1) {\Psi} (Z_2)
\label{Psi-Psi-ker}
\\
&&\hspace*{-20mm} - (1 - \alpha)^{2-\mathcal{N}} \big[{\Psi} ((1-\alpha) Z_1
+ \alpha Z_2) {\Psi} (Z_2) + {\Psi} (z_1) {\Psi} ((1-\alpha) Z_2 +
\alpha Z_1)\big] \bigg\} \, . \nonumber
\end{eqnarray}
The integrand has a pole at $\alpha=0$ but the linear combination of the
superfields vanishes for $\alpha\to 0$ so that the integral is convergent.

We remind that \re{Psi-Psi-ker} is valid in $\mathcal{N}=0$, $\mathcal{N}=1$
and $\mathcal{N}=2$ SYM theories. In a perfect agreement with our expectations,
\re{Psi-Psi-ker} coincides with the expression for the $SL(2|\mathcal{N})$
invariant operator \re{V-super} evaluated for $j_1=j_2=(3-\mathcal{N})/2$
corresponding to the conformal spin of the $\Psi-$superfield, Eq.~\re{Psi-j},
\be
\mathbb{H}_{\Psi\Psi}=\mathbb{V}^{(j_\Psi,j_\Psi)}\,.
\label{H-Psi-Psi-W}
\ee
Since the $\Psi-$superfield does not contain nonlocal fields, this kernel acts on
the subspace of Wilson operators only, $\mathbb{H}_{\Psi\Psi}^{\scriptscriptstyle
\rm W}=\Pi_{\scriptscriptstyle\rm W}\mathbb{H}_{\Psi\Psi}=\mathbb{H}_{\Psi\Psi}$.

Let us repeat a similar calculation in the $\Phi\Phi-$sector and obtain the
one-loop expression for the two-particle kernel $\mathbb{H}_{\Phi\Phi}$. As
before, our starting point is the matrix element of the light-cone operator
$\vev{0|\tr\{\Phi(Z_1)\Phi(Z_2)...\}|P}$. It receives one-loop corrections from
the Feynman diagrams similar to those shown in Fig.~\ref{HolomorphicKernel}. The
only difference is that the direction of the arrow for incoming and outgoing
lines should be flipped. As in the previous case, the annihilation diagram
(Fig.~\ref{HolomorphicKernel}c) vanishes, Eq.~\re{annihilation}, and the diagram
with the self-energy produces the one-loop anomalous dimension of the
$\Phi-$field. For the sum of three remaining diagrams, one gets (see Appendix~D3
for details)
\ba
\vev{0|\mathbb{O}_{\Phi\Phi}^{(1)}(Z_1,Z_2,...)|P}\!\!\!&=&\!\!\! -ig^2 N_c
\mu^{4-D}\!\!\int \frac{d^D k}{(2\pi)^D}\textrm{e}^{-i z_1(p_1-k)- iz_2
(p_2+k)}\int d^\mathcal{N} \pi
\,\textrm{e}^{-(\pi_1-\pi)_A\theta_1^A-(\pi_2+\pi)_A\theta_2^A} \nonumber
\\
&&\hspace*{-30mm}
\times\left[\frac{\delta^{(\mathcal{N})}\left(\pi-\pi_2\frac{k_+}{p_{2+}}\right)}
{k^2(p_1-k)^2}\frac{p_{1+}}{k_+}\left(\frac{p_{2+}}{p_{2+}\!+k_+}\right)^{2}
\!\!-\frac{\delta^{(\mathcal{N})}\left(\pi-\pi_1\frac{k_+}{p_{1+}}\right)}{k^2(p_2+k)^2}
\frac{p_{2+}}{k_+}\left(\frac{p_{1+}}{p_{1+}\!-k_+}\right)^{2} \right]\!\!,
\nonumber \\
{}
\label{Phi-Phi-int}
\ea
where the poles in $k_+$ are regularized using the Mandelstam-Leibbrandt
prescription \re{ML-def}. In comparison with \re{Psi-Psi-int}, the momenta of the
two incoming lines get interchanged inside the odd $\delta-$functions and the
factor $(...)^{2-\mathcal{N}}$ is modified. This makes the calculation much more
involved. Indeed, we expect from \re{ansatz-PhiPhi} that the two-particle kernel
$\mathbb{H}_{\Phi\Phi}$ should have a more complicated form as compared with
$\mathbb{H}_{\Psi\Psi}$.

The expression inside the square brackets in \re{Phi-Phi-int} can be rewritten
after some algebra in the following form
\ba
[\cdots]_{\mathcal{N}}&= &
\frac{\delta^{(\mathcal{N})}\left(\pi-\pi_1\frac{k_+}{p_{1+}}\right)}
{k^2(p_1-k)^2}\frac{p_{1+}^3}{k_+(p_{1+}-k_+)^2}
-\frac{\delta^{(\mathcal{N})}\left(\pi-\pi_2\frac{k_+}{p_{2+}}\right)}
{k^2(p_2+k)^2}\frac{p_{2+}^3}{k_+(p_{2+}+k_+)^2}
\nonumber \\
&-&\frac{(p_{1+}+p_{2+})^2}{(p_1-k)^2(p_2+k)^2}
\left[\frac{p_{1+}(p_{2+}+k_+)+p_{2+}(p_{1+}-k_+)}{(p_{1+}-k_+)^2(p_{2+}+k_+)^2}
\delta^{(\mathcal{N})}\left(\pi-\varpi\right) \right. +
{\mathcal{X}_\mathcal{N}}\bigg],
\label{aux1}
\ea
where the notation was introduced for 
\ba
\varpi_A &=& \pi_{1A}\frac{p_{2+}+k_+}{p_{1+}+p_{2+}} -
\pi_{2A}\frac{p_{1+}-k_+}{p_{1+}+p_{2+}} \nonumber
\\
\mathcal{X}_{\mathcal{N}=1} &=& \frac{{\pi_1 p_{2+}-\pi_2
p_{1+}}}{(p_{1+}-k_+)(p_{2+}+k_+)(p_{1+}+p_{2+})}
\nonumber\\
\mathcal{X}_{\mathcal{N}=2} &=& \frac{\varepsilon^{AB}(\pi_A-\varpi_A)(\pi_{1,B}
p_{2+}-\pi_{2,B} p_{1+})}{(p_{1+}-k_+)(p_{2+}+k_+)(p_{1+}+p_{2+})}\,.
\label{Pi-def}
\ea
For $\mathcal{N}=0$, one has $\mathcal{X}_{\mathcal{N}=0}=0$ and the odd
$\delta-$functions are replaced by $1$ in \re{aux1}. The first two terms in the
right-hand side\ of \re{aux1} depend on a single ``external'' momentum,
$P_1=(p_{1+},\pi_{1A})$ and $P_2=(p_{2+},\pi_{2A})$, respectively. This allows
one to perform the $k-$integration in \re{Phi-Phi-int} by making use of the identity
\re{PPmomentumIntegral}. Similarly, one replaces the integration variable $k_+' =
p_{2+}+k_+$ in the third term in \re{aux1} and performs the $k'-$integration with
a help of the identity \re{PPmomentumIntegral}. The details of the calculation can
be found in Appendix~D3.

The resulting expression for the Feynman integral in \re{Phi-Phi-int} is similar
to \re{Psi-Psi-div} and \re{Psi-Psi-vev}. Namely, the divergent part of
$\vev{0|\mathbb{O}_{\Phi\Phi}^{(1)}(Z_1,Z_2,...)|P}$ has the form of the
$\alpha-$integral with the integrand given by a rather lengthy expression.
Remarkably enough, it can be cast into the following form
\be
\mathbb{O}_{\Phi\Phi}^{(1)}(Z_1,Z_2,...)= - \frac{g^2 N_c}{(4 \pi)^2
}\frac{\mu^{2\varepsilon}}{\varepsilon}\bigg\{
\mathbb{V}_{\Phi\Phi}(1-\Pi_{\Phi\Phi}) +
\Delta_{\Phi\Phi}\bigg\}\mathbb{O}_{\Phi\Phi}^{(0)}(Z_1,Z_2,...)\,,
\label{Phi-Phi-div}
\ee
where the operators $\mathbb{V}_{\Phi\Phi}$, $\Pi_{\Phi\Phi}$ and
$\Delta_{\Phi\Phi}$ act on the superfields with the coordinates  $Z_1$ and $Z_2$.
They have the same, universal form for $\mathcal{N}=0$, $\mathcal{N}=1$ and
$\mathcal{N}=2$. The projector $\Pi_{\Phi\Phi}$ was already defined in
\re{Pi12-main}. The operator $\mathbb{V}_{\Phi\Psi}$ is given by
\ba
&&\mathbb{V}_{\Phi\Phi} \mathbb{O} (Z_1,Z_2...) = \int_0^1 \frac{d\alpha}{ \alpha
} \, \bigg\{ 2 \, \mathbb{O} (Z_1,Z_2,...)
\label{V-int}
\\
&& \hspace*{+20mm} - (1-\alpha)^{-2}\big[\mathbb{O}((1-\alpha) Z_1+\alpha
Z_{2},Z_2,...) + \mathbb{O}(Z_1,(1-\alpha)Z_2+ \alpha Z_{1},...)\big] \bigg\}
\nonumber \, .
\ea
One verifies that it coincides with the $SL(2|\mathcal{N})$ invariant operator
\re{V-super}, $\mathbb{V}_{\Phi\Phi}=\mathbb{V}^{(-1/2,-1/2)}$.
Eq.~\re{j=-1/2}. The integral in \re{V-int} diverges for $\alpha\to 1$ and,
therefore, the operator $\mathbb{V}_{\Phi\Phi}$ is well-defined only for the
operators $\mathbb{O}(Z_1,Z_2,...)$ which vanish sufficiently fast as $Z_1\to
Z_2$. It is easy to verify using \re{Pi12-main} that
\be
(1-\Pi_{\Phi\Phi})\mathbb{O}((1-\alpha) Z_1+\alpha Z_{2},Z_2,...) \sim
(1-\alpha)^2
\ee
as $\alpha\to 1$ and, as a consequence, $\mathbb{V}_{\Phi\Phi}(1-\Pi_{\Phi\Phi})$
is a well-defined integral operator. Finally, the operator $\Delta_{\Phi\Phi}$ is
defined as
\ba
\Delta_{\Phi\Phi} \mathbb{O} (Z_1,Z_2,...) &=& \left( 1 - \ft12 Z_{21}
\partial_{Z_2} \right)\bigg[\frac{\partial_{1+} - \partial_{2+}}{\partial_{1+} + \partial_{2+}}
\mathbb{O} (Z_1,Z_2,...)\bigg]\bigg|_{Z_1=Z_2} \nonumber
\\
&+&  \left( 1 - \ft12 Z_{12}
\partial_{Z_1} \right) \bigg[\frac{\partial_{2+} - \partial_{1+}}{\partial_{1+} +
\partial_{2+}} \mathbb{O} (Z_1,Z_2,...)\bigg]\bigg|_{Z_2=Z_1} \, ,
\label{Delta}
\ea
where $\partial_{k+} = \partial/\partial_{z_k}$ denotes the derivative with
respect to the light-cone coordinate, $Z_k=(z_k,\theta_k^A)$ and the notation was
introduced for $Z_{jk}=(z_j-z_k,\theta^A_j-\theta^A_k)$ and $ Z_{jk}\cdot
\partial_{Z_j} \equiv (z_j-z_k)  { \partial_ {z_j}} +
(\theta_j^A-\theta_k^A)\partial_{\theta_j^A}$ with $j,k=1,2$. In Eq.~\re{Delta},
one first evaluates the expressions inside the square brackets for $Z_1=Z_2$ (or
$Z_2=Z_1$) and applies the external derivative afterwards.

Matching \re{Phi-Phi-div} into \re{pole-H} we conclude that the two-particle
evolution kernel in the $\Phi\Phi-$sector is given by
\be
\mathbb{H}_{\Phi\Phi}{\Phi} (Z_1) {\Phi}
(Z_2)\bigg|_{\mathcal{N}=0,1,2}=\bigg\{
 \mathbb{V}^{(-1/2,-1/2)}(1-\Pi_{\Phi\Phi}) +
\Delta_{\Phi\Phi}\bigg\}{\Phi} (Z_1) {\Phi} (Z_2)\,,
\label{H-Phi-Phi-final}
\ee
where the operators $\mathbb{V}^{(-1/2,-1/2)}$, $\Pi_{\Phi\Phi}$ and
$\Delta_{\Phi\Phi}$ were given in Eqs.~\re{V-super}, \re{Pi12-main} and
\re{Delta}, respectively. Notice that the $\mathcal{N}-$dependence enters
\re{H-Phi-Phi-final} only through the dimension of the superspace
$Z=(z,\theta^A)$, with ${\scriptstyle A}=1,\ldots,\mathcal{N}$.

Comparing \re{H-Phi-Phi-final} with our ansatz for the two-particle kernel in the
$\Phi\Phi-$sector, Eq.~\re{ansatz-PhiPhi}, we find that \re{H-Phi-Phi-final}
contains the additional operator $\Delta_{\Phi\Phi}$. To understand its origin,
we recall that ${\Phi} (Z_1) {\Phi} (Z_2)$ is a generating function for
both Wilson operators and composite operators involving spurious fields. As was
explained in Sect.~3.2, the latter operators can be eliminated by implying the
projector $\Pi_{\scriptscriptstyle \rm W}$ to both sides of
\re{H-Phi-Phi-final}. According to its definition, Eqs.~\re{Wilson-gen-func} and
\re{Phi-Wilson}, the operator $\Pi_{\scriptscriptstyle \rm W}$ annihilates the
states $\mathbb{O} (Z_1,Z_2,...)$ which either do not depend on at least one of
the superspace coordinate $Z_k$ or are linear in $Z_k$. It is easy to see that each
term in the right-hand side\ of \re{Delta} verifies these conditions and, therefore,
\be
\Pi_{\scriptscriptstyle \rm W}\Delta_{\Phi\Phi} \mathbb{O} (Z_1,Z_2,...)=0\,.
\ee
This means that the operator $\Delta_{\Phi\Psi}$ does not affect Wilson operators
and only contributes to the scale dependence of spurious operators. Projecting
both sides of \re{H-Phi-Phi-final} onto the subspace of Wilson operators
according to \re{H-projected}, we find that the ``physical'' dilatation operator
in the $\Phi\Phi-$sector is given by
\be
\mathbb{H}_{\Phi\Phi}^{\scriptscriptstyle \rm W}\equiv\Pi_{\scriptscriptstyle\rm
W} \mathbb{H}_{\Phi\Phi}=\Pi_{\scriptscriptstyle\rm W}
\mathbb{V}^{(-1/2,-1/2)}(1-\Pi_{\Phi\Phi})\,. \label{H-Phi-Phi-W}
\ee
One verifies that $\mathbb{H}_{\Phi\Phi}^{\scriptscriptstyle \rm W}$ satisfies
\re{W2W} and coincides with $\Pi_{\scriptscriptstyle\rm W} \mathbb{H}^{\rm
(ansatz)}_{\Phi\Phi}$, Eq.~\re{ansatz-PhiPhi}. Thus, the evolution kernels
$\mathbb{H}_{\Phi\Phi}$ and $\mathbb{H}^{\rm (ansatz)}_{\Phi\Phi}$ are identical
on the subspace of Wilson operators.

\subsubsection{$\mathcal{N}=4$}

In the $\mathcal{N}=4$ SYM theory, there is only the $\Phi\Phi-$sector. To calculate
the corresponding evolution kernel $\mathbb{H}_{\Phi\Phi}$ one has to evaluate
the one-loop corrections to $\vev{0|\tr \{\Phi(Z_1)\Phi(Z_2)...\}|P}$. They are
given by the same Feynman diagrams in Fig.~\ref{HolomorphicKernel} as before. The
only difference is that for $\mathcal{N}=4$ the lines do not have arrows. In this
case, the diagrams in Fig.~\ref{HolomorphicKernel}(a) and (b) are identical and
only one of them has to be taken into account. The divergent part of the
self-energy diagram in Fig.~\ref{HolomorphicKernel}(e) is proportional to the
$\beta-$function in the $\mathcal{N}=4$ SYM and it
vanishes~\cite{SohWes81,BriLinNil83,Man83,HowSteow84} (see Appendix~D1). The
annihilation diagram in Fig.~\ref{HolomorphicKernel}(c) does not contribute by
the same reason as before: it is proportional to the holomorphic component of the
loop momentum $k=(k_1+ik_2)/\sqrt{2}$ and vanishes upon integration over $\int
dk_1dk_2$, Eq.~\re{annihilation}.

Applying the $\mathcal{N}=4$ Feynman rules (see Appendix~C) one finds that the
sum of the remaining two diagrams is given by the following lengthy expression
\ba
&& \vev{0|\mathbb{O}_{\Phi\Phi}^{(1)}(Z_1,Z_2,...)|P}=\frac{i}4g^2 N_c
\mu^{4-D}\!\!
\int \frac{d^D k}{(2\pi)^D}\frac{\textrm{e}^{-i z_1(p_1+k)- iz_2
(p_2-k)}}{(p_1+k)^2(p_2-k)^2}\int d^{4} \pi
\,\textrm{e}^{-(\pi_1+\pi)_A\theta_1^A-(\pi_2-\pi)_A\theta_2^A} \nonumber
\\
&&\quad\times\Bigg\{
\left(\frac{[p_1+k,p_2-k]}{((p_1+k)_+(p_{2}-k)_+)^2}+\frac{[p_1,p_2]}{(p_{1+}p_{2+})^2}\right)
\left(\frac{(p_1-p_2+2k)_+(p_1-p_2)_+}{(p_1+p_2)_+^2}+1\right) \nonumber
\\
& &\qquad -
\left(\frac{[p_2,k]}{(p_{2+}(p_2-k)_+)^2}+\frac{[p_1,k]}{(p_{1+}(p_1+k)_+)^2}\right)
\left(\frac{(2p_1+k)_+(2p_2-k)_+}{k_+^2}+1+\frac{4\bit{k}_\perp^2
p_{1+}p_{2+}}{k^2k_+^2}\right) \nonumber
\\
& &\qquad +
2\left(\frac{[p_2,k]}{(p_{2+}(p_2-k)_+)^2}+\frac{[p_1,k]}{(p_{1+}(p_1+k)_+)^2}
-\frac{[p_2,p_1+k]}{(p_{2+}(p_1+k)_+)^2}-\frac{[p_1,p_2-k]}{(p_{1+}(p_2-k)_+)^2}\right)
\Bigg\}. \nonumber
\\
{} \label{Phi-Phi-N=4}
\ea
Here the term involving transverse components of the loop momentum,
$\bit{k}_\perp^2$, comes from the diagram with triple coupling shown in
Fig.~\ref{HolomorphicKernel}(a) and the rest---from the diagram with quartic
coupling, Fig.~\ref{HolomorphicKernel}(d).

Eq.~\re{Phi-Phi-N=4} involves the square bracket between two (super)momenta
defined in \re{brackets}. Using its properties, the expression inside the curly
brackets in \re{Phi-Phi-N=4} can be simplified as described in Appendix~D4.
Remarkably enough, it can be brought to the very same form as in Eq.~\re{aux1}.
Namely, it is given by $[\cdots]_{\mathcal{N}=4}$ with
\be
\mathcal{X}_{\mathcal{N}=4} =\frac1{3!}\epsilon^{ABCD}
\frac{(\pi-\varpi)_{A}(\pi-\varpi)_{B}(\pi-\varpi)_{C}(\pi_{1,D} p_{2+}-\pi_{2,D}
p_{1+})}{(p_{1+}-k_+)(p_{2+}+k_+)(p_{1+}+p_{2+})}\,,
\label{X-N=4}
\ee
where the odd momentum $\varpi_A$ was defined in \re{Pi-def}. Eq.~\re{X-N=4}
generalizes the expression for $\mathcal{X}_{\mathcal{N}=1,2}$, Eq.~\re{Pi-def},
for $\mathcal{N}=4$. This suggests that \re{Phi-Phi-N=4} can be obtained from the
similar matrix element for $\mathcal{N}\le 2$, Eq.~\re{Phi-Phi-div}, by simply
extending the formulae to $\mathcal{N}=4$. We confirm this by an explicit
calculation of \re{Phi-Phi-N=4} in Appendix~D4. Thus, the one-loop evolution
kernel in the $\mathcal{N}=4$ SYM theory is given by
\be
\mathbb{H}_{\Phi\Phi}{\Phi} (Z_1) {\Phi}
(Z_2)\bigg|_{\mathcal{N}=4}=\bigg\{ \mathbb{V}^{(-1/2,-1/2)}(1-\Pi_{\Phi\Phi}) +
\Delta_{\Phi\Phi}\bigg\}{\Phi} (Z_1) {\Phi} (Z_2)\,,
\label{N=4-kernel}
\ee
where $Z=(z,\theta^A)$ with ${\scriptstyle A}=1,...,4$ and the operators
$\mathbb{V}^{(-1/2,-1/2)}$, $\Pi_{\Phi\Phi}$ and $\Delta_{\Phi\Phi}$ were
introduced in Eqs.~\re{V-super}, \re{Pi12-main} and \re{Delta}, respectively.

The operator $\mathbb{H}_{\Phi\Phi}$, Eq.~\re{N=4-kernel}, has the same form as
the evolution kernel for $\mathcal{N}\le 2$ in the $\Phi\Phi-$sector,
Eq.~\re{H-Phi-Phi-final}. In fact, the two operators would coincide if one
formally put $\mathcal{N}=4$ in \re{H-Phi-Phi-final}. As we will show in
Sect~4.3, this property is not accidental and is one of the consequences of a
general relation between the evolution kernels in the
$\mathcal{N}=4$ and $\mathcal{N}\le 2$ SYM theories. Finally, projecting
\re{N=4-kernel} onto the subspace of Wilson operators \re{H-projected} we
obtain the same expression for $\mathbb{H}_{\Phi\Psi}^{\scriptscriptstyle \rm
W}$ as before, Eq.~\re{H-Phi-Phi-W}.

\subsection{Mixed sector}

The two-particle kernel \re{N=4-kernel} allows one to construct the one-loop
dilatation operator in the $\mathcal{N}=4$ SYM theory. For $\mathcal{N}\le 2$
the two-particle kernel is given by a $2\times 2$ matrix \re{H-matrix}. Its
diagonal entries, $\mathbb{H}_{\Psi\Psi}$ and $\mathbb{H}_{\Phi\Phi}$, are
given by \re{H-Psi-Psi-W} and \re{H-Phi-Phi-W}. In this section, we calculate
the two-particle kernels in the $\Phi\Psi-$ and $\Psi\Phi-$sectors,
$\mathbb{H}_{\Phi\Psi}$ and $\mathbb{H}_{\Psi\Phi}$, respectively.

\begin{figure}[t]
\begin{center}
\mbox{
\begin{picture}(0,125)(240,0)
\put(3,0){\insertfig{16.5}{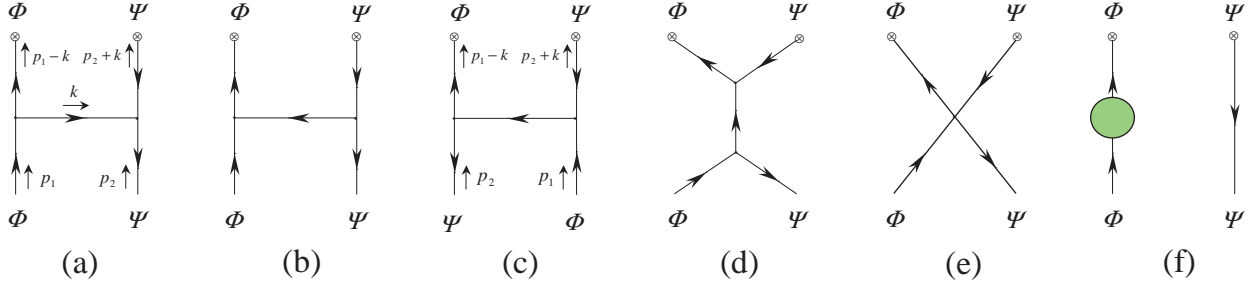}}
\end{picture}
}
\end{center}
\vspace*{-5mm} \caption{\label{MixedcKernel} Feynman diagrams contributing to the
one-loop dilatation operator in the $\Phi\Psi-$sector.}
\end{figure}

To start with, we examine one-loop corrections to the matrix element
$\vev{0|\tr\{\Phi(Z_1)\Psi(Z_2)...\}|P}$ defined by the Feynman diagrams shown
in Fig.~\ref{MixedcKernel}. As before, the self-energy diagram in
Fig.~\ref{MixedcKernel}(f) gives rise to the anomalous dimension of the
superfield while the annihilation diagram in Fig.~\ref{MixedcKernel}(d)
vanishes after integration over the transverse components of the loop momentum. The
diagrams in Fig.~\ref{MixedcKernel}(a) and (b) describe the transition
$\Phi\Psi\to \Phi\Psi$, the diagram in Fig.~\ref{MixedcKernel}(c) describes the
transition $\Phi\Psi\to \Psi\Phi$ and the diagram in Fig.~\ref{MixedcKernel}(c)
contributes to both.

For the sake of simplicity, we first consider the $\mathcal{N}=0$ theory. In
this case, the superspace does not have ``odd'' directions and coincides with
the light-cone, $Z=z$. Calculating the Feynman diagrams shown in
Fig.~\ref{MixedcKernel}(a), (b), (c) and (e), one finds that the one-loop
correction to the matrix element $\vev{\mathbb{O}_{\Phi\Psi}} \equiv
\vev{0|\tr\{\Phi(Z_1)\Psi(Z_2)...\}|P}$ can be split into a sum of two terms
corresponding to the $\Phi\Psi\to \Phi\Psi$ and $\Phi\Psi\to \Psi\Phi$
transitions. The details of calculations can be found in Appendix~D5. The
final result for the one-loop correction to $\vev{\mathbb{O}_{\Phi\Psi}}$
in the channel $\Phi\Psi\to \Phi\Psi$ is given by
\begin{eqnarray}
\vev{\mathbb{O}_{\Phi\Psi}^{(1)}(Z_1,Z_2,...)} \!\!\!&\stackrel{\Phi\Psi\to
\Phi\Psi}{=}&\!\!\! -\frac{g^2 N_c}{(4
\pi)^2}\frac{\mu^{2\varepsilon}}{\varepsilon}
\label{Phi-Psi-a}\\
&& \hspace*{-30mm}\times \Big\{ \mathbb{V}_{\Phi\Psi} \left( 1 - {\Pi}_{\Phi\Psi}
\right) \vev{\mathbb{O}_{\Phi\Psi}^{(0)}(Z_1,Z_2,...)} +
\Delta_{\Phi\Psi}^{({\cal N} = 0)}\vev{\mathbb{O}_{\Phi\Psi}^{(0)}(Z_1,Z_2,...)}
\Big\} \, , \nonumber
\end{eqnarray}
and in the channel $\Phi\Psi\to \Psi\Phi$
\ba
\vev{\mathbb{O}_{\Phi\Psi}^{(1)}(Z_1,Z_2,...)} \!\!\!&\stackrel{\Phi\Psi\to
\Psi\Phi}{=}&\!\!\! -\frac{g^2 N_c}{(4
\pi)^2}\frac{\mu^{2\varepsilon}}{\varepsilon}
\label{Phi-Psi-b}\\
&&\hspace*{-30mm}\times \left\{ \mathbb{W}_{\Phi\Psi} \left( 1 -
{\Pi}_{\Phi\Psi}\right) \vev{\mathbb{O}_{\Phi\Psi}^{(0)}(Z_1,Z_2,...)} -
\Delta_{\Phi\Psi}^{({\cal N} =
0)}\vev{\mathbb{O}_{\Psi\Phi}^{(0)}(Z_2,Z_1,...)}\right\} \, . \nonumber
\ea
Here $\vev{\mathbb{O}_{\Psi\Phi}^{(0)}(Z_2,Z_1,...)}\equiv \vev{0|\tr
\{\Psi(Z_2)\Phi(Z_1)... \}|P}$ and the superscript $\scriptstyle (0)$ indicates
the Born level approximation, Eq.~\re{Born}, that is, the product of the plane
waves. In Eq.~\re{Phi-Psi-b}, the notation was introduced for the integral
operators $\mathbb{V}_{\Phi\Psi}$ and $\mathbb{W}_{\Phi\Psi}$
\ba
\mathbb{V}_{\Phi\Psi} \mathbb{O}_{\Phi\Psi}(Z_1,Z_2,...) &=& \int_0^1 \frac{d
\alpha}{\alpha}\bigg[ 2\, {\mathbb{O}_{\Phi\Psi}(Z_1,Z_2,...)}
\\
&&\hspace*{-30mm} -(1-\alpha)^2 {\mathbb{O}_{\Phi\Psi}(Z_1,\alpha Z_1 +
(1-\alpha)Z_2,...)} - (1-\alpha)^{-2}{\mathbb{O}_{\Phi\Psi} ( (1-\alpha)Z_1 +
\alpha Z_2,Z_2,...)} \bigg]
\nonumber \\
\mathbb{W}_{\Phi\Psi} \mathbb{O}_{\Phi\Psi}(Z_1,Z_2,...) &=& -\int_0^1 d \alpha
\frac{\alpha^3}{(1-\alpha)^2} \, \mathbb{O}_{\Psi\Phi}((1-\alpha) Z_1 + \alpha
Z_2,Z_2,...)\,.
\ea
As before, they only act on the first two arguments of a test function
$\mathbb{O}(Z_1,Z_2,...)$. Notice that the operator $\mathbb{W}_{\Phi\Psi}$
interchanges the superfields inside the trace. Comparison with \re{V-super} and
\re{ansatz-V-ex} allows one to identify these operators as $
\mathbb{V}_{\Phi\Psi}=\mathbb{V}^{(-1/2,j_\Psi)}$ and $
\mathbb{W}_{\Phi\Psi}=-\mathbb{V}_{\rm ex}^{(-1/2,j_\Psi)}$ with $j_\Psi=3/2$ for
$\mathcal{N}=0$. The operator ${\Pi}_{\Phi\Psi}$ is the projector  defined in
Eq.~\re{projector-phi-psi}. Finally, $\Delta_{\Phi\Psi}^{({\cal N} = 0)}$ is the
following operator
\begin{equation}
\Delta_{\Phi\Psi}^{({\cal N} = 0)} \mathbb{O}(Z_1 , Z_2) =   \left(2 - Z_{12}
\partial_{Z_2} \right) \left[\frac{\partial_{1+} }{
\partial_{1+} + \partial_{2+} }
\mathbb{O}(Z_1 , Z_2) \right] \bigg|_{Z_1=Z_2}\, ,
\label{Delta-mixed-N=0}
\end{equation}
where $\partial_{k+} = \partial/\partial_{z_k}$ is the light-cone derivative and
the operator $ Z_{12} \partial_{Z_2}$ is applied to the square bracket evaluated
for $Z_1=Z_2$. Notice that in Eq.~\re{Phi-Psi-b} the operator
$\Delta_{\Phi\Psi}^{({\cal N} = 0)}$ is applied to the matrix element with the
arguments $Z_1$ and $Z_2$ interchanged, that is $\mathbb{O}(Z_1 , Z_2) =
\vev{\mathbb{O}_{\Psi\Phi}^{(0)}(Z_2,Z_1,...)}$.

The total one-loop correction to $\vev{\mathbb{O}_{\Phi\Psi}^{(1)}(Z_1,Z_2,...)}$
is a sum of the two expressions, Eqs.~\re{Phi-Psi-a} and \re{Phi-Psi-b}. Its
matching into \re{pole-H} yields
the two-particle evolution kernel in the $\Phi\Psi-$sector in the $\mathcal{N}=0$
theory
\ba
\mathbb{H}_{\Phi\Psi} \Phi(Z_1)\Psi(Z_2) \!\!\!& = &\!\!\!
(\mathbb{V}^{(-1/2,j_\Psi)}-\mathbb{V}_{\rm ex}^{(-1/2,j_\Psi)}) \left( 1 -
{\Pi}_{\Phi\Psi} \right)\Phi(Z_1)\Psi(Z_2)
\nonumber \\[2mm]
\!\!\!&+&\!\!\! \Delta_{\Phi\Psi} \left(\Phi(Z_1)\Psi(Z_2)-\Psi(Z_2)\Phi(Z_1)
\right) .
\label{H-Phi-Psi}
\ea
This relation follows from the explicit evaluation of the Feynman diagrams in
the $\mathcal{N}=0$ theory shown in Fig.~\ref{MixedcKernel}. Going over through
the calculation of the same diagrams in the $\mathcal{N}=1$ and $\mathcal{N}=2$
theories one finds that the evolution kernel $\mathbb{H}_{\Phi\Psi}$ is given
by the same expression \re{H-Phi-Psi} with $j_\Psi$ taking the value $j_{\Psi}
= (3 - \mathcal{N})/2$ which depends on the number of supercharges. Also, the
superspace acquires extra ``odd'' dimensions, $Z=(z,\theta^A)$ with
${\scriptstyle A}=1,\ldots,\mathcal{N}$, and the operator $\Delta_{\Phi\Psi}$
is given for an arbitrary $\mathcal{N}$ by
\begin{equation}
\Delta_{\Phi\Psi} \mathbb{O}(Z_1 , Z_2) =  \left(2-\mathcal{N} - Z_{12}
\partial_{Z_2} \right) \left[\frac{\partial_{1+} }{
\partial_{1+} + \partial_{2+} }
\mathbb{O}(Z_1 , Z_2) \right] \bigg|_{Z_1=Z_2}\,.
\end{equation}
This operator has the same meaning as the operator $\Delta_{\Phi\Phi}$,
Eq.~\re{Delta}. It contributes to the scale dependence of composite operators
involving nonlocal fields and has a vanishing projection onto the subspace of
Wilson operators $\Pi_{\scriptscriptstyle \rm W}\Delta_{\Phi\Psi} \mathbb{O}
(Z_1,Z_2,...)=0\,.$ Therefore, in agreement with our expectations
\re{H-matrix}, the one-loop evolution kernel for Wilson operators in the
$\Phi\Psi-$sector is given by
\be
\mathbb{H}^{\scriptscriptstyle \rm W}_{\Phi\Psi}=\Pi_{\scriptscriptstyle \rm
W}\mathbb{H}_{\Phi\Psi}=\Pi_{\scriptscriptstyle \rm W}
\left[\mathbb{V}^{(-1/2,j_\Psi)}-\mathbb{V}_{\rm ex}^{(-1/2,j_\Psi)}\right]
\left( 1 - {\Pi}_{\Phi\Psi} \right).
\label{H-PhiPsi-fin}
\ee

To identify the evolution kernel in the $\Psi\Phi-$sector one has to calculate
one-loop corrections to the matrix element
$\vev{0|\tr\{\Psi(Z_1)\Phi(Z_2)...\}|P}$. The only difference with the previous
case is that one has to interchange the two superfields inside the trace.
Denoting the corresponding permutation operator as $\mathbb{P}_{\Phi\Psi}$
\be
{\Psi}(Z_1){\Phi}(Z_2)=\mathbb{P}_{\Phi\Psi}
{\Phi}(Z_1){\Psi}(Z_2)\,,
\ee
one finds that the evolution kernel in the $\Psi\Phi-$sector is related to
\re{H-PhiPsi-fin} as
\be
\mathbb{H}^{\scriptscriptstyle \rm W}_{\Psi\Phi} =\mathbb{P}_{\Phi\Psi}
\mathbb{H}^{\scriptscriptstyle \rm
W}_{\Phi\Psi}\mathbb{P}_{\Phi\Psi}=\Pi_{\scriptscriptstyle \rm W}
\left[\mathbb{V}^{(j_\Psi,-1/2)}-\mathbb{V}_{\rm ex}^{(j_\Psi,-1/2)}\right]
\left( 1 - {\Pi}_{\Psi\Phi} \right)\,,
\label{P-H-P}
\ee
where the operators $\mathbb{V}^{(j_\Psi,-1/2)}$, $\mathbb{V}_{\rm
ex}^{(j_\Psi,-1/2)}$ and ${\Pi}_{\Psi\Phi}$ are given by Eqs.~\re{V-super},
\re{V-ex-gen-SL2N} and \re{projector-psi-phi}, respectively, with
$j_\Psi=(3-\mathcal{N})/2$.

\subsection{Relation between $\mathcal{N}=4$ and $\mathcal{N}\le 2$}

According to \re{H-Phi-Phi-final} and \re{N=4-kernel}, the one-loop evolution
kernel in the $\Phi\Phi-$sector has the same, universal form in the SYM
theories with $\mathcal{N}=4$ and $\mathcal{N}\le 2$. To understand the origin
of this property we remind that the SYM theories with different number of
supercharges $\mathcal{N}$ are related to each other via the reduction
procedure described in Sect.~2.1.

In the Mandelstam formulation, the decomposition of the $\mathcal{N}=4$
superfield over the $\mathcal{N}=2$ superfields looks as follows (see
Eqs.~\re{Phi2-from-Phi4} and \re{MandelstamPhi})
\be
{\Phi}^{(4)} (z n_\mu, \theta^A,\theta^3,\theta^4 )=
{\Phi}^{(2)}(Z)+\theta^3 {\Psi}^{(2)}_{\scriptscriptstyle \rm
WZ}(Z)-\theta^4\,
\partial_+^{-1} \bar D^1\bar D^2 \bar{\Psi}^{(2)}_{\scriptscriptstyle \rm WZ}(Z)
- \theta^3\theta^4\,{\Psi}^{(2)}(Z) \, .
\label{M-N4=N2}
\ee
where $Z=(z,\theta^A)$ with ${\scriptstyle A}=1,2$. We would like to stress
that as long as one retains in \re{M-N4=N2} the contribution of the Wess-Zumino
superfields, $\Psi^{(2)}_{\scriptscriptstyle \rm WZ}$ and
$\bar\Psi^{(2)}_{\scriptscriptstyle \rm WZ}$, the dilatation operators in the
$\mathcal{N}=4$ SYM theory and the $\mathcal{N}=2$ SYM theory coupled to the
Wess-Zumino superfields are identical. In particular, substituting \re{M-N4=N2}
into \re{N=4-kernel} and comparing the coefficients in front of $\theta^3$ and
$\theta^4$ in both sides of \re{N=4-kernel}, one can identify the two-particle
kernels in the various sectors including $\Phi^{(2)}\Phi^{(2)}-$,
$\Psi^{(2)}\Psi^{(2)}-$, $\Psi^{(2)}\Phi^{(2)}-$ and
$\Phi^{(2)}\Psi^{(2)}-$sectors. In general, these kernels should be different
from the same kernels in the $\mathcal{N}=2$ SYM theory since the former
receive a nontrivial contribution from the Wess-Zumino superfields. Therefore,
in order to derive the evolution kernels in the $\mathcal{N}=2$ theory from the
one in the $\mathcal{N}=4$, Eq.~\re{N=4-kernel}, via the truncation procedure
one has to eliminate from the latter kernel the contribution of the superfields
$\Psi^{(2)}_{\scriptscriptstyle \rm WZ}$ and
$\bar\Psi^{(2)}_{\scriptscriptstyle \rm WZ}$.

For this purpose, it is not sufficient to put $\Psi^{(2)}_{\scriptscriptstyle \rm
WZ}=\bar\Psi^{(2)}_{\scriptscriptstyle \rm WZ}=0$ in \re{N=4-kernel} and
\re{M-N4=N2}, since the Wess-Zumino superfields could propagate along the
internal line in Fig.~\ref{HolomorphicKernel}(a)--(c) and inside the loop in
Fig.~\ref{HolomorphicKernel}(e). In the latter case, the Wess-Zumino superfields
contribute to the self-energy and their elimination affects the $\beta-$function
of the SYM theory (see Eq.~\re{Sigma-N=2}).
In the former case, since the superfields $\Psi^{(2)}_{\scriptscriptstyle \rm
WZ}$ and $\bar\Psi^{(2)}_{\scriptscriptstyle \rm WZ}$ are fermionic, they could
couple to bosonic superfields only in pairs and, therefore, can contribute
starting from the two-loop level. Thus, going over from the $\mathcal{N}=4$ to
$\mathcal{N}=2$ SYM theory, one can safely put $\Psi^{(2)}_{\scriptscriptstyle
\rm WZ}=\bar\Psi^{(2)}_{\scriptscriptstyle \rm WZ}=0$ in \re{M-N4=N2}, adjust
the value of the $\beta-$function and apply \re{N=4-kernel} to evaluate the
one-loop dilatation operator in the $\mathcal{N}=2$ SYM.

Let us apply the reduction procedure to reproduce the $\mathcal{N}=2$
two-particle evolution kernels in different sectors. To obtain the
$\mathcal{N}=2$ kernel in the $\Phi\Phi-$sector, one puts
$\theta^3_k=\theta^4_k=0$ (with $k=1,2$) in both sides of \re{N=4-kernel}.
According to \re{M-N4=N2} the product of the superfields reduces to
${\Phi}^{(2)}(Z_1){\Phi}^{(2)}(Z_2)$ leading to
\be
\mathbb{H}^{(4)}_{\Phi\Phi}\left[{\Phi}^{(2)}(Z_1){\Phi}^{(2)}(Z_2)\right]
=
\mathbb{H}^{(2)}_{\Phi\Phi}\left[{\Phi}^{(2)}(Z_1){\Phi}^{(2)}(Z_2)\right].
\label{tr-pr-1}
\ee
The operator $\mathbb{H}^{(2)}_{\Phi\Phi}$ defined in this way takes the same
form as before, Eq.~\re{N=4-kernel}, but the number of odd dimensions in the
superspace is reduced from $\mathcal{N}=4$ to $\mathcal{N}=2$,
$Z_k=(z_k,\theta^1_k,\theta^2_k)$. As a result, $\mathbb{H}^{(2)}_{\Phi\Phi}$
coincides with the one-loop $\mathcal{N}=2$ evolution kernel in the
$\Phi\Phi-$sector, Eq.~\re{H-Phi-Phi-final}.

In a similar manner, to obtain the $\mathcal{N}=2$ evolution kernel in the
$\Psi\Psi-$sector one has to retain in \re{M-N4=N2} the contribution of the
$\Psi-$superfield. One substitutes ${\Phi}(Z_k)=-(\theta\!\cdot\! \theta)_k
{\Psi}^{(2)}(Z_k)$ with $(\theta\!\cdot\!\theta)_k\equiv\theta_k^3\theta_k^4$
into \re{N=4-kernel} and gets
\be
\mathbb{H}^{(4)}_{\Phi\Phi}\left[\prod_{k=1,2}(\theta\cdot \theta)_k\,
{\Psi}^{(2)}(Z_1){\Psi}^{(2)}(Z_2)\right]
= \prod_{k=1,2}(\theta\!\cdot\! \theta)_k
\,
\mathbb{H}^{(2)}_{\Psi\Psi}\left[{\Psi}^{(2)}(Z_1){\Psi}^{(2)}(Z_2)\right],
\label{tr-pr-2}
\ee
where the superscripts $\scriptstyle (2)$ and $\scriptstyle (4)$ refer to the
underlying SYM theory. In the left-hand side\ of this relation, we take into
account that the state $(\theta\!\cdot\! \theta)_1(\theta\!\cdot\! \theta)_2$
is annihilated by the $\mathcal{N}=4$ operators $\Pi_{\Phi\Phi}$ and
$\Delta_{\Phi\Phi}$, given by Eqs.~\re{Pi12-main} and \re{Delta}, and get
\be
\mathbb{V}^{(-1/2,-1/2)}\left[\prod_{k=1,2}(\theta\!\cdot\! \theta)_k\,
{\Psi}^{(2)}(Z_1){\Psi}^{(2)}(Z_2)\right]=
\prod_{k=1,2}(\theta\!\cdot\! \theta)_k\,\mathbb{V}^{(1/2,1/2)}\left[
{\Psi}^{(2)}(Z_1){\Psi}^{(2)}(Z_2)\right]\,.
\ee
Here the upper indices of the $\mathbb{V}-$operator, Eq.~\re{V-super}, are
modified because $(\theta_\alpha\!\cdot\! \theta_\alpha)_1(\theta\!\cdot\!
\theta)_2=\alpha^2(\theta\!\cdot\! \theta)_1(\theta\!\cdot\! \theta)_2 $ for
$\theta_{1\alpha}^A=(1-\alpha)\theta_1^A+\alpha\theta_2^A$ brings an
additional factor $\alpha^2$. One verifies that the operator
$\mathbb{H}^{(2)}_{\Psi\Psi}=\mathbb{V}^{(1/2,1/2)}$ coincides with the
$\mathcal{N}=2$ evolution kernel in the $\Psi\Psi-$sector, Eq.~\re{Psi-Psi-ker}.

To obtain the $\mathcal{N}=2$ evolution kernel in the mixed $\Psi\Phi-$ and
$\Phi\Psi-$sectors, one examines the product of the $\mathcal{N}=4$ superfields
\re{M-N4=N2}, ${\Phi}^{(4)}(Z_1){\Phi}^{(4)}(Z_2)$, and retains the
terms involving the product of the $\mathcal{N}=2$ superfields
${\Phi}^{(2)}(Z_1){\Psi}^{(2)}(Z_2)$ and
${\Psi}^{(2)}(Z_1){\Phi}^{(2)}(Z_2)$. Their substitution into the
left-hand side\ of \re{N=4-kernel} yields
\ba
 \lefteqn{\mathbb{H}_{\Phi\Phi}^{(4)} \left[(\theta\!\cdot\! \theta)_2
{\Phi}^{(2)}(Z_1){\Psi}^{(2)}(Z_2) + (\theta\!\cdot\! \theta)_1
{\Psi}^{(2)}(Z_1){\Phi}^{(2)}(Z_2) \right]}
\nonumber \\
 && \qquad =(\theta\!\cdot\! \theta)_2\cdot \mathbb{H}_{\Phi\Psi}^{(2)} \left[
{\Phi}^{(2)}(Z_1){\Psi}^{(2)}(Z_2)\right]+(\theta\!\cdot\!
\theta)_1\cdot\mathbb{H}_{\Psi\Phi}^{(2)}
 \left[{\Psi}^{(2)}(Z_1){\Phi}^{(2)}(Z_2)\right]+ \ldots\,,
\label{H4-2-red}
\ea
where the ellipses denote terms quadratic in the odd variables,
$\theta_1^3\theta_2^4$ and $\theta_2^3\theta_1^4$. These terms describe the
transitions $\Psi_{\scriptscriptstyle \rm WZ}
\bar\Psi_{\scriptscriptstyle \rm WZ}\to\Psi\Phi, \Phi\Psi$ which do not survive
the truncation procedure.
Replacing $\mathbb{H}_{\Phi\Phi}^{(4)}$ in \re{H4-2-red} by its expression
\re{N=4-kernel} and going over through a lengthy calculation, one matches the
result into the right-hand side\ of \re{H4-2-red} and verifies that
$\mathbb{H}_{\Phi\Psi}^{(2)}$ indeed coincides with \re{H-Phi-Psi}.

The truncation procedure explained above for $\mathcal{N}=2$ can be continued to
produce the evolution kernels in the $\mathcal{N}=1$ and $\mathcal{N}=0$ SYM
theories, Eqs.~\re{H-Phi-Phi-final} and \re{Psi-Psi-ker}. To this end, one
follows the steps described in Sect.~2.1 and expands the $\mathcal{N}=2$
superfields in terms of the $\mathcal{N}=1$ and, eventually, $\mathcal{N}=0$
(super)fields. As before, this generates additional Wess-Zumino superfields
which modify the $\beta-$function of the SYM theory but do not affect the
one-loop evolution kernel.

Thus, the two-particle evolution kernels $\mathbb{H}_{\Phi\Phi}$,
$\mathbb{H}_{\Psi\Psi}$, $\mathbb{H}_{\Psi\Phi}$ and $\mathbb{H}_{\Phi\Psi}$ in
the SYM theory with $\mathcal{N}=0,1,2$ supercharges can be derived from the
kernel $\mathbb{H}_{\Phi\Phi}^{\scriptscriptstyle (4)}$ in the $\mathcal{N}=4$
theory, Eq.~\re{N=4-kernel}, through the truncation procedure by eliminating
the contribution of the Wess-Zumino superfields.

\section{Anomalous dimensions of Wilson operators}

Let us demonstrate the relation of our approach based on non-local light-cone
operators with the conventional one that deals with local Wilson operators. To
this end, we will show how the obtained expressions for the one-loop dilatation
operator allow one to evaluate anomalous dimensions of various Wilson operators
in $\mathcal{N}-$extended SYM theories.

We recall that the Wilson operators of the maximal Lorentz spin, or simply
quasipartonic operators, are local gauge-invariant single-trace operators built
from transverse components of the strength tensor $F_{+\perp}\equiv n^\mu
F_{\mu\perp}$, ``good'' components of fermions, $\psi_+^A$ and $\bar\psi_{+A}$,
scalar fields, $\phi^{AB}$ and $\bar\phi_{AB}$, and covariant derivatives
$D_+\equiv n^\mu D_\mu$ acting on these fields. In the light-cone formalism, in
the light-like gauge $A_+(x)=0$, the same operators are constructed from gauge
fields, $\partial_+A$ and $\partial_+\bar A$, Grassman fields, $\lambda^A$ and
$\bar \lambda_A$, complex scalars, $\phi^{AB}$ and $\bar\phi_{AB}$, and
light-cone derivatives $n^\mu D_\mu\equiv \partial_+$. The relation between the
two sets of fields looks as follows. The gauge strength tensor $F_{+ \perp}=(F_{+
x},F_{+ y})$ and its dual $\widetilde F_{\mu\nu} = \ft12
\varepsilon_{\mu\nu\rho\lambda}F^{\rho\lambda}$ are expressed in terms of the
helicity $\pm 1$ gauge fields \re{gauge-perp}
\be
F_{+x}=\widetilde F_{+ y}=\ft{1}{\sqrt{2}} (\partial_+ A + \partial_+ \bar
A)\,,\qquad F_{+y}=-\widetilde F_{+ x}=-\ft{i}{\sqrt{2}} (\partial_+ A -
\partial_+
\bar A)\,.
\ee
The ``good'' components of Majorana fermions, $\psi_+^A$ and $\bar\psi_{+A}$, are
expressed in terms of helicity $\pm 1/2$ Grassmann fields (see
Eqs.~\re{psi-Weyl}, \re{LCprojection} and \re{lambda-i})
\be
\psi_+^{A}=\sqrt[4]{2}\left(\!\!\begin{array}{c}
\lambda^A \\
0 \\
0 \\
i\bar\lambda_A
\end{array}\!\!\right)\,,\qquad
\bar\psi_{+A}=-\sqrt[4]{2}\left(
0 , \lambda^A , i\bar\lambda_A , 0 \right)\,.
\ee
Using these relations, one can establish the correspondence between the Wilson
operators in the covariant and light-cone formulations. As an example, we
present expressions for a few twist-two operators in the $\mathcal{N}=4$ theory:
parity even/odd fermion operators
\begin{eqnarray}
O^{\rm q}_N(0) \!\!\!&=&\!\!\! \tr \{\bar\psi_A \gamma^{+} (iD_+)^{N-1}\psi^A\} \
\ = 2i^{N-2}\tr \{
\bar\lambda_A \partial_+^{N-1}\lambda^A
+ \lambda^A \partial_+^{N-1}\bar\lambda_A \}\,,
\label{tw=2-q}
\\[3mm]
\widetilde O^{\rm q}_N(0) \!\!\!&=&\!\!\! \tr \{
\bar\psi_A \gamma^{+} \gamma^5 (iD_+)^{N-1}\psi^A
\} = 2i^{N-2} \tr \{
\bar\lambda_A \partial_+^{N-1}\lambda^A
- \lambda^A \partial_+^{N-1}\bar\lambda_A \} \,, \nonumber
\end{eqnarray}
parity even/odd gauge field operators
\begin{eqnarray}
O^{\rm g}_N(0) \!\!\!&=&\!\!\! \tr \{ F_{+\nu} (iD_+)^{N-2}F_{\nu +}\} \ =
i^{N-2}\tr \{
\partial_+ A\, \partial_+^{N-1} \bar A + \partial_+\bar A\, \partial_+^{N-1}A
\}\,,
\label{tw=2-g}
\\[2mm]
\widetilde O^{\rm g}_N(0) \!\!\!&=&\!\!\! \tr \{F_{+\nu} (iD_+)^{N-2} i
\widetilde F_{\nu +}\} =i^{N-2} \tr \{
\partial_+ A\, \partial_+^{N-1}\bar A
-
\partial_+ \bar  A\, \partial_+^{N-1}A
\}\,, \nonumber
\end{eqnarray}
and scalar operators
\begin{equation}
O^{\rm s}_N(0) = \tr \{\bar\phi_{AB} \,(i\partial_+)^{N}\phi^{AB}\}\,.
\end{equation}
In the light-cone formalism, one obtains the Wilson operators by expanding the
nonlocal light-cone operators \re{Chiral-sector} -- \re{Mixed-sector} in powers
of even, $z_k$, and odd variables, $\theta^A_k$, Eqs.~\re{OW-W} and \re{OW-OS}.
The light-cone operators satisfy the evolution equation \re{EQ} with the
one-loop dilatation operator given in the multi-color limit by
\re{H-multi-color} and \re{ansatz-gen-H12}. To reconstruct the mixing matrix
for the Wilson operators, one has to substitute \re{OW-W} and \re{OW-OS} into
the evolution equation \re{EQ} and equate the coefficients in front of
different powers of $z$'s and $\theta$'s in both sides of \re{EQ}. We
illustrate below this procedure by calculating the mixing matrices for various
Wilson operators in the SYM theories with $\mathcal{N}=0,1,2,4$.

\subsection{Wilson operators in $\mathcal{N}=0$ theory}

In the $\mathcal{N}=0$ theory, that is, pure gluodynamics with the $SU(N_c)$
gauge group, the light-cone fields are given by (see Eq.~\re{M=0-field})
\be
{\Phi} (z) = \partial_+^{-1} A(z)\,,\qquad {\Psi} (z) = - \partial_+
\bar A(z)\,,
\ee
with $A(z)$ and $\bar A(z)=A^*(z)$ being the gauge fields of helicity $+1$ and
$-1$, respectively.  The conventional local Wilson operators arise from the
Taylor expansion of the light-cone operators $\mathbb{O}(z_1,\ldots,z_L)$ in
the light-cone separations. For the light-cone operators built only from
$\Psi-$ or $\Phi-$fields, Eq.~\re{Chiral-sector} and \re{Chiral-sector-1}, the
corresponding Wilson operators belong to the sector of the aligned-helicity
gluon operators.

\subsubsection*{$\Psi\Psi-$sector}

The product of two ${\Psi}-$fields can be expanded as
\begin{equation}
{\Psi} (z_1) {\Psi} (z_2) = \sum_{j_1, j_2 \geq 0} \frac{z_1^{j_1}}{j_1
!} \frac{z_2^{j_2}}{j_2 !} O_{j_1 j_2} (0) = \sum_{j = 0}^\infty
\frac{1}{j!}\sum_{k = 0}^j
 \left( {j \atop k} \right) z_1^k z_2^{j - k} O_{k, j - k} (0) \, ,
\label{PsiPsi-Taylor}
\end{equation}
where the notation was introduced for the aligned-helicity Wilson operators
\begin{equation}
O_{j_1 j_2} (0) =
\partial_+^{j_1 + 1} \bar{A} (0) \, \partial_+^{j_2 + 1} \bar{A} (0)
=\partial_{z_1}^{j_1} \partial_{z_2}^{j_2}{\Psi} (z_1) {\Psi}
(z_2)\Big|_{z_i = 0}
\label{aligned-h}
\end{equation}
with $\bar A(0)=\bar A^a(0) t^a$ being a matrix of dimension $N_c$.

To one-loop order, $O_{j_1 j_2} (0)$ mix under renormalization with the
operators $O_{k_1 k_2} (0)$ carrying the same Lorentz spin $j = j_1 +
j_2=k_1+k_2$. The corresponding mixing matrix $V_{j_1 j_2}^{k_1 k_2}$ is
related to the two-particle dilatation operator in the $\Psi\Psi-$sector,
Eq.~\re{Psi-Psi-ker}, as
\be
\mathbb{H}_{\Psi\Psi}\,O_{j_1 j_2}(0)=\sum_{ {k_1+k_2=j_1+j_2}}V_{j_1 j_2}^{k_1
k_2} \, O_{k_1 k_2}(0)\,.
\label{PsiPsi-V}
\ee
For a given $j=j_1+j_2$, there are $(j+1)$ operators \re{aligned-h} so that the
mixing matrix has dimension $(j+1)$. Since the Lorentz spin takes the values
$0\le j < \infty$, the matrix $V_{j_1 j_2}^{k_1 k_2}$ may have an arbitrary
large size. To find this matrix, one substitutes the expansion
\re{PsiPsi-Taylor} into the expression for the one-loop dilatation operator
$\mathbb{H}_{\Psi\Psi}$ at $\mathcal{N}=0$, Eq.~\re{Psi-Psi-ker}, and equates
the coefficients in front of $z_1^{j_1}z_2^{j_2}$
\be
\mathbb{H}_{\Psi\Psi}O_{j_1 j_2} = \int_0^1\frac{d\alpha}{\alpha} \left[ 2
\partial_{z_1}^{j_1}\partial_{z_2}^{j_2} - \bar\alpha^{j_1 + 2}
(\alpha\partial_{z_1}\!+\partial_{z_2})^{j_2}\partial_{z_1}^{j_1} -
\bar\alpha^{j_2 + 2}
(\alpha\partial_{z_2}\!+\partial_{z_1})^{j_1}\partial_{z_2}^{j_2} \right]
{\Psi} (z_1) {\Psi} (z_2) \Big|_{z_i = 0} \, ,
\label{VV}
\ee
where $\bar\alpha\equiv 1-\alpha$. This relation establishes the correspondence
between the mixing matrix for local Wilson operators and the evolution kernel
for nonlocal light-cone operators. Matching \re{VV} into \re{PsiPsi-V} with a
help of \re{aligned-h}, one calculates the mixing matrix
\begin{eqnarray}
\label{MaxGlMixingMatrix}
V_{j_1 j_2}^{k_1 k_2} \!\!\!&=&\!\!\! \delta_{j_1 k_1} \delta_{j_2 k_2} \left[
\psi (j_1 + 3) + \psi (j_2 + 3) - 2 \psi (1)
\right]
\\
&-&\!\!\! \delta_{j_1 + j_2 , k_1 + k_2} \left[ \frac{\theta_{j_2 , k_2}}{j_2 -
k_2} \frac{j_2!}{k_2!} \frac{(j_1 + 2)!}{(k_1 + 2)!}  + \frac{\theta_{j_1 ,
k_1}}{j_1 - k_1} \frac{j_1!}{k_1!} \frac{(j_2 + 2)!}{(k_2 + 2)!}  \right] \, .
\nonumber
\end{eqnarray}
The eigenvalues of this matrix determine the spectrum of anomalous dimensions of
the aligned-helicity Wilson operators \re{aligned-h}.

According to \re{PsiPsi-V}, the matrix $V_{j_1 j_2}^{k_1 k_2}$ defines a
representation of the dilatation operator on the space spanned by the Wilson
operators \re{aligned-h}. The choice of the basis of local operators in this
space is not unique. In order to reveal symmetries of the mixing matrix imposed
by the $SL(2)$ invariance of the dilatation operator, one switches to the basis
of conformal operators~\cite{BraKorMul03}. The conformal gauge operators are
linear combinations of the operators \re{aligned-h}
\begin{equation}
\mathcal{O}_j (0) = \sum_{k = 0}^j c_{jk} {O}_{j - k, k} =
\partial_+ \bar{A} (0)
\left( \stackrel{\scriptscriptstyle\rightarrow}{\partial}_+ +
\stackrel{\scriptscriptstyle\leftarrow}{\partial}_+ \right)^j C_j^{5/2} \left(
\frac{ \stackrel{\scriptscriptstyle\rightarrow}{\partial}_+ -
\stackrel{\scriptscriptstyle\leftarrow}{\partial}_+ }{
\stackrel{\scriptscriptstyle\rightarrow}{\partial}_+ +
\stackrel{\scriptscriptstyle\leftarrow}{\partial}_+ } \right)
\partial_+ \bar{A} (0)
\, ,
\label{Conf-op}
\end{equation}
which are expressed in terms of the Gegenbauer polynomials. The expansion
coefficients $c_{jk}$ are fixed from the condition that $\mathcal{O}_j (0)$
have to be the lowest-weight vectors in the tensor product of two $SL(2)$
moduli. In the conformal basis, the expansion of nonlocal light-cone operator
\re{PsiPsi-Taylor} looks as follows
\be
{\Psi} (z_1) {\Psi} (z_2) =  \sum_{j = 0}^\infty c_j    z_{21}^j \int_0^1 d
\alpha \, (\alpha\bar\alpha)^{j + 2} {\cal O}_{j } (\bar \alpha z_1+ \alpha
z_{2}) \, ,
\label{conf-expansion}
\ee
where $c_j = {12 (2 j + 5) }/{ {\Gamma} (j + 3) } $ and $z_{21}=z_2-z_1$. A
unique feature of the conformal operators is that the mixing matrix
\re{PsiPsi-V} is diagonal in this basis
\be
\mathbb{H}_{\Psi\Psi}\,\mathcal{O}_j (0)=\gamma_{\Psi\Psi}(j+3)
\, \mathcal{O}_j (0)\,.
\label{PsiPsi-V1}
\ee
The expansion coefficients $c_{jk}$ entering \re{Conf-op} define (left)
eigenstates of the mixing matrix and the corresponding anomalous dimension
$\gamma_{\Psi\Psi}(j)$, given below n Eq.\ (\ref{anom-dim0}), can be
calculated using \re{MaxGlMixingMatrix}.%
\footnote{Strictly speaking, $\gamma_{\Psi\Psi}(j)$ is the eigenvalue of the
two-particle dilatation operator rather than anomalous dimension. The two are
related to each other, see Eqs.~\re{spectrum} and \re{Sch-eq} below.}

There exist a much simpler way of calculating $\gamma_{\Psi\Psi}(j)$. The
conformal operators \re{Conf-op} can be written in the following form
\be
\mathcal{O}_j (0) = a_j\,  \partial_+ \bar{A}\, \partial_+^{j + 1} \bar{A} (0) +
b_j\,\partial_+ (\partial_+ \bar{A}\, \partial_+^{j } \bar{A} (0))+\cdots\,,
\label{conf-op-exp}
\ee
where the expansion goes over local operators involving total derivatives and
$a_j$, $b_j$, $...$ are some coefficients. The conformal invariance allows one to
reconstruct the whole sum out of the first term only. One can neglect all
operators with the total derivatives by going over to the so-called
\textsl{forward limit}. It amounts to taking the forward matrix element of the
conformal operators with respect to some reference state
\be
\mathcal{O}_j (0) \to \vev {P| \mathcal{O}_j (0) |P} =a_j\, \vev {P|  \partial_+
\bar{A}\, \partial_+^{j + 1} \bar{A} (0) |P}\,,
\ee
since $\vev{P|\partial_+(\ldots)|P}=0$. This does not affect the anomalous
dimension \re{PsiPsi-V1}, but allows one to substitute the conformal operator
inside the forward matrix element by a simple operator $\partial_+ \bar{A}\,
\partial_+^{j + 1} \bar{A} (0)$. We will accept this strategy
in the remainder of this section.

The expansion \re{conf-expansion} looks in the forward limit as
\begin{equation}
{\Psi} (z_1) {\Psi} (z_2) \stackrel{\rm fw}{=} \sum_{j = 0}^\infty
\frac{z_{21}^j}{j!}\, \partial_+ \bar{A}\, \partial_+^{j + 1} \bar{A} (0) \, .
\label{forward-limit}
\end{equation}
Hereafter $\stackrel{\rm fw}{=}$ means that the relation is only valid upon
taking the forward matrix element, that is, up to contribution of Wilson
operators involving total derivatives. Substituting \re{forward-limit} into
\re{Psi-Psi-ker} one finds after a simple calculation
\be
\mathbb{H}_{\Psi\Psi}{\Psi} (z_1) {\Psi} (z_2) \stackrel{\rm fw}{=} \sum_{j =
0}^\infty \frac{z_{21}^j}{j!}\gamma_{\Psi\Psi}(j+3)\, \partial_+ \bar{A}\,
\partial_+^{j + 1} \bar{A} (0) \,,
\label{H-PsiPsi-N=0-ex}
\ee
where the anomalous dimension is given by
\be
\gamma_{\Psi\Psi}(j+3)  =2\int_0^1\frac{d\alpha}{\alpha}(1-\bar\alpha^{2+j}) =2
\big[\psi(j+3)-\psi(1)\big]\,.
\label{anom-dim0}
\ee
Here $j + 3$ is the two-particle $SL(2)$ conformal spin \re{Sl2-Casimir},
$\mathbb{J}_{12} = j + j_1+j_2$ for $j_1=j_2=j_\Psi=3/2$. Eqs.~\re{PsiPsi-V1} and
\re{anom-dim0} are an agreement with \re{H-PsiPsi-J} for $\mathcal{N}=0$.
Comparing the coefficients in front of $z_{21}^j$ in both sides of
\re{H-PsiPsi-N=0-ex} we conclude that
\be
\bigg[\mathbb{H}_{\Psi\Psi}-\gamma_{\Psi\Psi}(j+3)\bigg]\,\partial_+\bar A \,
\partial_+^{j +1} \bar A(0) \stackrel{\rm fw}{=} 0\,.
\label{anom-dim-N=0}
\ee
As was already mentioned, the operator $\mathbb{H}_{\Psi\Psi}$ can be mapped
into a two-particle Hamiltonian of the $SL(2)$ Heisenberg magnet of spin
$j_\Psi=3/2$~\cite{BraDerMan98,BraDerKorMan99,Bel99}.

\subsubsection*{$\Phi\Phi-$sector}

The scale dependence of the operator $\Phi(z_1)\Phi(z_2)$ is driven to one-loop
order by the dilatation operator $\mathbb{H}_{\Phi\Phi}$, Eq.~\re{H-Phi-Phi-final}.
In distinction with the previous case, the first two terms of the expansion of
the field $\Phi(z)$ around $z=0$ involve nonlocal, spurious field components
\be
\Phi(z)= \sum_{k=0,1} z^k\, \partial_+^{k-1} A(0) + \sum_{k=2}^\infty z^k\,
\partial_+^{k-1} A(0)  =\Phi_{\rm sp}(z) + \Phi_{\rm\scriptscriptstyle W}(z)\,,
\ee
with $(\Phi_{\rm\scriptscriptstyle W}(z))^* = -z^2\,\Psi(z)$. Substituting
$\Phi = \Phi_{\rm sp} + \Phi_{\rm\scriptscriptstyle W}$ into
\re{H-Phi-Phi-final}, one can find anomalous dimensions for different
components arising in the product $\Phi(z_1)\Phi(z_2)$. Going over to the
forward limit one gets
\begin{equation}
\Phi (z_1) \Phi (z_2) \stackrel{\rm fw}{=}   \sum_{j = 0}^\infty
\frac{z_{21}^j}{j!} \partial_+A \, \partial_+^{j - 3} A (0) \, ,
\label{PhiPhi-Taylor-exp}
\end{equation}
where the terms with $j\le 3$ and $j > 3$ correspond to spurious and Wilson
operators, respectively.

The one-loop dilatation operator \re{H-Phi-Phi-final} involves the projector
$\Pi_{\Phi\Phi}$, Eq.~\re{Pi12-main}. The action of the operator $(1 -
\Pi_{\Phi\Phi})$ on the product $\Phi (z_1) \Phi (z_2)$ amounts to subtracting
the first two terms in the expansion \re{PhiPhi-Taylor-exp}
\begin{equation}
(1 - \Pi_{\Phi\Phi}) \Phi (z_1) \Phi (z_2) \stackrel{\rm fw}{=} \sum_{j =
2}^\infty \frac{z_{21}^j}{j!} \partial_+A  \, \partial_+^{j - 3} A (0) \, .
\label{PhiPhi-proj-exp}
\end{equation}
In addition, one finds that the expansion of the addendum $\Delta_{\Phi\Phi} \Phi
(z_1) \Phi (z_2)$, Eq.~\re{Delta}, around $z_{12}=0$ only involves operators with
total derivatives and, therefore, it vanishes in the forward limit
\begin{equation}
\Delta_{\Phi\Phi} \Phi (z_1) \Phi (z_2) \stackrel{\rm fw}{=} 0 \, .
\label{Delta=0}
\end{equation}
Substituting \re{PhiPhi-Taylor-exp} into \re{H-Phi-Phi-final} and taking into
account the last two relations we find
\begin{equation}
\mathbb{H}_{\Phi\Phi} \Phi (z_1) \Phi (z_2) \stackrel{\rm fw}{=}    \sum_{j =
2}^\infty \frac{z_{21}^j}{j!}\gamma_{\Psi\Psi}(j-1)\,  \partial_+A \,
\partial_+^{j - 3} A(0) \,,
\end{equation}
with $\gamma_{\Psi\Psi}(j)$ defined in \re{anom-dim0}. Comparing the coefficients
in front of $z_{21}^j$ in both sides of this relation, we conclude that
\be
\bigg[\mathbb{H}_{\Phi\Phi}-\gamma_{\Psi\Psi}(j-1) \theta(j-1)\bigg]\,\partial_+A
\,
\partial_+^{j - 3} A(0) \stackrel{\rm fw}{=} 0\,.
\label{PhiPhi-V}
\ee
We recall that the two-particle $SL(2)$ conformal spin in the $\Phi\Phi-$sector,
Eq.~\re{Sl2-Casimir}, equals $\mathbb{J}_{12} = j + 2j_\Phi$ with $j_\Phi=-1/2$.
One observes that \re{PhiPhi-V} is in an agreement with \re{H-PhiPhi-J}.

It follows from \re{PhiPhi-V} and \re{anom-dim-N=0} that the anomalous
dimensions of the Wilson operators $\partial_+A \,\partial_+^{j-3} A(0)$ (with
$j\ge 4$) and complex conjugated operators $\partial_+\bar A
\,\partial_+^{j-3}\bar A(0)$ coincide, as it should be. Nonlocal operators
$\partial_+A \,\partial_+^{j-3} A(0)$ with $j=0,1,2$ have vanishing anomalous
dimensions, while for $j=3$ the anomalous dimension of the operator $\partial_+
A \, A (0)$ equals $\gamma(2)=2$.

\subsubsection*{$\Phi\Psi-$ and $\Psi\Phi-$sectors}

Let us turn to the mixed sector $\Phi\Psi$ and go right away to the forward limit
\be
\Phi(z_1)\Psi(z_2) \stackrel{\rm fw}{=} - \sum_{j = 0}^\infty \frac{z_{21}^j}{j!}
\partial_+A\,\partial_+^{j-1} \bar{A} (0)\,,
\label{PhiPsi-expansion}
\ee
with $z_{21}=z_2-z_1$. This expansion involves spurious $(j=0,1)$ and Wilson
operators $(j\ge 2)$ both built from the gauge fields of opposite helicity. The
one-loop dilatation operator for $\Phi(z_1)\Psi(z_2)$ is given by
\re{H-Phi-Psi}. It involves the projector $\Pi_{\Phi\Psi}$,
Eq.~\re{projector-phi-psi}, which eliminates spurious operators in the
right-hand side\ of \re{PhiPsi-expansion}
\begin{equation}
(1 - \Pi_{\Phi\Psi}) \Phi (z_1) \Psi (z_2) \stackrel{\rm fw}{=} - \sum_{j \ge
2} \frac{z_{21}^j}{j!} \partial_+A \, \partial_+^{j-1} \bar{A} (0) \, .
\label{1-Pi-PhiPsi}
\end{equation}
One substitutes \re{PhiPsi-expansion} into Eqs.~\re{ansatz-gen-H12} and \re{V}
and takes into account \re{1-Pi-PhiPsi} to get
\begin{equation}
\mathbb{V}^{(-1/2, 3/2)} (1 - \Pi_{\Phi\Psi}) \Phi (z_1) \Psi (z_2) \stackrel{\rm
fw}{=} -\sum_{j \ge 2} \frac{z_{21}^j}{j!} \gamma_{\Phi\Psi}(j+1)
\partial_+A \,
\partial_+^{j-1} \bar{A} (0) \, ,
\label{V-PhiPsi}
\end{equation}
and analogously for the exchanged kernel, Eqs.~\re{ansatz-gen-H12} and
\re{V-ex-super},
\begin{equation}
\mathbb{V}^{(-1/2, 3/2)}_{\rm ex} (1 - \Pi_{\Phi\Psi}) \Phi (z_1) \Psi (z_2)
\stackrel{\rm fw}{=} -\sum_{j \ge 2} \frac{z_{21}^j}{j!}\gamma^{\rm
(ex)}_{\Phi\Psi}(j+1)
\partial_+\bar{A}\,\partial_+^{j-1} A (0)
\, .
\label{V-ex-PhiPsi}
\end{equation}
The anomalous dimensions entering into these relations are given by
\ba
\gamma_{\Phi\Psi}(j) &=& \psi (j + 2) + \psi (j - 2) - 2 \psi (1)\,,
\\
\gamma^{\rm (ex)}_{\Phi\Psi}(j) &=& \frac{6}{(j - 2)(j-1) j (j + 1)} =
\frac{\Gamma(j-2)}{\Gamma(j+2)}\Gamma(4)\,. \nonumber
\label{gamma-gamma}
\ea
Combining together \re{V-PhiPsi} and \re{V-ex-PhiPsi}, we obtain from
\re{ansatz-gen-H12} for $j \ge 2$
\be
\mathbb{H}_{\Phi\Psi} \partial_+A \,
\partial_+^{j-1} \bar{A} (0) \stackrel{\rm fw}{=}\gamma_{\Phi\Psi}(j+1)\,\partial_+A \,
\partial_+^{j-1} \bar{A} (0) -\gamma^{\rm (ex)}_{\Phi\Psi}(j+1)\,\partial_+\bar{A}\,
\partial_+^{j-1}  A (0)\,.
\label{H-PhiPsi-N=0}
\ee
As in the previous case, Eq.~\re{Delta=0}, the operator $\Delta_{\Phi\Psi}$,
Eq.~\re{Delta-mixed-N=0}, does not contribute to anomalous dimensions of the
Wilson operators.

Repeating a similar analysis in the $\Psi\Phi-$sector, one finds that in virtue of
\re{P-H-P}, $\mathbb{H}_{\Psi\Phi}\partial_+\bar{A}\,\partial_+^{j-1}A$ is given
by the same expression with the fields $A$ and $\bar{A}$ interchanged in the
right-hand side\ of \re{H-PhiPsi-N=0}. Let us rewrite \re{H-PhiPsi-N=0} in the following
form (for $j\ge 2$)
\be
\bigg[\mathbb{H}_{\Phi\Psi}-\left(\gamma_{\Phi\Psi}(j+1) - \gamma^{\rm
(ex)}_{\Phi\Psi}(j+1)\mathbb{P}_{\Phi\Psi}\right) \bigg]\partial_+A \,
\partial_+^{j-1} \bar{A} (0)\stackrel{\rm fw}{=} 0\,,
\label{H-PhiPsi-op-N=0}
\ee
where the permutation operator $\mathbb{P}_{\Phi\Psi}$ interchanges the gauge
fields, $\mathbb{P}_{\Phi\Psi}\partial_+A \,\partial_+^{j-1}
\bar{A}=\partial_+\bar{A} \,
\partial_+^{j-1}A$. The operators $\partial_+A \, \partial_+^{j-1} \bar{A} (0)$
carry the conformal spin  $\mathbb{J}_{12} = j + j_\Psi + j_\Phi$ with
$j_\Phi=-1/2$ and $j_\Psi=3/2$. Setting $j+1=\mathbb{J}_{12}$ in
\re{H-PhiPsi-op-N=0} one recovers \re{H-PhiPsi-J} for $\mathcal{N}=0$.

According to \re{H-PhiPsi-N=0}, the Wilson operators $\partial_+A \,
\partial_+^{j-1} \bar{A} (0)$  (with $j\ge 2$) mix under renormalization with
the operators $\partial_+\bar{A} \,\partial_+^{j-1} A (0)$. One can resolve
the mixing by considering their linear combinations $\partial_+\bar{A}
\,\partial_+^{j-1}A(0)\pm \partial_+ A\,\partial_+^{j-1}\bar{A}(0)$, which
diagonalize the permutation operator $\mathbb{P}_{\Phi\Psi}$. In the special
case of the twist-two operators, Eq.~\re{tw=2-g}, one finds from \re{H-PhiPsi-op-N=0}
\be
\bigg[\mathbb{H}_{\Phi\Psi}-\gamma_{\mathcal{N}=0}(j) \bigg] \tr\{\partial_+A \,
\partial_+^{j-1} \bar{A} (0)\}
\stackrel{\rm fw}{=} 0\,,
\label{H-N=0-anom-dim}
\ee
with the anomalous dimension
\begin{equation}
\gamma_{\mathcal{N}=0}(j) = \psi (j + 3) + \psi (j -1) - 2 \psi (1) - \frac{6(-
1)^{j}}{(j+2)(j + 1)j(j - 1)}\, .
\label{gamma-Regge}
\end{equation}
For even/odd $j$, Eq.~\re{H-N=0-anom-dim} defines the anomalous dimensions of the
parity even/odd operators \re{tw=2-g}, $\gamma_{O^{\rm g}(j)}$ and
$\gamma_{\widetilde{O}^{\rm g}(j)}$, respectively, which are in agreement with
the known results~\cite{GroWil73,BukFroKurLip85}. Eq.~\re{gamma-Regge} can be
obtained from the general relation \re{H-PhiPsi-J} by taking into account that
$c_{\mathcal{N}}=2$ for $\mathcal{N}=0$,
$\mathbb{J}_{\Phi\Psi}=c_\mathcal{N}-1+j=j+1$ and $\mathbb{P}_{\Phi\Psi}=(-1)^j$.

\subsection{Wilson operators in $\mathcal{N}=1$ and $\mathcal{N}=2$ theories}

Let us now extend the analysis to supersymmetric gauge theories and start with a
simplest supersymmetric extension of gluodynamics, the $\mathcal{N} = 1$ SYM
theory. The light-cone superfields are
\begin{eqnarray*}
\Phi(Z_1)=\partial_+^{-1} A(z_1) + \theta_1 \partial_+^{-1}
\bar\lambda(z_1)\,,\qquad \Psi(Z_2) =-\lambda(z_2) + \theta_2 \partial_+ \bar
A(z_2)\,,
\end{eqnarray*}
with $Z_k=(z_k,\theta_k)$ (for $k=1,2$). The product of superfields can be
expanded in both $z$'s and $\theta$'s. Expansion in powers of the light-cone
variables $z_1$ and $z_2$ generates Wilson operators of arbitrary Lorentz spin
while the expansion in powers of $\theta_1$ and $\theta_2$ produces operators of
various partonic content. Supersymmetry imposes restrictions on the mixing
matrices of these operators. As before, to simplify analysis, we shall take the
forward limit and neglect operators involving total \textsl{light-cone}
derivatives.

\subsubsection*{$\Psi\Psi-$sector}

In the $\Psi\Psi-$sector, the product of two superfields admits the following
expansion in terms of local operators in the forward limit
\begin{equation}
\Psi (Z_1) \Psi (Z_2) \stackrel{\rm fw}{=} \sum_{j = 0}^\infty
\frac{z_{21}^j}{j!} \left\{ \lambda \partial_+^j \lambda(0) + \theta_2\!\cdot\!
\lambda
\partial_+^{j + 1} \bar{A}(0)
-
\theta_1\!\cdot\! \partial_+\bar{A}\, \partial_+^{j } \lambda(0)
+ \theta_1 \theta_2\!\cdot\!\partial_+\bar{A}\, \partial_+^{j + 1} \bar{A}(0)
\right\} \, .
\label{PsiPsi-exp-fw}
\end{equation}
The scale dependence of this product is driven to one-loop order by the
dilatation operator $\mathbb{H}_{\Psi\Psi}$, Eq.~\re{Psi-Psi-ker}. Substitution
of \re{PsiPsi-exp-fw} into \re{Psi-Psi-ker} yields
\begin{eqnarray}
\mathbb{H}_{\Psi\Psi} \Psi (Z_1) \Psi (Z_2) \stackrel{\rm fw}{=} \sum_{j =
0}^\infty \frac{z_{21}^j}{j!} \Big\{ \gamma_{\rm qq} (j) \, \lambda
\partial_+^j \lambda \!\!\!&+&\!\!\! \theta_2 \Big( \gamma_{\rm qg} (j)
\, \lambda\, \partial_+^{j + 1} \bar{A} + \gamma^{\rm (ex)}_{\rm qg} (j)
\, \partial_+\bar{A}\, \partial_+^{j } \lambda \Big)
\label{H-PsiPsi-ex}
\\
-\theta_1 \Big( \gamma_{\rm qg} (j) \, \partial_+\bar{A}\, \partial_+^{j } \lambda
\!\!\!&+&\!\!\! \gamma^{\rm (ex)}_{\rm qg} (j) \, \lambda\, \partial_+^{j + 1}
\bar{A} \Big) + \theta_1 \theta_2 \gamma_{\rm gg} (j) \, \partial_+\bar{A}\,
\partial_+^{j + 1}
\bar{A} \Big\} \, , \nonumber
\end{eqnarray}
where the anomalous dimensions are
\be
\begin{array}{ll}
  \gamma_{\rm qq} (j) = 2 \psi (j + 2) - 2\psi (1)\,,
  & \quad \gamma_{\rm gg} (j) = 2 \psi (j + 3) - 2 \psi (1)\,, \\[3mm]
  \gamma_{\rm qg} (j) =\psi (j + 3) + \psi (j + 2) - 2 \psi (1)
  \,,
  & \quad  \gamma^{\rm (ex)}_{\rm qg} (j) = {\displaystyle \frac{1}{j + 2}}\, . \\
\end{array}
\label{N=1-anom-dim-gauge0}
\ee
Equating the coefficients in front of an even number of $\theta$'s in both
sides of \re{H-PsiPsi-ex}, one gets the expressions for anomalous dimensions
of the maximal-helicity gauge field operators
\be
\big[\mathbb{H}_{\Psi\Psi}-\gamma_{\rm gg} (j)\big]\partial_+\bar{A}\,
\partial_+^{j + 1}\bar{A}(0)\stackrel{\rm fw}{=}0\,,
\label{N=1-anom-dim-gauge}
\ee
and maximal-helicity gaugino operators
\be
\big[\mathbb{H}_{\Psi\Psi}-\gamma_{\rm qq} (j)\big]\lambda\,
\partial_+^j \lambda(0)\stackrel{\rm fw}{=}0\,.
\label{N=1-anom-dim}
\ee
These relations are in agreement with the known
results~\cite{BukFroKurLip85,BelMul99,Kir04}. Notice that the anomalous
dimension of the operator $\partial_+\bar{A}\, \partial_+^{j + 1}\bar{A}(0)$
is the same as in the $\mathcal{N}=0$ theory, Eq.~\re{PsiPsi-V1}.

Comparing the terms linear in $\theta$'s in both sides of \re{H-PsiPsi-ex}, one
identifies the mixing matrix for the operators $\partial_+\bar{A}\,
\partial_+^{j } \lambda$ and $\lambda \partial_+^{j + 1} \bar{A}$. Its
diagonalization reveals that the operators
\begin{equation}
\lambda\, \partial_+^{j + 1} \bar{A} -
\partial_+\bar{A}\, \partial_+^{j} \lambda
\, , \qquad \lambda\, \partial_+^{j + 1} \bar{A} +
\partial_+\bar{A}\, \partial_+^{j} \lambda
\end{equation}
have an autonomous scale dependence in the forward limit and possess the
eigenvalues
\ba
&& \gamma_{\rm qg} (j)-\gamma^{\rm (ex)}_{\rm qg} (j)= \gamma_{\rm qq} (j)= 2
\psi (j + 2) - 2
\psi(1)
\nonumber \\[2mm]
&& \gamma_{\rm qg} (j)+\gamma^{\rm (ex)}_{\rm qg} (j)=\gamma_{\rm gg} (j)=2
\psi (j + 3) - 2
\psi(1)\, ,
\ea
respectively (see, e.g., \cite{BukFroKurLip85,BelMul99}).

In the $\mathcal{N} = 2$ SYM theory, the analysis goes along the same lines but
it is slightly lengthier due to the presence of an extra fermionic direction in
the superspace, $Z=(z,\theta^A)$ (with $A=1,2$). The $\mathcal{N}=2$ light-cone
superfields are given by \re{M=2-field} and involve an additional complex scalar
field $\phi$. For the product of two superfields, we find in the forward limit
\begin{eqnarray}
\Psi (Z_1) \Psi (Z_2) \stackrel{\rm fw}{=} \sum_{j = 0}^\infty
\frac{z_{21}^j}{j!} \Big\{ \!\!\!\!\!&-&\!\!\!  \phi\, \partial^j_+ \phi
-\widetilde\theta_{1A} \widetilde\theta_{2B} \, \lambda^{\{A}
\partial_+^j \lambda^{B\}}  + (\theta_1 \!\cdot\! \theta_1)(
\theta_2\!\cdot\!\theta_2) \,\partial_+\bar{A}\,\partial_+^{j + 1} \bar{A}
\label{PsiPsi-N=2-ex}
\\
&& \hspace*{-20mm} +i (\theta_2\!\cdot\!\theta_2) \phi \partial^{j + 1}_+
\bar{A} + i( \theta_1\!\cdot\!\theta_1) \, \partial_+\bar{A} \partial^{j}_+
\phi - (\theta_1\!\cdot\!\theta_2)\,\varepsilon_{AB} \lambda^A
\partial_+^j \lambda^B + \dots
\Big\} \, , \nonumber
\end{eqnarray}
where ellipses stand for fermionic operators built from the gaugino and scalar
fields. Here $\scriptstyle \{ AB \} = \ft12 (AB + BA)$ denotes the symmetrization
with respect to the $SU(2)$ indices and notations were introduced for $(\theta
\cdot\theta') \equiv \ft12 \varepsilon_{AB} \theta^A \theta^{'B}$ and
$\widetilde\theta_A \equiv \varepsilon_{AB} \theta^B$.

As before, we substitute \re{PsiPsi-N=2-ex} into \re{Psi-Psi-ker} for
$\mathcal{N}=2$ and evaluate $\mathbb{H}_{\Psi\Psi} \Psi (Z_1) \Psi (Z_2)$ in the
forward limit. Matching the coefficients in front of powers of $\theta$'s, we
evaluate the anomalous dimensions of various Wilson operators in the
$\mathcal{N}=2$ SYM theory \cite{KouRos82}. In this manner, one finds that the
anomalous dimensions of the gauge field operators, $\bar{A}\,\partial_+^{j + 2}
\bar{A}$, and gaugino operators in the triplet $SU (2)$ representation, $\lambda^{\{A}
\partial_+^j \lambda^{B\}}$, are the same as in the $\mathcal{N}=1$ theory,
Eqs.~\re{N=1-anom-dim-gauge0} and \re{N=1-anom-dim-gauge}, respectively. For the
operators built from two scalars one finds
\begin{equation}
\mathbb{H}_{\Psi\Psi} \phi\, \partial^j_+ \phi(0) \stackrel{\rm fw}{=} 2 \left[
\psi (j + 1) - \psi (1) \right] \phi \,\partial^j_+ \phi(0) \, .
\label{N=2-scal-scal}
\end{equation}
The remaining three operators, $\phi\, \partial_+^{j + 1}
\bar{A}(0)$, $\partial_+\bar{A}\,\partial_+^{j} \phi(0)$ and
$\varepsilon_{AB} \lambda^A \partial_+^j \lambda^B(0)$, mix under
renormalization. For instance,
\be
\mathbb{H}_{\Psi\Psi}\varepsilon_{AB} \lambda^A
\partial_+^j \lambda^B(0) \stackrel{\rm fw}{=} 2 \left[
\psi (j + 2) - \psi (1) \right] \varepsilon_{AB} \lambda^A
\partial_+^j \lambda^B +\frac{2i}{j+2}\left(\phi \,\partial_+^{j + 1} \bar{A}
+
\partial_+\bar{A}\, \partial_+^{j}\phi \right)\,.
\ee
Diagonalizing the corresponding $3\times 3$ mixing matrix one constructs three
operators
\begin{eqnarray}
O^{(1)}_j(0) \!\!\!&=&\!\!\!    \phi\, \partial_+^{j + 1} \bar{A}(0) +
\partial_+\bar{A}\,\partial_+^{j} \phi(0) +i \varepsilon_{AB}
\lambda^A
\partial_+^j \lambda^B (0)
\, , \nonumber \\[2mm]
O^{(2)}_j(0) \!\!\!&=&\!\!\!  \phi \,\partial_+^{j + 1} \bar{A}(0) -
\partial_+\bar{A}\, \partial_+^{j} \phi(0) \, , \label{OOO}
\\
O^{(3)}_j(0) \!\!\!&=&\!\!\!  \phi\, \partial_+^{j + 1} \bar{A}(0) +
\partial_+\bar{A}\,\partial_+^{j} \phi(0)-\frac{j+2}{j+1}\,
i\varepsilon_{AB} \lambda^A
\partial_+^j \lambda^B (0)\, . \nonumber
\end{eqnarray}
They have an autonomous scale dependence in the forward limit
\begin{equation}
\mathbb{H}_{\Psi\Psi} O^{(n)}_j(0) \stackrel{\rm fw}{=} 2 \left[ \psi (j + n) -
\psi (1) \right] O^{(n)}_j(0) \, ,\qquad (n=1,2,3).
\end{equation}
Thus, in agreement with our expectations, Eq.~\re{H-PsiPsi-J}, the anomalous
dimensions of Wilson operators in the $\Psi\Psi-$sector in the $\mathcal{N}=1$
and $\mathcal{N}=2$ SYM theories are given by the same universal function
$\gamma_{\Psi\Psi}(j)$, Eq.~\re{anom-dim0}, with the argument determined by the
conformal spin of the Wilson operator.

It is straightforward to extend the above analysis to the $\Phi\Phi-$sector.
The Wilson operators in this sector can be obtained from those in the
$\Psi\Psi-$sector by substituting gauge field, gaugino and scalar by complex
conjugated fields. This does not affect their anomalous dimensions and leads to
\re{H-PhiPhi-J}.

We would like to stress that the operators \re{OOO} have an autonomous scale
dependence only in the forward limit. Beyond this limit they mix under
renormalization with Wilson operators involving total derivatives. The
corresponding mixing matrix takes a triangular form and its non-diagonal elements
are fixed by the $SL(2)$ invariance. As in the $\mathcal{N}=0$ case,
Eqs.~\re{conf-op-exp} and \re{Conf-op}, taking the mixing into account amounts to
replacing Wilson operators in the right-hand side\ of \re{OOO} by conformal
operators~\cite{BraKorMul03}. The resulting operators are primaries of the
$SL(2|\mathcal{N})$ group and we shall refer to them as superconformal operators.

\subsubsection*{$\Phi\Psi-$sector}

Let us consider Wilson operators in the mixed $\Phi\Psi-$ sector in $\mathcal{N}
= 1$ SYM theory. The scale dependence of $\Phi (Z_1) \Psi (Z_2)$ is governed to
one-loop order by the dilatation operator $\mathbb{H}_{\Psi\Phi}$ defined in
\re{ansatz-gen-H12}. For the product of two light-cone superfields one gets in
the forward limit
\begin{equation}
\Phi (Z_1) \Psi (Z_2) \stackrel{\rm fw}{=} \sum_{j = 0}^\infty
\frac{z_{21}^j}{j!} \left\{ -\partial_+A \,\partial_+^{j - 2} \lambda  +
\theta_2\,
\partial_+ A\, \partial_+^{j-1} \bar{A} + \theta_1\,
\bar\lambda\,
\partial_+^{j - 1} \lambda + \theta_1 \theta_2\, \bar\lambda\,
\partial^{j} \bar{A}\right\} \, .
\label{PsiPhi-ex-N=1}
\end{equation}
In this expansion, odd (even) powers of $\theta$'s are accompanied by bosonic
(fermionic) operators. In Eq.~\re{PsiPhi-ex-N=1}, the first few terms with
$j\le 2$ involve spurious operators. The latter are eliminated by the projector
$\Pi_{\Phi\Psi}$ , Eq.~\re{projector-phi-psi}
\be
(1-\Pi_{\Phi\Psi})\Phi(Z_1)\Psi(Z_2)\stackrel{\rm fw}{=} \theta_1\sum_{j\ge 1}
\frac{z_{21}^j}{j!}\bar\lambda\partial_+^{j-1}\lambda+ \theta_2\left(-z_{21}
\bar\lambda\lambda+\sum_{j\ge 2} \frac{z_{21}^j}{j!} \partial_+A\,\partial_+^{j-1}
\bar A\right) +\ldots\,,
\label{ann-1}
\ee
where ellipses denote the contribution of fermionic operators $\partial_+A
\,\partial_+^{j - 2} \lambda$ and $\bar\lambda\, \partial^{j + 2} \bar{A}$.
In a similar manner one obtains
\be
(1-\Pi_{\Psi\Phi})\Psi(Z_1)\Phi(Z_2)\stackrel{\rm fw}{=}
\theta_1\left(\sum_{j\ge 2} \frac{z_{21}^j}{j!}\partial_+\bar A\,\partial_+^{j-1}
A -z_{21}\lambda\bar\lambda \right) +\theta_2\sum_{j\ge 1} \frac{z_{21}^j}{j!}
\lambda\partial_+^{j-1}\bar\lambda +\ldots\,.
\label{ann-2}
\ee
The subsequent analysis goes through the same steps as in Sect.~5.1. Namely, one
substitutes \re{PsiPhi-ex-N=1} into \re{ansatz-gen-H12} and \re{V}, takes into
account \re{ann-1} and \re{ann-2} and evaluates
$\mathbb{H}_{\Phi\Psi}\Phi(Z_1)\Psi(Z_2)$. One extracts the scale dependence of
Wilson operators by comparing the coefficients in front of $\theta_1$ and
$\theta_2$.

In this way, one obtains for the gauge field Wilson operators (for $j \ge 2$)
\ba
\mathbb{H}_{\Phi\Psi}\,\partial_+A \,\partial_+^{j-1} \bar
A(0)\!\!\!&\stackrel{\rm fw}{=}&\!\!\! \gamma_{\Phi\Psi}(j+1)\partial_+ A
\,\partial_+^{j-1}\bar A(0)-\gamma^{\rm (ex)}_{\Phi\Psi}(j+1)\partial_+\bar A
\,\partial_+^{j-1} A(0)
\label{H-PhiPsi-N=1}
\\[2mm]
\!\!\!&-&\!\!\! \frac{\bar\lambda
\partial_+^{j-1}\lambda(0)}{j-1} - \frac{2\,\lambda
\partial_+^{j-1}\bar\lambda(0)}{(j-1)j(j+1)}\,,
\nonumber
\ea
where $\gamma_{\Phi\Psi}(j+1)$ and $\gamma^{\rm (ex)}_{\Phi\Psi}(j+1)$ the same
as in the $\mathcal{N}=0$ theory, Eq.~\re{H-PhiPsi-N=0}, and for gaugino
operators (for $j \ge 1$)
\ba
\mathbb{H}_{\Phi\Psi}\,\bar\lambda\,\partial_+^{j-1}
\,\lambda(0)\!\!\!&\stackrel{\rm fw}{=}&\!\!\!
\left[\psi(j+2)+\psi(j)-2\psi(1)\right]
\bar\lambda \,\partial_+^{j-1}\lambda(0) +\delta_{j 1}\frac13
\left(\bar\lambda\lambda(0)+\lambda\bar\lambda(0)\right)
\label{H-lambda-barlambda}\\[2mm]
\!\!\!&-&\!\!\! \theta_{j, 2}\left(\frac1{j+2}
\partial_+A \,\partial_+^{j-1}\bar
A(0)+
\frac{2}{j(j+1)(j+2)}\partial_+\bar A \,
\partial_+^{j-1}A(0)\right). \nonumber
\ea
In Eq.~\re{H-lambda-barlambda}, the $\theta-$function in front of the gauge field
operators ensures that spurious operators $A \,\partial_+\bar A(0)$ and $\bar A
\,\partial_+A(0)$ (for $j=1$) do not mix with the Wilson gaugino operators. The
contribution $\sim\delta_{j,1}$ is generated by the terms $z_{21}
\bar\lambda\lambda$ and $z_{21}\lambda\bar\lambda$ in the expression for
the projector, Eqs.~\re{ann-1} and \re{ann-2}. In the conventional approach it
comes from the Feynman (annihilation) diagram in which gaugino and antigaugino
are first annihilated into a gluon and, then, are produced back,
$\lambda\bar\lambda+\bar\lambda\lambda=it^a f^{abc}\lambda^b\bar\lambda^c$.

For gaugino operators with no derivatives one gets from \re{H-lambda-barlambda}
\be
\mathbb{H}_{\Phi\Psi}\,\bar\lambda\lambda(0)=\frac{11}6\,\bar\lambda \lambda(0) +
\frac13\,\lambda\bar\lambda(0)\,.
\label{H-barl-l}
\ee
For $j\ge 2$ the gaugino operators $\partial_+^{j-1}\bar\lambda \,\lambda(0)$ mix
with the gauge field operators, Eq.~\re{H-lambda-barlambda}. Diagonalizing the
$2\times 2$ mixing matrix one finds that the following operators have autonomous
evolution in the forward limit
\ba
{\cal O}^{(1)}_j &=& \tr \{\partial_+ A \partial_+^{j-1} \bar A \} + \tr
\{\bar\lambda\,\partial_+^{j-1} \lambda \}\,,
\label{O-oper}
\\
{\cal O}^{(2)}_j &=& \tr \{\partial_+ A \partial_+^{j-1} \bar A \}
-\frac{j+2}{j-1} \tr \{\bar\lambda\,\partial_+^{j-1} \lambda \}\,.
\nonumber
\ea
For even/odd $j$, these operators are expressed in terms of the parity even/odd
twist-two operators \re{tw=2-q} and \re{tw=2-g}. The superconformal operators
\re{O-oper} diagonalize the $\mathcal{N}=1$ dilatation operator
\be
\bigg[\mathbb{H}_{\Phi\Psi}-\gamma_{\mathcal{N}=1}(j+n-1) \bigg]{\cal
O}^{(n)}_{j}(0)\stackrel{\rm fw}{=} 0\,,
\label{H-N=1-anom-dim}
\ee
with $n=1,2$ and their anomalous dimension is given by
\begin{equation}
\gamma_{\mathcal{N}=1}(j) = \psi (j + 2) + \psi (j -1) - 2 \psi (1) - \frac{2 (-
1)^j}{(j + 1)j(j -1)} \, .
\end{equation}
Eq.~\re{H-N=1-anom-dim} is in a perfect agreement with the general relation
\re{H-PhiPsi-J}. To reproduce \re{H-PhiPsi-J} one has to take into account that
$c_{\mathcal{N}}=3/2$ for $\mathcal{N}=1$,
$\mathbb{J}_{\Phi\Psi}=c_\mathcal{N}-1+j=j+1/2$ and
$\mathbb{P}_{\Phi\Psi}=(-1)^j$.

It is straightforward to extend the above analysis to the $\mathcal{N}=2$
SYM theory. The expressions for the mixing matrices of Wilson operators in the
$\Phi\Psi-$sector become more involved due to a larger number of operators
involved. To save space, we only present explicit expressions for the
renormalization group evolution of the scalar operators $\bar\phi\,
\partial_+^j \phi(0)$. For $j=0$, scalar operators with no derivatives form
a closed sector
\be
\mathbb{H}_{\Phi\Psi}\,\bar\phi\phi(0) = \frac32 \bar\phi\phi(0) - \frac12
\phi\bar\phi(0)\,.
\label{H-barf-f}
\ee
For $j=1$ they only mix with the gaugino operators with no derivatives. For $j\ge
2$ they mix  both with the gauge field operators and the $SU(2)$ singlet gaugino
operators
\begin{equation}
\mathbb{H}_{\Phi\Psi}\,\bar\phi\,\partial_+^j \phi(0)\stackrel{\rm fw}{=} 2
\left[ \psi (j + 1) - \psi (1) \right] \bar\phi \partial^j \phi +
\frac{\partial_+A\partial_+^{j-1} \bar{A}+\partial_+\bar A
\partial_+^{j-1} {A}}{(j + 1)(j + 2)} + \frac{\bar\lambda_A \partial^{j - 1}
\lambda^A}{j + 1} \, .
\end{equation}
Diagonalizing the $3\times 3$ mixing matrix, one finds that the following
operators have an autonomous evolution in the forward limit
\ba
{\cal O}^{(1)}_j &=& \tr \{\partial_+ A \partial_+^{j-1} \bar A \} + \tr
\{\bar\lambda\,\partial_+^{j-1} \lambda \}-\tr \{\bar\phi\,\partial_+^j \phi\}\,,
\nonumber \\[2mm]
{\cal O}^{(2)}_j &=& \tr \{\partial_+ A \partial_+^{j-1} \bar A \} -\frac{1}{j-1}
\tr \{\bar\lambda\,\partial_+^{j-1} \lambda \}+\frac{j+1}{j-1}\tr
\{\bar\phi\,\partial_+^j \phi\}\,,
\\
{\cal O}^{(3)}_j &=& \tr \{\partial_+ A \partial_+^{j-1} \bar A \}
-\frac{j+2}{j-1} \tr \{\bar\lambda\,\partial_+^{j-1} \lambda
\}-\frac{(j+1)(j+2)}{(j-1)j}\tr \{\bar\phi\,\partial_+^j \phi\}\,. \nonumber
\ea
The superconformal operators ${\cal O}^{(n)}_j$ (with $n=1,2,3$) diagonalize the
dilatation operator $\mathbb{H}_{\Phi\Psi}$, Eq.~\re{H-PhiPsi-J}, and satisfy the
same relation as before, Eq.~\re{H-N=1-anom-dim}, with the corresponding
anomalous dimension $\gamma_{\mathcal{N}=2}(j+n-1)$ defined as
\be
\gamma_{\mathcal{N}=2}(j) = \psi( j+1) +\psi( j-1) - 2\,\psi(1)-{\frac {
 ( -1 ) ^{j}}{j ( j-1) }}\,.
\ee
To match \re{H-PhiPsi-J} one has to take into account that $c_{\mathcal{N}}=1$
for $\mathcal{N}=2$, $\mathbb{J}_{\Phi\Psi}=c_\mathcal{N}-1+j=j$ and
$\mathbb{P}_{\Phi\Psi}=(-1)^j$.

\subsection{Wilson operators in $\mathcal{N}=4$ theory}

In the $\mathcal{N}=4$ SYM theory, \textsl{all} two-particle quasipartonic
operators belong to the $\Phi\Phi-$sector. As before they can be obtained as
coefficients in the expansion of the product of two light-cone superfields
$\Phi(Z_1)\Phi(Z_2)$ in powers of $z$'s and $\theta$'s. The  scale dependence of
$\Phi(Z_1)\Phi(Z_2)$ is driven to one-loop by the dilatation operator
\re{H-4-matrix}. This operator has a much simpler form for $\mathcal{N}=4$ as
compared with the SYM theories with less supersymmetries, Eq.~\re{H-matrix}. This
simplicity gets lost as soon as one replaces the light-cone superfield $\Phi(Z)$
by its explicit expression \re{M=4-field} and switches from nonlocal light-cone
operators $\Phi(Z_1)\Phi(Z_2)$ to local Wilson operators built from gaugino,
scalar and gauge fields. A complete one-loop dilatation operator in the
$\mathcal{N}=4$ theory acting on the space spanned by Wilson operators has been
constructed in Ref.~\cite{BeiSta03}. Going over from nonlocal light-cone
operators to local Wilson operators, one finds that the obtained expressions for
the $\mathcal{N}=4$ dilatation operator, Eqs.~\re{H-4-matrix} and \re{N=4-kernel}
(see also \re{H-diag-1} below), agree with the results of Ref.~\cite{BeiSta03}.

As we have seen in Sect.~4.3, the dilatation operator for $\mathcal{N}\le 2$ can
be derived from the $\mathcal{N}=4$ dilatation operator through the truncation
procedure. According to \re{tr-pr-1} and \re{tr-pr-2}, the anomalous dimensions
of Wilson operators in the $\Phi\Phi-$ and $\Psi\Psi-$sectors for $\mathcal{N}\le
2$ coincide with anomalous dimensions of the same operators in the
$\mathcal{N}=4$ theory. In particular, this is the case for the maximal helicity
gauge field operators $\partial_+ A\,\partial_+^{j-2} A(0)$ and $\partial_+ \bar
A\,\partial_+^{j-2}\bar A(0)$. In the $\mathcal{N}=4$ theory these operators have
the same anomalous dimension as in the SYM theories with $\mathcal{N}=0,1,2$,
Eqs.~\re{PhiPhi-V} and \re{N=1-anom-dim-gauge}.

Let us examine Wilson operators built from scalar fields $\phi^{AB}$ and
$\bar\phi_{AB}$ with no derivatives. Such operators have a minimal possible
scaling dimension and could only mix to one-loop order with themselves. It is
convenient to switch from the complex fields $\phi^{AB}$ and $\bar \phi_{AB}=
\left( \phi^{AB} \right)^* = \ft12\varepsilon_{ABCD} \phi^{CD}$ to six real
scalars $\phi_j$ (with $j=1,\ldots,6$) defined as
\begin{equation}
\phi_j(z) = \frac1{2\sqrt{2}}\, {\Sigma}_j^{AB} \bar \phi_{AB}(z) \,,
\label{real-scalar}
\end{equation}
where ${\Sigma}_j^{AB}$ is a chiral block of six-dimensional Euclidean
Dirac matrices, expressed by means of 't Hooft symbols~\cite{BelDerKorMan03}.
According to \re{M=4-field}, the scalar field is related to the $\mathcal{N}=4$
superfield as
\be
\phi_j(z) = \frac{i}{2\sqrt{2}} {\Sigma}_j^{AB}
\partial_{\theta^A}\partial_{\theta^B} {\Phi}(z, \theta^A , 0)|_{\theta^A =
0}\,.
\ee
This allows one to deduce the scale dependence of the scalar operator
$\phi_{j_1}(0)\phi_{j_2}(0)$ from the scale dependence of the nonlocal light-cone
operator $\Phi(Z_1)\Phi(Z_2)$
\begin{equation}
\mathbb{H}_{\Phi\Phi}[\phi_{j_1} \phi_{j_2}(0)]= - \frac18 \left(
{\Sigma}_{j_1}^{AB}
\partial_{\theta_1^A}
\partial_{\theta_1^B} \right) \left( {\Sigma}_{j_2}^{CD}
\partial_{\theta_2^C} \partial_{\theta_2^D} \right)
\mathbb{H}_{\Phi\Phi} \Phi(Z_1)\Phi(Z_2)\bigg|_{Z_1 = Z_2 = 0}
\!\!
=
\! \sum_{k_1k_2}
V{}_{j_1j_2}^{k_1k_2} \ \phi_{k_1}\phi_{k_2}(0) \, .
\end{equation}
Replacing $\mathbb{H}_{\Phi\Phi} \Phi(Z_1)\Phi(Z_2)$ by its expression
\re{N=4-kernel} and calculating Grassmann derivatives one finds after some
algebra
\begin{equation}
V{}_{j_1j_2}^{k_1k_2} = \delta_{j_1}^{k_1} \delta_{j_2}^{k_2} + \ft12\,
\delta_{j_1j_2}\delta^{k_1k_2}- \delta_{j_1}^{k_2} \delta_{j_2}^{k_1}
\, . \label{V}
\end{equation}
This relation defines the one-loop mixing matrix the scalar operators
$\phi_{j_1}(0)\phi_{j_2}(0)$ in the multi-color limit~\cite{MinZar02}. As
was shown in~\cite{MinZar02}, the mixing matrix \re{V} possesses a hidden
symmetry---it can be mapped into a Hamiltonian of a completely integrable
Heisenberg $SO(6)\sim SU(4)$ spin chain.

As a last example, we examine the scale dependence of the gauge field operator
$\partial_+ A\,\partial_+^{j-1} \bar A(0)$. To this end one relates the gauge
fields with the $\mathcal{N}=4$ light-cone superfield \re{M=4-field}
\be
\partial_+ A(z) = \partial_{z}^2\, {\Phi}(z,0)\,,\qquad
\partial_+ \bar A(z) = - \textrm{d}_{\theta}\,{\Phi}(z,\theta^A)\big|_{\theta^A=0}\,,
\ee
where $\textrm{d}_\theta\equiv \ft1{4!} \varepsilon^{ABCD}
\partial_{\theta^A}\partial_{\theta^B}\partial_{\theta^C}\partial_{\theta^D}$.
Then, the evolution equation for the gauge field operator $\partial_+
A\,\partial_+^{j-1} \bar A(0)$ can be derived from \re{N=4-kernel} as
\be
\mathbb{H}_{\Phi\Phi}[\partial_+ A\,\partial_+^{j-1} \bar A(0)]=-
\left(\partial_{z_1}^2\partial_{z_2}^{j-2} \textrm{d}_{\theta_2}\right)
\mathbb{H}_{\Phi\Phi} \Phi(Z_1)\Phi(Z_2)\big|_{Z_1=Z_2=0}\,.
\label{tw-2-kernel}
\ee
Calculating the derivatives one obtains in the forward limit (for $j\ge 2$)
\ba
\mathbb{H}_{\Phi\Phi}\,\partial_+A \, \partial_+^{j-1}\bar A(0)
\!\!\!&\stackrel{\rm fw}{=}&\!\!\! \gamma_{\Phi\Psi}(j+1) \partial_+
A\,\partial_+^{j-1} \bar A -\gamma^{\rm (ex)}_{\Phi\Psi}(j+1)
\partial_+\bar A \,\partial_+^{j-1} A
\label{PhiPhi-AA-N=4}
\\[2mm]
\!\!\!&-&\!\!\!
\frac{\bar\lambda_A \partial_+^{j-1}\lambda^A}{j-1}
-
\frac{2\,\lambda^A \partial_+^{j-1}\bar\lambda_A}{(j-1)j(j+1)}
-
\frac{\bar\phi_{AB}\partial_+^j \phi^{AB}}{2j(j-1)}\,, \nonumber
\ea
where the anomalous dimensions in front of gauge field operators are the same as
in \re{gamma-gamma}. Let us compare \re{PhiPhi-AA-N=4} with similar relations
in the $\mathcal{N}=0$ and $\mathcal{N}=1$ theories, Eqs.~\re{H-PhiPsi-N=0} and
\re{H-PhiPsi-N=1}, respectively. We observe that \re{PhiPhi-AA-N=4} stays intact
as one goes over from $\mathcal{N}=4$ down to $\mathcal{N}=0$. The only
difference is that the contribution of ``unwanted'' fields has to be removed in
the right-hand side\ of \re{PhiPhi-AA-N=4}. This property is yet another
manifestation of the truncation procedure described in Sect.~4.3.

The aforementioned scalar and gauge field operators, $\phi_{j_1} \phi_{j_2}(0)$
and $\partial_+ A\,\partial_+^{j-1} \bar A(0)$, respectively, represent two
special examples of two-particle (twist-two) Wilson operators in the
$\mathcal{N}=4$ theory \cite{KotLip02}. A complete classification of such
operators has been worked out in Ref.~\cite{BelDerKorMan03}. Diagonalizing their
mixing matrix one can construct two-particle superconformal Wilson operators in
the $\mathcal{N}=4$ SYM theory with autonomous scale dependence. It is a
straightforward but tedious exercise to verify that using the obtained expression
for the dilatation operator \re{N=4-kernel} one reproduces the results of
Ref.~\cite{BelDerKorMan03}. One finds from \re{H-PhiPhi-J} that, in agreement
with Ref.~\cite{BeiSta03}, the one-loop anomalous dimensions of all
superconformal quasipartonic operators in the $\mathcal{N}=4$ theory are given by
the same universal function $2\left[\psi(\mathbb{J}_{\Phi\Phi})-\psi(1)\right]$
with the superconformal spin $\mathbb{J}_{\Phi\Phi}$ depending on the operator
under consideration.

\section{Hidden symmetries of the dilatation operator}

So far we discussed mostly the two-particle dilatation operators in different
sectors in the SYM theories, Eqs.~\re{H-matrix} and \re{H-4-matrix}. These
operators govern the scale dependence of the product of two light-cone
superfields, $\Phi(Z_1)\Phi(Z_2)$, $\Phi(Z_1)\Psi(Z_2)$, $\Psi(Z_1)\Phi(Z_2)$ and
$\Psi(Z_1)\Psi(Z_2)$, and allow us to construct the one-loop dilatation operator
$\mathbb{H}$ acting on arbitrary multi-particle operators, Eqs.~\re{Mixed-sector}
-- \re{Chiral-sector}, in the multi-color limit.

The nonlocal light-cone operators $\mathbb{O}(Z_1,\ldots,Z_L)$,
Eqs.~\re{Mixed-sector} -- \re{Chiral-sector}, satisfy the evolution equation
\re{EQ}. To solve it, one has to diagonalize the integral operator $\mathbb{H}$,
Eq.~\re{H-multi-color},
\be
\mathbb{H}\, \Psi_q(Z_1,\ldots Z_L) = (\mathbb{H}_{12}+\ldots+\mathbb{H}_{L1})\,
\Psi_q(Z_1,\ldots Z_L) = E_q \Psi_q(Z_1,\ldots Z_L)\,,
\label{Sch-eq}
\ee
where quantum numbers $q$ parameterize all possible solutions. Then, a general
solution to \re{EQ} takes the form
\be
\mathbb{O}(Z_1,\ldots,Z_L)=\sum_q \Psi_q(Z_1,\ldots Z_L)\, \mathcal{O}_q(0)\,,
\ee
with the expansion coefficients $\mathcal{O}_q(0)$ being local composite
operators. It follows from  \re{EQ} that to one-loop order, the operators
$\mathcal{O}_q(0)$ have an autonomous scale dependence and satisfy the
evolution equation
\be
\mu \frac{d}{d\mu}\mathcal{O}_q(0) =-\gamma_q\,
\mathcal{O}_q(0)\,, \qquad \gamma_q=\frac{g^2
N_c}{8\pi^2}\left(E_q+L\gamma_\mathcal{N}^{(0)}\right),
\label{spectrum}
\ee
where the one-loop anomalous dimension of the superfields
$\gamma_\mathcal{N}^{\scriptscriptstyle (0)}=\beta_0/(2N_c)$ is proportional to
the one-loop beta-function, $\gamma_\mathcal{N}^{\scriptscriptstyle (0)}= -
{11}/{6}, - 3/2 , - 1, 0$ for $\mathcal{N}=0,1,2,4$, respectively.

Equation~\re{spectrum} determines the spectrum of anomalous dimensions of Wilson
operators in the SYM theory built from $L$ fundamental fields. For $L=2$ one has
$\mathbb{H}=2 \mathbb{H}_{12}$ and, therefore, $E_q$ is twice the eigenvalue of
two-particle dilatation operators, Eqs.~\re{H-PsiPsi-J} -- \re{H-PsiPhi-J}. It
turns out that for some operators the dilatation operator $\mathbb{H}$ can be
mapped into a Hamiltonian of integrable lattice models so that their anomalous
dimensions are in the one-to-one correspondence with the energy spectrum of these
models.

\subsection{XXX Heisenberg (super)spin chain}

For the light-cone operators built from $\Phi-$superfields,
Eq.~\re{Chiral-sector} the dilatation operator takes the form \re{H-multi-color}
with the two-particle kernel given by the dilatation operator in the
$\Phi\Phi-$sector, $\mathbb{H}_{\Phi\Phi}$, Eqs.~\re{ansatz-PhiPhi} and
\re{H-PhiPhi-J}
\be
\mathbb{H}\,\mathbb{O}_{\Phi\ldots\Phi} (Z_1, \ldots, Z_L) =\sum_{k=1}^L
2\left[
\psi(\mathbb{J}_{k,k+1})-\psi(1)
\right] \theta(\mathbb{J}_{k,k+1})\,\mathbb{O}_{\Phi\ldots\Phi} (Z_1, \ldots,
Z_L)\,.
\label{H-diag-1}
\ee
Here $\mathbb{J}_{k,k+1}$ is the $SL(2|\mathcal{N})$ superconformal spin in the
sector $\Phi(Z_k)\Phi(Z_{k+1})$ defined in Eqs.~\re{J-ab} and \re{Casimir} and
the periodic boundary conditions are imposed,
$\mathbb{J}_{L,L+1}=\mathbb{J}_{L,1}$. We recall that in the $\mathcal{N}=4$
theory all quasipartonic operators reside inside the nonlocal light-cone operators
$\mathbb{O}_{\Phi\ldots\Phi} (Z_1, \ldots, Z_L)$ and, therefore, \re{H-diag-1}
defines a complete one-loop $\mathcal{N}=4$ dilatation operator in the
multi-color limit.

For $\mathcal{N}\le 2$ one has to consider in addition the light-cone operators
involving $\Psi-$superfields. For the light-cone operators built from
$\Psi-$superfields only, Eq.~\re{Chiral-sector-1}, the one-loop dilatation
operator is given in the multi-color limit by the sum over two-particle
dilatation operators in the $\Psi\Psi-$sector, $\mathbb{H}_{\Psi\Psi}$,
Eqs.~\re{ansatz-PsiPsi} and \re{H-PsiPsi-J},
\be
\mathbb{H}\,\mathbb{O}_{\Psi\ldots\Psi} (Z_1, \ldots, Z_L) =\sum_{k=1}^L
2\left[\psi(\mathbb{J}_{k,k+1})-\psi(1)\right]\,\mathbb{O}_{\Psi\ldots\Psi} (Z_1,
\ldots, Z_L)\,,
\label{H-diag-2}
\ee
where $\mathbb{J}_{k,k+1}$ is the $SL(2|\mathcal{N})$ superconformal spin in the
sector $\Psi(Z_k)\Psi(Z_{k+1})$ and $\mathbb{J}_{L,L+1}=\mathbb{J}_{L,1}$.

The dilatation operators \re{H-diag-1} and \re{H-diag-2} have a hidden
symmetry~\cite{BelDerKorMan04}. In both cases, the operator $\mathbb{H}$ can be
identified as a Hamiltonian of a completely integrable XXX Heisenberg spin magnet
with the $SL(2|\mathcal{N})$ symmetry~\cite{Kulish,Frahm,DKK}. The length of the
spin chain equals $L$ and the (super)spin in each site is defined by the
superconformal spin of the corresponding superfield $j_\Phi=-1/2$ and
$j_\Psi=(3-\mathcal{N})/2$. As a result, the Schr\"odinger equation \re{Sch-eq}
for the Hamiltonians \re{H-diag-1} and \re{H-diag-2} can be solved exactly by the
Quantum Inverse Scattering Method~\cite{QISM} and the spectrum of the anomalous
dimensions of the light-cone operators \re{Chiral-sector} and
\re{Chiral-sector-1} can be calculated by the Bethe Ansatz technique.

For the ``mixed'' light-cone operators \re{Mixed-sector} built from both
${\Phi}-$ and ${\Psi}-$superfields in the $\mathcal{N}\le 2$ theory, the
dilatation operator has a more complicated form as compared with \re{H-diag-1}
and \re{H-diag-2}. The reason for this is that the operators \re{Mixed-sector}
could mix with other light-cone operators containing the same number of ${\Phi}-$
and ${\Psi}-$superfields but ordered differently inside the trace. In the
expression for the two-particle evolution kernels \re{Mixed-sector} such mixing
is described by the ``exchange'' interaction, Eq.~\re{V-ex-super}. Its impact on
the properties of the dilatation operator has been thoroughly studied in context
of the $\mathcal{N}=0$ theory in Refs.~\cite{BraDerKorMan99,Bel99}. It was found
that the exchange interaction breaks integrability symmetry of the one-loop
dilatation operator and modifies the scaling properties of the operators
\re{Mixed-sector} in a very peculiar way---it lifts the degeneracy in the
``energy'' levels of $\mathbb{H}$ and generates a finite mass gap in the spectrum
of anomalous dimensions of Wilson operators with large conformal spin.

Another manifestation of the same phenomenon comes from the analysis of the
dilatation operator in the sector of Wilson operators with the minimal scaling
dimension. As we show in the next section, ``nonintegrable'' addenda to the
dilatation operator in the $\mathcal{N}-$extended SYM theory modify the mixing
matrix for these operators in such a way that it can be mapped into a Hamiltonian
of the XXZ Heisenberg magnet with the anisotropy parameter depending on
$\mathcal{N}$. For $\mathcal{N}=2$ a similar observation has been made in
Ref.~\cite{DiVTan04}. It is interesting to note that XXZ spin chains have
previously emerged in QCD in the studies of high-energy (Regge) asymptotics of
multi-gluonic scattering amplitudes in the double-logarithmic
approximation~\cite{Shuvaev,DerKir01}.

\subsection{XXZ Heisenberg spin chain}

The Wilson operators with minimal scaling dimension are built from the
fundamental fields with the lowest dimension, that is, from the gauge field
strength $\partial_+ A$ and $\partial_+\bar A$ for $\mathcal{N}=0$, from the
gaugino fields  $\lambda$ and $\bar\lambda$ for $\mathcal{N}=1$, from the
complex scalar fields $\phi$ and $\bar\phi$ for $\mathcal{N}=2$ and, finally,
from the real scalars $\phi_j$ for $\mathcal{N}=4$. They do not contain
additional light-cone derivatives and define a closed sector at one-loop
order.

We recall that in the $\mathcal{N}=4$ theory the two-particle mixing matrix for
the scalar operators with no derivatives is given by \re{V} and it can be mapped
into the XXX Heisenberg spin chain with the $SO(6)$ symmetry. It turns out that
in the SYM theories with $\mathcal{N}\le 2$, the mixing matrix for the Wilson
operators with a minimal scaling dimension built from the fields mentioned above
can be mapped into an XXZ Heisenberg spin chain.

Notice that for $\mathcal{N}\le 2$ the fundamental fields with the lowest
dimension are the lowest components of the light-cone superfield $\Psi(Z)$ and
the highest components of $\Phi(Z)$, Eqs.~\re{M=0-field} -- \re{M=2-field}. To
write down the mixing matrix it is convenient to associate with them two
spin$-1/2$ states
\be
\ket{\uparrow} = \{ \partial_+ \bar A(0), \lambda(0), \phi(0) \}\,,\qquad
\ket{\downarrow} = \{ \partial_+ A(0),  \bar\lambda(0), \bar\phi(0) \}\,,
\ee
where the three entries inside the curly brackets correspond to $\mathcal{N}=0,1,2$,
respectively. Then, multi-particle single-trace Wilson operators of minimal
dimension can be mapped into the spin states. For example, the state
$\ket{\uparrow\downarrow\dots\uparrow}$ gives rise to the operators
$\tr\{\partial_+
\bar A\partial_+ A\dots\partial_+ \bar A\}$ for $\mathcal{N}=0$,
$\tr\{\lambda\bar\lambda\dots\lambda\}$ for $\mathcal{N}=1$ and
$\tr\{\phi\bar\phi\dots\phi\}$ for $\mathcal{N}=2$.

Let us examine the action of the one-loop dilatation operator on the
two-particle Wilson operators defined by the states $\ket{\uparrow\uparrow}$,
$\ket{\uparrow\downarrow}$, $\ket{\downarrow\uparrow}$ and
$\ket{\downarrow\downarrow}$. By definition, the operators
$\ket{\uparrow\uparrow}=\{\partial_+\bar A\partial_+\bar
A,\lambda\lambda,\phi\phi\}$ belong to the $\Psi\Psi-$sector and have the lowest
conformal spin possible in this sector, $\mathbb{J}_{\Psi\Psi}=3-\mathcal{N}$.
Their anomalous dimension is given by \re{H-PsiPsi-J} for
$\mathbb{J}_{\Psi\Psi}=3-\mathcal{N}$
\be
\mathbb{H}_{12} \,{\ket{\uparrow\uparrow\dots}}=2\big[\psi(3-\mathcal{N})-\psi(1)
\big]\ket{\uparrow\uparrow\dots}\,,
\label{H-spin1}
\ee
where the subscript indicates that the dilatation operator acts on the first two
spin states. One verifies that for $\mathcal{N}=0,1,2$ this relation is in
agreement with \re{anom-dim-N=0}, \re{N=1-anom-dim} and \re{N=2-scal-scal},
respectively. The complex conjugated operators
$\ket{\downarrow\downarrow}=\{\partial_+A\partial_+
A,\bar\lambda\bar\lambda,\bar\phi\bar\phi\}$ have the same anomalous dimension as
${\ket{\uparrow\uparrow}}$ and, therefore,
\be
\mathbb{H}_{12}
\,{\ket{\downarrow\downarrow\dots}}=2\big[\psi(3-\mathcal{N})-\psi(1)
\big]\ket{\downarrow\downarrow\dots}\,.
\ee
In a similar manner, the operators $\ket{\uparrow\downarrow}=\{\partial_+\bar
A\partial_+ A,\lambda\bar\lambda,\phi\bar\phi\}$ belong to the $\Psi\Phi-$sector
and have the lowest conformal spin possible in this sector,
$\mathbb{J}_{12}=c_\mathcal{N}+1=3-\mathcal{N}/2$, Eq.~\re{H-PhiPsi-J}.
Therefore, one finds from \re{H-PhiPsi-J}
\be
\mathbb{H}_{12}
\,{\ket{\uparrow\downarrow\dots}}=\big[\psi(5-\mathcal{N})-\psi(1)
\big]\ket{\uparrow\downarrow\dots}-\frac{(-1)^\mathcal{N}}{4-\mathcal{N}}
\ket{\downarrow\uparrow\dots}\,,
\label{H-up-down}
\ee
where
$\mathbb{P}_{\Phi\Psi}\ket{\uparrow\downarrow}=(-1)^\mathcal{N}\ket{\downarrow\uparrow}$
since for $\mathcal{N}=1$ the corresponding (gaugino) fields have Fermi
statistics. For $\mathcal{N}=0,1,2$ Eq.~\re{H-up-down} is in agreement with
\re{H-PhiPsi-N=0}, \re{H-barl-l} and \re{H-barf-f}, respectively. For complex
conjugated operators $\ket{\downarrow\uparrow}=\{\partial_+A\partial_+
\bar A,\bar\lambda\lambda,\bar\phi\phi\}$ one gets
\be
\mathbb{H}_{12}
\,{\ket{\downarrow\uparrow\dots}}=\big[\psi(5-\mathcal{N})-\psi(1)
\big]\ket{\downarrow\uparrow\dots}-\frac{(-1)^\mathcal{N}}{4-\mathcal{N}}
\ket{\uparrow\downarrow\dots}\,.
\label{H-spin4}
\ee
Combining together \re{H-spin1} -- \re{H-spin4} one can write the operator
$\mathbb{H}_{12}$ as the Hamiltonian of a XXZ Heisenberg spin$-1/2$ magnet
\be
\label{H-XXZ}
\mathbb{H}_{12} = H_x \sigma^x\otimes \sigma^x+ H_x \sigma^y \otimes\sigma^y+ H_z
\sigma^z\otimes \sigma^z + H_0 \II\otimes \II
\ee
where the Pauli matrices act on the spin states as
\be
\sigma^-\ket{\uparrow}=\ket{\downarrow}\,,\quad \sigma^+ \ket{\uparrow}=0\,,\quad
\sigma^z \ket{\uparrow}=\ket{\uparrow}\,,\quad \sigma^+
\ket{\downarrow}=\ket{\uparrow}\,,\quad \sigma^- \ket{\downarrow}=0\,,\quad
\sigma^z \ket{\downarrow}=-\ket{\downarrow}
\ee
with $\sigma^\pm=(\sigma^x\pm i\sigma^y)/2$. Matching \re{H-XXZ} into
\re{H-spin1} -- \re{H-spin4} one gets
\ba
H_x&=&- {\frac { \left( -1 \right) ^{\mathcal{N}}}{2(4-\mathcal{N})}}
\nonumber \\
H_z&=&\psi \left( 3-\mathcal{N} \right) -\frac 12\,\psi \left( 5-\mathcal{N}
\right) -\frac 12\,\psi(1)
\\
H_0&=&\psi \left( 3-\mathcal{N} \right) + \frac 12\,\psi \left( 5-\mathcal{N}
\right) - \frac 32\,\psi(1)\,. \nonumber
\ea
Using these relations one finds the
anisotropy parameter $\Delta=H_z/H_x$
\be
\Delta_{\mathcal{N}=0}=-\frac{11}3\,,\qquad
\Delta_{\mathcal{N}=1}=\frac12\,,\qquad  \Delta_{\mathcal{N}=2}= 3\,.
\label{anisotropy}
\ee
For $\mathcal{N}=2$ these expressions are in agreement with the results of
Ref.~\cite{DiVTan04}.

For Wilson operators with the minimal scaling dimension built from $L$
fundamental field, the one-loop dilatation operator is given in the multi-color
limit by \re{H-multi-color} with the two-particle kernels $\mathbb{H}_{k,k+1}$
defined in \re{H-XXZ}. It coincides with the Hamiltonian of the XXZ Heisenberg
spin$-1/2$ chain of the length $L$ and the anisotropy parameter
$\Delta_\mathcal{N}$ given by \re{anisotropy}. According to \re{spectrum} and
\re{Sch-eq}, the ground state of the magnet corresponds the Wilson operator with
the minimal anomalous dimension. It is well known~\cite{QISM} that in the
thermodynamical limit $L\to \infty$ its properties depend on the value of the
anisotropy parameter $\Delta_\mathcal{N}$:
\begin{itemize}
\item For $\Delta_\mathcal{N}\ge 1$, or equivalently $\mathcal{N}=2$, the ground
state is ferromagnetic and it is separated from the rest of the spectrum by a
mass gap;

\item For $-1 \le \Delta_\mathcal{N} < 1$, or equivalently $\mathcal{N}=1$, the
ground state is antiferromagnetic, and the spectrum is gapless;

\item For $\Delta_\mathcal{N} < -1$, or equivalently $\mathcal{N}=0$, the ground
state is antiferromagnetic, and there is a mass gap.
\end{itemize}

We should mention that appearance of the XXZ Heisenberg magnet in the sector of
Wilson operators with the minimal scaling dimension is not a unique feature of
gauge theories~\cite{Roiban}. The same structure will appear in a field theory
containing complex fields, say $\varphi$ and $\bar\varphi$, provided that the
two-particle transitions are $\varphi\varphi\to\varphi\varphi$,
$\bar\varphi\bar\varphi\to\bar\varphi\bar\varphi$ and $
(\varphi\bar\varphi,\bar\varphi\varphi)\to (\varphi\bar\varphi\,,
\bar\varphi\varphi)$ and the corresponding mixing matrix has real entries.

\section{Conclusions}
\label{Conclusion}

In this paper we employed the light-cone formalism to construct the one-loop
dilatation operator, which governs the scale dependence of Wilson operators of
the maximal Lorentz spin, in all $\mathcal{N}-$extended SYM theories . The
advantage of this formalism is that it provides a unifying superfield description
of SYM theories with different number of supercharges $\mathcal{N}=0,1,2,4$. The
$\mathcal{N} = 4$ SYM theory is formulated in terms of a single chiral superfield
$\Phi(x,\theta^A)$ which describes all propagating modes in the model, while in
SYM theories with less supersymmetry $\mathcal{N}\le 2$ there are two chiral
superfields, $\Phi(x,\theta^A)$ and $\Psi(x,\theta^A)$, each describing half of
the propagating modes.

We demonstrated that the one-loop dilatation operator takes a remarkably simple
form when realized on the space spanned by single-trace products of the
superfields separated by light-like distances. The latter operators serve as
generating functions for Wilson operators of the maximal Lorentz spin and the
scale dependence of the two are in the one-to-one correspondence with each
other. In the maximally supersymmetric, $\mathcal{N}=4$ theory all nonlocal
light-cone operators are built from a single superfield,
Eq.~\re{Chiral-sector}, while for $\mathcal{N}=0,1,2$ one has to distinguish
three different types of such operators, Eqs.~\re{Chiral-sector} --
\re{Mixed-sector}. For the nonlocal light-cone operators, the full
superconformal $SU(2,2|\mathcal{N})$ group is reduced to its ``collinear''
$SL(2|\mathcal{N})$ subgroup. The superconformal invariance allowed us to
determine the one-loop dilatation operator up to some scalar functions. We
deduced their form from previous QCD calculations of anomalous dimensions of
the maximal helicity Wilson operators and confirmed the resulting expressions
by explicit superspace calculation of one-loop kernels entering the evolution
(Callan-Symanzik) equation for nonlocal light-cone operators.

The superspace formalism allowed us to establish an intricate relation between
the one-loop dilatation operators in the SYM theories with $\mathcal{N}\leq 2$
supercharges and the maximally supersymmmetric $\mathcal{N}=4$ theory: all of the
former can be deduced from the latter by merely truncating the number of
fermionic directions in superspace. In the light-cone approach, the
$\mathcal{N}=4$ theory can be reformulated as a SYM theory with $\mathcal{N}\leq
2$ supercharges coupled to additional Wess-Zumino supermultiplets. The above
relation between the dilatation operators is a consequence of vanishing
contribution of the Wess-Zumino supermultiplets to two-particle connected Feynman
diagrams contributing to the one-loop evolution kernels. They do contribute
however to the self-energy diagrams resulting into distinct beta-functions in SYM
theories with different $\mathcal{N}$.

We found that the dilatation operator in the sector of light-cone operators built
only form the $\Phi-$superfields, Eq.~\re{Chiral-sector}, has the same, universal
form in all SYM theories. It can be identified in the multi-color limit as a
Hamiltonian of the $SL(2|\mathcal{N})$ Heisenberg spin chain of length equal to
the number of superfields involved~\cite{BelDerKorMan04}. For $\mathcal{N} = 4$
this implies that, in agreement with the findings of Ref.~\cite{BeiSta03}, the
one-loop dilatation operator is completely integrable. For $\mathcal{N}=0,1,2$
the one-loop dilatation operator possesses the $SL(2|\mathcal{N})$ integrability
in the sector of light-cone operators built from $\Phi-$ and $\Psi-$superfields
only, Eqs.~\re{Chiral-sector} and \re{Chiral-sector-1}, respectively. At the same
time, for ``mixed'' light-cone operators built from both superfields,
Eq.~\re{Mixed-sector}, the dilatation operator receives an additional
contribution from the exchange interaction between the $\Phi-$ and
$\Psi-$superfields which breaks its integrability. Thus, in distinction with the
$\mathcal{N}=4$ theory, the dilatation operator in the SYM theories with
$\mathcal{N}\le 2$ supercharges is integrable only for the light-cone operators
\re{Chiral-sector} and \re{Chiral-sector-1}. To understand the reason for this,
we notice that the mixing matrices for Wilson operators at $\mathcal{N}=4$ and
$\mathcal{N}\le 2$ are related to each other through the truncation procedure:
the mixing matrix for $\mathcal{N}=0$ is a minor of the same matrix for
$\mathcal{N}=1$ which in its turn is a minor of the $\mathcal{N}=2$ matrix and so
on. Going over from $\mathcal{N}=4$ down to $\mathcal{N}=0$ one replaces some of
its entries by $0$ and, therefore, breaks integrability of the whole matrix.
Still, integrability survives in its blocks corresponding to the Wilson operator
generated by nonlocal light-cone operators \re{Chiral-sector} and
\re{Chiral-sector-1}.

\subsection*{Acknowledgements}

We would like to thank V. Braun, A. Gorsky and A. Vainshtein for illuminating
discussions and V. Korepin for very helpful correspondence. This work was
supported in part by the grant 00-01-005-00 of the Russian Foundation for
Fundamental Research (A.M. and S.D.), by the Sofya Kovalevskaya programme of
Alexander von Humboldt Foundation (A.M.).



\appendix

\setcounter{section}{0} \setcounter{equation}{0}
\renewcommand{\theequation}{\Alph{section}.\arabic{equation}}

\section{Some useful formulae}

In this Appendix we specify the conventions used throughout the paper.

\subsection*{A1.\quad Light-cone coordinates}

For our purposes, it is convenient to split four-dimensional vectors
$x_\mu=(x_0,x_1,x_2,x_3)$ into longitudinal light-cone components $x_\pm =(x_0\pm
x_3)/\sqrt{2}$ and (anti)holomorphic transverse components
$x=(x_1+ix_2)/\sqrt{2}$ and $\bar x=x^*$. In these notations, the scalar product
looks as
\begin{equation}
x^\mu y_\mu = x_+ y_- + x_- y_+ - x \bar y - \bar x y\,.
\end{equation}
We also define the derivatives
\begin{equation}
\label{Derivatives}
\partial_+
\equiv \frac{\partial}{\partial
x_-}=\ft{1}{\sqrt{2}}(\partial_{x_0}-\partial_{x_3}) \,,\qquad
\partial
\equiv \frac{\partial}{\partial \bar{x}} = \ft{1}{\sqrt{2}}(\partial_{x_1} +
i\partial_{x_2}) \, , \qquad
\bar\partial = (\partial)^*
\, ,
\end{equation}
so that $\partial_+ x_-=\partial \bar x =  \bar \partial x =1$ and $\partial_+
x_+ = \partial x =  \bar \partial \bar x =0$. The action of the SYM theory on the
light-cone, Eqs.~\re{N=0-M} -- \re{N=4-M}, involves a nonlocal operator
$\partial_+^{-1}$. It is defined in the momentum representation using the
Mandelstam-Leibbrandt prescription \cite{Man83}
\begin{equation}
\partial_+^{- 1} f (x)
= i \int \frac{d^4 k}{(2 \pi)^4} \frac{{\rm e}^{- i k \cdot
x}}{[k_+]_{\scriptscriptstyle\rm ML}} \widetilde f (k) \, ,
\label{ML-pre}
\end{equation}
with the causal prescription for the pole in the momentum space
\begin{equation}
\label{ML-def}
\frac{1}{[k_+]_{\scriptscriptstyle\rm ML}} \equiv \frac{1}{k_+ + i 0 \cdot
k_-}=\frac{k_-}{k_+k_- + i 0} \, .
\end{equation}

\subsection*{A2.\quad Light-cone spinors}

The four-component Majorana spinors (both the gaugino fields and the odd
generators of the superconformal group) are composed from two Weyl spinors
\begin{equation}
\psi= \left(\begin{array}{c} \lambda_\alpha\\\bar\lambda^{\dot\alpha}\end{array}
\right) \, ,\qquad\bar\psi \equiv \psi^\dagger \gamma^0 =
(\lambda^{\alpha}\,,\bar\lambda_{\dot\alpha})\,,
\label{psi-Weyl}
\end{equation}
where the Weyl indices are lowered/raised according to the rules
\begin{equation}
{\lambda}^\alpha = \varepsilon^{\alpha \beta} {\lambda}_\beta \, ,
\qquad
\bar{\lambda}_{\dot\alpha}
= \varepsilon_{\dot\alpha \dot\beta} \bar{\lambda}^{\dot\beta} \, , \qquad
{\lambda}_\alpha = {\lambda}^\beta \varepsilon_{\beta\alpha} \, ,
\qquad
\bar{\lambda}^{\dot\alpha}
=
\bar{\lambda}_{\dot\beta} \varepsilon^{\dot\beta \dot\alpha}
\, ,
\end{equation}
with the Levi-Civita tensor normalized as $ \varepsilon^{12} = \varepsilon_{12} =
- \varepsilon^{\dot{1} \dot{2}} = - \varepsilon_{\dot{1} \dot{2}} = 1$. Complex
conjugation acts on the covariant Weyl spinors as
\begin{equation}
\label{CovWeylconj}
(\bar{\lambda}^{\dot{\alpha}})^\ast = \lambda^\alpha \, , \qquad
(\lambda_\alpha)^\ast = \bar{\lambda}_{\dot{\alpha}} \, ,
\end{equation}
and the product of two spinors obeys
\begin{equation}
\left( \lambda_{1 \alpha} \lambda_2^\alpha \right)^\ast = \left( \lambda_2^\alpha
\right)^\ast \left( \lambda_{1 \alpha} \right)^\ast =
\bar{\lambda}_2^{\dot\alpha}\, \bar{\lambda}_{1 \dot\alpha}
\, .
\end{equation}
In the Weyl basis \re{psi-Weyl}, the Dirac matrices admit the representation
\begin{equation}
\label{4Dmatrices}
\gamma^\mu = \left(
\begin{array}{cc}
0 & [\bar\sigma^\mu{}]_{\alpha \dot\beta}
\\{}
[\sigma^\mu{}]^{\; \dot\alpha \beta} & 0
\end{array}
\right) \, , \qquad \gamma^5 = i \gamma^0 \gamma^1 \gamma^2 \gamma^3 = \left(
\begin{array}{rr}
1 & 0
\\
0 & - 1
\end{array}
\right) ,
\end{equation}
where $\sigma^\mu = (1 , \bit{\sigma})$ and $\bar\sigma^\mu = (1 , -
\bit{\sigma})$ involve the conventional vector of Pauli $\bit{\sigma}-$matrices.

In the light-cone formalism, one splits Majorana spinors into the ``good'' and
``bad'' components using \re{good-bad}. In the Weyl representation \re{psi-Weyl},
the former is given by
\begin{equation}
\label{LCprojection}
{\lambda}_{+\alpha}
=
\ft12 \bar\sigma^-{}_{\alpha\dot\beta} \, \sigma^{+ \; \dot\beta\gamma}
\lambda_\gamma
=
\left(\begin{array}{c} \lambda_1\\ 0\end{array} \right)
\, , \qquad
\bar{\lambda}_+^{\dot\alpha}
=
\ft12 \sigma^{- \; \dot\alpha\beta} \, \bar\sigma^+{}_{\beta\dot\gamma}
\bar\lambda^{\dot\gamma}
=
\left(\begin{array}{c} 0 \\ -\bar\lambda_{\dot 1}
\end{array} \right) \,,
\end{equation}
with $\bar\lambda_{\dot 1}=-\bar\lambda^{\dot 2}$. The remaining components of
the Weyl spinors ${\lambda}_{\alpha}$ and $\bar{\lambda}^{\dot\alpha}$ define
the ``bad'' spinors. Thus, the ``good'' and ``bad'' spinors,
${\lambda}_{\pm\alpha}$ and $\bar{\lambda}_\pm^{\dot\alpha}$, can be described
by a single complex Grassmann number without specifying Lorentz indices. To
simplify formulae for the components of superfields $\Phi$ and $\Psi$ building
up the light-cone actions in Sect.~2.1 and 2.2, it is convenient to rescale the
covariant components of the gaugino field as
\begin{equation}
\lambda_1^A \equiv \sqrt[4]{2} \, \lambda^A \, , \qquad
\bar\lambda_{{\dot 1} A} \equiv - i \sqrt[4]{2} \, \bar\lambda_A
\, .
\label{lambda-i}
\end{equation}
With such a convention the infinitesimal supersymmetric variations of the
fields $X=(A,\lambda^A,  \phi^{AB})$
\begin{equation}
\delta_{\rm Q} X = [\xi^A \mathbf{Q}_A , X] + [\bar\xi_A \bar{\mathbf{Q}}^A ,
X] \, ,
\end{equation}
with rescaled generators
\be
\mathbf{Q}_{1 A} \equiv -i\sqrt[4]{8} \, \mathbf{Q}{}_A \, , \qquad
\bar{\mathbf{Q}}{}_{\dot 1}^A \equiv -  \sqrt[4]{8} \, \bar{\mathbf{Q}}{}^A \, ,
\ee
and Grassmann transformation parameters, $\xi^{1 A} = - \xi^A/\sqrt[4]{8}$ and
$\bar \xi^{\dot 1}_A = i \bar\xi_A/\sqrt[4]{8}$,
result in a relatively simple formulae, say for $\mathcal{N} = 4$ SYM (see Ref.\
\cite{BelDerKorMan03}),
\ba
\nonumber
& & \delta_{\rm Q} A =  \xi^A \bar\lambda_A \, , \\
& & \delta_{\rm Q} \lambda^A =  - \left(
\partial^+ \bar{A} \right) \xi^A - i \left( \partial^+ \phi^{AB} \right)
\bar\xi_B \, , \label{N4supervariation} \\
\nonumber & & \delta_{\rm Q} \phi^{AB} = i \left( \xi^A \lambda^B - \xi^B
\lambda^A - \varepsilon^{ABCD} \bar\xi_C \bar\lambda_D \right) \, .
\ea
Finally, one introduces rescaled fermionic parameters in light-cone superspace
in terms of components of covariant Weyl coordinates,
\begin{equation}
\label{RedefLCweyl}
\theta^A \equiv \sqrt[4]{8} \, \theta^{1 A} \, , \qquad
\bar\theta_A \equiv i \sqrt[4]{8} \, \bar\theta^{\dot 1}_A
\, ,
\end{equation}
so that the realization of superconformal generators in superspace has a concise
form, Eq.\ (\ref{sl2}). Due to the presence of additional factors in the right-hand side\
of \re{lambda-i} and \re{RedefLCweyl}, complex conjugation acts on the odd
variables $\chi=(\lambda^A, \theta^A)$ and $\bar\chi=(\bar\lambda_A,
\bar\theta_A)$ as
\begin{equation}
\chi^\ast = - i \bar\chi \, , \qquad
\bar\chi^\ast = - i \chi
\label{cc-rule1}
\, ,
\end{equation}
while for their product one has
\be
(\chi_1 \chi_2)^\ast = \chi_2^\ast \chi_1^\ast = - \bar\chi_2
\bar\chi_1 =
\bar\chi_1 \bar\chi_2
\, , \qquad (\bar\chi_1 \chi_2)^\ast = \chi_2^\ast \bar\chi_1^\ast = -
\bar\chi_2 \chi_1 = \chi_1 \bar\chi_2 \, .
\label{cc-rule2}
\ee

\subsection*{A3.\quad Grassmann integration}

The integration measure over Grassmann variables is normalized as
\begin{equation}
\label{NormalizationSuperspace}
\int d^{\cal N} \theta\, \theta^1 \ldots \theta^{\cal N} = \int d^{\cal N}
\bar\theta \, \bar\theta_1 \ldots \bar\theta_{\cal N} = 1 \, .
\end{equation}
Performing calculation of Feynman diagrams in the momentum representation, we
apply the Fourier transformation to the superfield, Eq.~\re{Super-Fourier}. The
inverse Fourier transformation is defined as
\begin{equation}
\widetilde {\Phi} (p, \pi_A) = \int d^D x \int d^\mathcal{N} \theta\
\textrm{e}^{- i \, p \cdot x \, - \pi_A \theta^A} {\Phi} (x, \theta^A) \, ,
\label{Inverse-Fourier}
\end{equation}
where $\pi_A$ is the momentum conjugated to the odd coordinates $\theta^A$. To
establish the normalization of the integration measure over Grassmann valued
momenta $\pi_A$, one computes sequentially the Fourier transform and its inverse,
i.e., ${\Phi} (\theta^A) = \int d^{\cal N} \pi \int d^{\cal N} \theta_1 \,
{\Phi} (\theta^A_1) \, \exp \pi \cdot (\theta - \theta_1)$. Taking into
account \re{NormalizationSuperspace}, one finds that
\begin{equation}
\int d^{\mathcal{N}} \pi \, \pi_1 \ldots \pi_{\mathcal{N}} =
(-1)^{\mathcal{N}(\mathcal{N} - 1)/2} \, .
\end{equation}
For the odd variables the delta-functions in the coordinate and momentum space
are defined as
\begin{eqnarray}
\delta^{(\mathcal{N})}(\pi) &\equiv& \int d^\mathcal{N} \theta \ \textrm{e}^{-
\pi_A \theta^A} = \pi_\mathcal{N} \ldots \pi_1\,, \nonumber
\\
\delta^{(\mathcal{N})}(\theta) &\equiv& \int d^\mathcal{N} \pi
\,\textrm{e}^{\pi_A \theta^A} = \theta^1 \dots \theta^{\cal N}\,.
\label{delta-function}
\end{eqnarray}
They satisfy \re{delta-pi} together with $\int d^\mathcal{N}
\theta\,\delta^{(\mathcal{N})}(\theta-\theta_1)
{\Phi}(x,\theta^A)={\Phi}(x,\theta_1^A)$.

Going over to the momentum representation in \re{Q-op} we find that the products
of differential operators in the right-hand side of \re{Q-op} are replaced by
products of momenta $p$ and $\pi_A$ which we denote as
\begin{eqnarray}
&& (p_1,p_2) = p_1 p_{2+} - p_2 p_{1+} \, , \qquad  (p_1,p_2)^* = \bar p_1 p_{2+}
- \bar p_2 p_{1+} \, ,
\nonumber \\
&& [p_1,p_2] = \prod_{A=1}^{\mathcal{N}} \left( \pi_{1,A}p_{2+} - \pi_{2,A}
p_{1+} \right) \equiv \left( \pi_{1, 1} p_2^+ - \pi_{2, 1} p_1^+ \right) \dots
\left( \pi_{1, {\cal N}} p_2^+ - \pi_{2, {\cal N}} p_1^+ \right),
\label{brackets}
\end{eqnarray}
where the ordering of fermionic momenta is such that the factors with larger
$\scriptstyle A$ appear to the right. These brackets have the following
properties
\begin{eqnarray}
&& (k + x p_1, p_1) = (k, p_1) = - (p_1,k) \, , \nonumber
\\[2mm]
&& [k+x p_1,p_1] = [k,p_1]  = (-1)^\mathcal{N} [p_1,k]
\end{eqnarray}
with $x$ arbitrary c-number. The square bracket can be expressed in terms of
momentum delta-function, Eq.~\re{delta-function},
\begin{eqnarray}
[p_1,p_2] &=& (-1)^{\mathcal{N}(\mathcal{N}-1)/2}  (p_{2+})^{{\cal N}}
\delta^{(\mathcal{N})} \left( \pi_1 - \pi_2 \frac{p_{1+}}{p_{2+}} \right)
\nonumber
\\
&=& (-1)^{\mathcal{N}(\mathcal{N}+1)/2}   (p_{1+})^\mathcal{N}
\delta^{(\mathcal{N})} \left( \pi_2 - \pi_1 \frac{p_{2+}}{p_{1+}} \right) \, ,
\label{pp-bracket}
\end{eqnarray}
where the delta-function in momentum $\pi-$space is defined in
\re{delta-function}.

\subsection*{A4.\quad Mandelstam's approach}

To establish the relation between the expressions for the light-cone SYM actions,
Eq.~\re{N=0-M} and \re{N=4-M}, with those proposed by Mandelstam in
Ref.~\cite{Man83}, one introduces the operator
\be
\mathrm{D}_A = \partial_{\theta^A} - \theta^A \partial_+\,, \qquad
\{\mathrm{D}_A,\mathrm{D}_B \}= -2\partial_+ \delta_{AB}\,.
\ee
In distinction with $D_{{\scriptscriptstyle \rm M}, A}$, Eq.~\re{derivatives-M},
this operator is not covariant under the $SU(\mathcal{N})$ rotations of
$\theta-$coordinates, generated by the charges $T_B{}^A$, Eq.~\re{sl2}. It is
straightforward to verify that for arbitrary light-cone scalar superfield
$\Phi(x_\mu,\theta^A)$ the following relation holds true (no summation over
$\scriptstyle A$!)
\be
\mathrm{D}_A\big( \mathrm{D}_A \Phi_1  \mathrm{D}_A \Phi_2 \big) =
-\partial_+\Phi_1\,\partial_{\theta^A}\Phi_2 +\partial_{\theta^A}\Phi_1\,
\partial_+\Phi_2 \,.
\ee
Then, comparison with \re{Q-op} yields
\be
 [\Phi_1 , \Phi_2] = (-1)^{\mathcal{N}(\mathcal{N} - 1)/2}\ \mathrm{D}\, \big( \mathrm{D}
\Phi_1 \, \mathrm{D} \Phi_2 \big)\,,
\label{M-bracket}
\ee
where $\mathrm{D}\equiv\mathrm{D}_1\ldots\mathrm{D}_\mathcal{N}$. The light-cone
action \re{N=0-M} involves two superfields $\Psi(x_\mu,\theta^A)$ and
$\Phi(x_\mu,\theta^A)$. Let us use the following ansatz for the former field
\be
\Psi(x,\theta^A) =
\partial_+^{2-\mathcal{N}}\mathrm{D}\, \bar\Phi(x,\theta^A)\,,
\label{M-bar-Phi}
\ee
or equivalently
$
\bar\Phi(x,\theta^A) = (-1)^{\mathcal{N}(\mathcal{N} + 1)/2} \partial_+^{-2}
\mathrm{D} \, \Psi (x,\theta^A)$. Substituting \re{M-bar-Phi} into \re{N=0-M} and
taking into account \re{M-bracket}, one obtains the light-cone action of the SYM
theory in terms of the superfields $\Phi(x,\theta^A)$ and $\bar\Phi(x,\theta^A)$
which coincides with the light-cone actions proposed in
Refs.~\cite{Man83,Smith85}.

\section{Projectors}

To define the one-loop dilatation operator we introduced the $SL(2|\mathcal{N})$
invariant projection operators, Eqs.~\re{condition-phi} and
\re{pro-mixed-sector}. As in Sect.~3.2.2, we shall restrict ourselves to the
$SL(2)$ case and make use of the $SL(2|\mathcal{N})$ invariance to ``lift'' the
resulting expressions from the light-cone to the superspace, Eq.~\re{lift1}.

To start with one considers a nonlocal light-cone operator
$\mathbb{O}_{j_1j_2}(z_1,z_2)$. It belongs to the tensor product of two $SL(2)$
moduli labelled by the conformal spins $j_1$ and $j_2$ which can be decomposed
over the irreducible components as $ \mathcal{V}_{j_1}^{\scriptscriptstyle \rm
SL(2)} \otimes \mathcal{V}^{\scriptscriptstyle \rm SL(2)}_{j_2} =
\bigoplus_{j_{12}} \mathcal{V}_{j_{12}}^{\scriptscriptstyle \rm SL(2)}$ with the
conformal spin $j_{12}=j_1+j_2+n$ for $n=0,1,\ldots$. Let us introduce the
operator $\Pi_m^{(j_1,j_2)}$ that projects $\mathbb{O}_{j_1j_2}(z_1,z_2)$ onto
the $SL(2)$ moduli $\mathcal{V}_{j_{12}}^{\scriptscriptstyle \rm SL(2)}$ with
$j_{12}=j_1+j_2+m$.

By the definition, $\Pi_m^{(j_1,j_2)}$ is the $SL(2)$ invariant operator
satisfying $\Pi_m^{(j_1,j_2)}\Pi_n^{(j_1,j_2)}=\delta_{mn}\Pi_m^{(j_1,j_2)}$. The
$SL(2)$ invariance fixes the form of integral operator $\Pi_m^{(j_1,j_2)}$ up to
some scalar function $f(\xi)$, Eq.~\re{general-SL2}. The $SL(2)$ integrals in
\re{general-SL2} can be simplified with a help of the identity~\cite{DKM02}
\be
\int [\mathcal{D}w]_j (z_1-\bar w)^{-2j+x} (z_2-\bar w)^{-x} {\Phi}(w) =
\frac{\Gamma(2j)\e^{-i\pi j}}{\Gamma(2j-x)\Gamma(x)}\int_0^1d\alpha\,
\bar\alpha^{2j-x-1} \alpha^{x-1}\,{\Phi}(\bar\alpha z_1+\alpha z_2)
\label{SL2-identity}
\ee
where $\bar\alpha=1-\alpha$. To this end, one uses the integral representation
$f(\xi) = \int_C \frac{dx}{2\pi i}\, \xi^{-x} \tilde f(x)$ interchanges the
order of integration in \re{general-SL2} and applies \re{SL2-identity}
consequently. In this way, one obtains from \re{general-SL2}
\be
{\Pi}_m^{(j_1, j_2)} \mathbb{O}_{j_1j_2}(z_1,z_2) = \int_0^1 d\alpha\int_0^1
d\beta\,
\bar\alpha^{2j_1-2}\bar\beta^{2j_2-2}
\varphi_m\left(\frac{\alpha\beta}{\bar\alpha\bar\beta}\right)\mathbb{O}_{j_1j_2}(\bar\alpha
z_1+\alpha z_2,\bar\beta z_2+\beta z_1)\,,
\label{Pi-phi}
\ee
where $\bar\alpha=1-\alpha$, $\bar\beta=1-\beta$ and the function $\varphi_m$ is
related to the function of anharmonic ratio entering \re{general-SL2} as
\be
f(\xi)=\int_0^1 d\alpha\int_0^1 d\beta\,
\bar\alpha^{2j_1-2}\bar\beta^{2j_2-2}(\bar\alpha+\xi\alpha)^{-2j_1}
\varphi_m\left(\frac{\alpha\beta}{\bar\alpha\bar\beta}\right)\,.
\label{f-phi}
\ee
Important difference between the functions $f(\xi)$ and $\varphi_m$ is that the
latter is a distribution.

To determine the function $\varphi_m$ entering \re{Pi-phi} it is sufficient to
examine the action of $\Pi_m^{(j_1,j_2)}$ on the lowest weight in the $SL(2)$
module $\mathcal{V}_{j_{12}}^{\scriptscriptstyle \rm SL(2)}$ which are given by
$(z_1-z_2)^n\equiv z_{12}^n$. Then, replacing $\mathbb{O} (z_1 , z_2)$ in
\re{Pi-phi} by ${O}^{(n)} (z_1 , z_2) =z_{12}^n$ and taking into account that
${\Pi}_m^{(j_1, j_2)} {O}^{(n)}=\delta_{nm}{O}^{(n)}$ one obtains
\begin{equation}
\label{ActionOfProjector}
[ {\Pi}_m^{(j_1, j_2)} {O}^{(n)} ] (z_1 , z_2) = z_{12}^n \int_0^1 d \alpha
\int_0^1 d \beta \, \bar\alpha^{2 j_1 - 2} \bar\beta^{2 j_2 - 2} \,
{\varphi}_m(\zeta) (1 - \alpha - \beta)^n =\delta_{mn} z_{12}^n \, .
\end{equation}
where $ \zeta \equiv {\alpha \beta}/{(\bar\alpha \bar\beta)}\,.$ Solving this
relation for $m=0$ and $m=1$ one gets
\begin{eqnarray}
\label{Projectors}
{\varphi}_0(\zeta) \!\!\!&=&\!\!\! \frac{ {\Gamma} (2 j_1 + 2 j_2) }{
{\Gamma} (2 j_1) {\Gamma} (2 j_2) } \delta (\zeta - 1)
\, , \\
{\varphi}_1(\zeta) \!\!\!&=&\!\!\! \frac{ {\Gamma} (2 j_1 + 2 j_2 + 2) }{
{\Gamma} (2 j_1 + 1) {\Gamma} (2 j_2 + 1) } \frac{d}{d \zeta} \delta
(\zeta - 1) - (2j_1 + 2j_2 + 1) \frac{ {\Gamma} (2 j_1 + 2 j_2) }{
{\Gamma} (2 j_1) {\Gamma} (2 j_2) } \delta (\zeta - 1) \, . \nonumber
\end{eqnarray}
Let us define the following operator
\begin{equation}
\widetilde{\mathbb{O}}_{j_1j_2}(z_1,z_2) = (1- {\Pi}_0^{(j_1, j_2)}-
{\Pi}_1^{(j_1, j_2)})\mathbb{O}_{j_1j_2}(z_1,z_2)\,.
\label{sum-projectors}
\end{equation}
By the construction, it receives contribution from the $SL(2)$ moduli
$\mathcal{V}_{j_{12}}^{\scriptscriptstyle \rm SL(2)}$ with the conformal spins
$j_{12}=j_1+j_2+m$ and $m\ge 2$. A distinguished property of the states belonging
to these moduli is that for $z_{12}\to 0$ they have the same asymptotic behavior
as the lowest weight $(z_1-z_2)^m$ leading to
\begin{equation}
\widetilde{\mathbb{O}}_{j_1j_2}(z_1,z_2) \sim (z_1-z_2)^2
\label{vanish}
\end{equation}
for $z_1-z_2\to 0$. Let us substitute  \re{Projectors} into \re{Pi-phi} and
examine the explicit expressions for the projectors ${\Pi}_{m=0,1}^{(j_1, j_2)}$
for special values of the conformal spins:

\medskip

\noindent $\bullet$ ${(j_1=j_2=-1/2)}$: To define the action of projectors
(\ref{Projectors}) on an arbitrary function $\mathbb{O}(z_1,z_2)$ one regularizes
the integrand in \re{Pi-phi} by setting $j_1=j_2=-1/2 + \varepsilon$ and takes
the limit $\varepsilon\to 0$. Then
\begin{eqnarray}
[ {\Pi}_0^{(-1/2, -1/2)} \mathbb{O} ] (z_1,z_2) \!\!\!&=&\!\!\! \lim_{\varepsilon
\to 0} \frac{{\Gamma} (4 \varepsilon - 2)}{{\Gamma}^2 (2 \varepsilon -
1)} \int_0^1 d \alpha \, \left( \alpha \bar\alpha \right)^{2 \varepsilon - 2}
\mathbb{O} \left(\alpha z_1 + \bar\alpha z_2 , \alpha z_1 +
\bar\alpha z_2 \right) \nonumber
\\
&=&\!\!\! \frac{1}{2} \left( 1 - \frac{1}{2} z_{12} \partial_{z_1} \right)
\mathbb{O}( z_1 , z_1 ) + \frac{1}{2} \left( 1 - \frac{1}{2} z_{21}
\partial_{z_2} \right) \mathbb{O} ( z_2 , z_2 ) \, .
\end{eqnarray}
In the similar manner, one finds for the projector ${\Pi}_1^{(-1/2, -1/2)}$
\begin{eqnarray}
[ {\Pi}_1^{(-1/2, -1/2)} \mathbb{O}] (z_1 , z_2) = \!\!\!&-&\!\!\! \frac{1}{4}
z_{12} \left. \partial_{z}   \mathbb{O} ( z_2 , z )\right|_{z = z_2} -
\frac{1}{4} z_{21} \left. \partial_{z}\mathbb{O}( z , z_2 ) \right|_{z = z_2}
\\
\!\!\!&-&\!\!\! \frac{1}{4} z_{12} \left. \partial_{z}  \mathbb{O}( z_1 , z )
\right|_{z = z_1} - \frac{1}{4} z_{21} \left. \partial_{z} \mathbb{O} ( z , z_1
)\right|_{z = z_1} \, . \nonumber
\end{eqnarray}
Combining together the last two expressions we define the $SL(2)$ projector
\begin{equation}
{\Pi}_{{\Phi}{\Phi}} \equiv {\Pi}_0^{(-1/2, -1/2)} + {\Pi}_1^{(-1/2,
-1/2)} \, ,
\end{equation}
which enters into \re{sum-projectors} and leads to \re{vanish}. Going over from
the light-cone to the superspace we apply \re{lift1} and replace simultaneously
the light-cone derivatives by derivatives in the superspace $z_{12}\partial_{z}
\to Z_{12}\cdot \partial_{Z}$. In this way, we arrive at the expression for the
$SL(2|\mathcal{N})$ invariant projector ${\Pi}_{{\Phi}{\Phi}}$ given in
\re{Pi12-main}.

\medskip

\noindent $\bullet$ ($j_1=-1/2$, $j_2\ge 1/2$): To be specific we first choose
$j_2 = 3/2$ which corresponds to the conformal spin of the
${\Psi}-$(super)field in the $\mathcal{N}=0$ theory. As before, we
regularize the integrand in \re{Pi-phi} by setting $j_1=-1/2+\varepsilon$ for
$\varepsilon\to 0$. Making use of \re{Projectors}, one evaluates the projectors
\begin{eqnarray}
{} [ {\Pi}_0^{(-1/2, 3/2)} \mathbb{O} ] (z_1 , z_2) &=& \left( 1 - \frac{1}{2}
z_{12}
\partial_{z_2} \right) \mathbb{O}(z_2 , z_2) \, ,
\\
{} [ {\Pi}_1^{(-1/2, 3/2)} \mathbb{O}] (z_1 , z_2) &=& \frac{1}{2} z_{12} \left.
\partial_{z} \mathbb{O}(z_2 , z )\right|_{z = z_2} + \frac{3}{2} z_{12} \left.
\partial_{z} \mathbb{O}(z , z_2 )\right|_{z = z_2} \, . \nonumber
\end{eqnarray}
Their sum ${\Pi}_{{\Phi}{\Psi}}  = {\Pi}_0^{(-1/2,
3/2)}+{\Pi}_1^{(-1/2, 3/2)}$ is given by
\begin{equation}
[ {\Pi}_{{\Phi}{\Psi}}  \mathbb{O}] (z_1 , z_2) = \mathbb{O}(z_2 , z_2)
+ z_{12} \left. \partial_{z}  \mathbb{O} (z , z_2 )\right|_{z = z_2} \, .
\label{spec-proj}
\end{equation}
Repeating the calculation for other (half)integer positive $j_2$ one can verify
that the projector ${\Pi}_{{\Phi}{\Psi}}$ does not depend on $j_2$ and
is given by \re{spec-proj}. Going over from the light-cone to the superspace one
arrives at \re{projector-phi-psi}.

\section{Feynman diagram technique}

To develop the Feynman diagram technique, we introduce a generating functional in
the SYM theory (for $\mathcal{N}\le 2$)
\begin{equation}
\label{Gen-func}
{\rm e}^{W [J , \bar J]} = \int {\cal D} {\Phi} \, {\cal D} {\Psi}\,
{\rm e}^{ i S_{\cal N} + i \int d^4 x \int d^{\cal N} \theta \left( J^a (x,
\theta^A) {\Phi}^a (x, \theta^A) +
\bar J^a (x, \theta^A) {\Psi}^a (x, \theta^A)
\right) } \, .
\end{equation}
For $\mathcal{N}=4$, the theory is formulated in terms of a single superfield, so
that there is no integration over ${\Psi}$ and corresponding sources
$\bar{J}$ are absent. Connected correlation functions can be calculated from $W
[J,\bar J]$ as
\begin{equation}
\label{CorrFunctions}
\frac{\delta}{i \delta J^{a_1} (x_1 , \theta^A_1)} \frac{\delta}{i \delta
\bar{J}^{a_2} (x_2 , \theta^A_2)} \dots \frac{\delta}{i \delta J^{a_N} (x_N ,
\theta^A_N)} W [J, \bar J] = \langle {\Phi}^{a_1} (x_1 , \theta^A_1)
{\Psi}^{a_2} (x_2 , \theta^A_2) \dots {\Phi}^{a_N} (x_N , \theta^A_N)
\rangle \, ,
\end{equation}
where the functional derivatives are defined as
\begin{equation}
\frac{ \delta J^{a} (x , \theta^A) }{ \delta J^{b} (x' , {\theta'}^A) } =
\delta^{ab} \delta^{(4)} (x - x') \delta^{({\cal N})} (\theta - \theta') \, .
\end{equation}
It is proves convenient to perform calculation of the Feynman diagrams in the
momentum representation. To this end, we apply the Fourier transformation in the
superspace \re{Super-Fourier} and switch from the superfields ${\Phi}^a
(x_\mu,\theta^A)$ and ${\Psi}^a(x_\mu,\theta^A)$ to the conjugated fields
$\widetilde {\Phi}^a (p_\mu, \pi_A)$ and $\widetilde {\Psi}^a (p_\mu,
\pi_A)$, respectively.

In order to distinguish the lines corresponding to two species of chiral
superfields, i.e., $\widetilde {\Phi}$ and $\widetilde {\Psi}$, we will
denote them as incoming and outgoing lines, respectively, and indicate the
momentum flow by a small arrow
\begin{equation}
\widetilde {\Phi} (p,\pi) \ = \ \
\parbox[t][6mm][b]{17mm}{\insertfig{1.5}{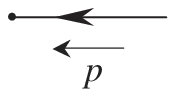}}\, , \qquad
\widetilde {\Psi} (p,\pi) \ = \ \
\parbox[t][6mm][b]{17mm}{\insertfig{1.5}{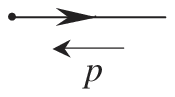}}
\, .
\end{equation}
In what follows we shall use the convention that the external momenta flow
\textsl{into} vertices in the Feynman diagrams.

Since the Lagrangian of the SYM theory in the Mandelstam formalism is invariant
under translations in $x_\mu$ and $\theta^A$, the correlation functions in the
momentum representations are proportional to the product of delta-functions in
even and odd momenta. In particular, the free propagator of the superfield can be
easily found from the generating functional \re{Gen-func} as
\begin{equation}
\langle \widetilde {\Phi}^a (p_1, \pi_1) \widetilde {\Psi}^b (p_2, \pi_2) \rangle
= \sigma_\mathcal{N}\frac{i \delta^{ab}}{p_1^2 + i 0} (2 \pi)^4 \delta^{(4)} (p_1
+ p_2) \delta^{({\cal N})} (\pi_1 + \pi_2) \, ,
\label{propagator}
\end{equation}
where the notation was introduced for the signature factor
\begin{equation}
\sigma_\mathcal{N}=(- 1)^{{\cal N}({\cal N} + 1)/2}\,,
\label{sigma}
\end{equation}
so that $\sigma_0=1$ and $\sigma_1=\sigma_2=-1$. The interaction vertices
${\Gamma}$ are identified as amputated Green functions, Eq.\
(\ref{CorrFunctions}), transformed into the momentum space,
\begin{eqnarray}
&& \langle \widetilde {\Phi}^{a_1} (p_1,\pi_1) \widetilde {\Psi}^{a_2}
(p_2,\pi_2) \dots \widetilde {\Phi}^{a_L} (p_L,\pi_L) \rangle \equiv (2
\pi)^4 \delta^{(4)} \bigg( \sum_{k = 1}^L p_k \bigg) \delta^{({\cal N})} \bigg(
\sum_{k = 1}^L \pi_k \bigg)
\\[-3mm]
&& \qquad\qquad\qquad\qquad \times \bigg( \prod_{j=1}^L
\frac{i\sigma_\mathcal{N}}{p_j^2} \bigg) {\Gamma}^{a_1 a_2\ldots
a_L}(p_1,\pi_1; p_2,\pi_2;\ldots; p_L,\pi_L) \, . \nonumber
\end{eqnarray}

\noindent \underline{\bf\large Feynman rules for ${\cal N} = 0, 1, 2$}:

\vspace*{-5mm}
\begin{eqnarray*}
\parbox[c][25mm]{30mm}{
\insertfig{2.5}{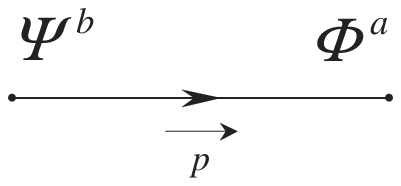} } \!\!\!\!\!\!\!&=& \sigma_\mathcal{N}\frac{i
\delta^{ab}}{p^2 + i0}
\\[-6mm]
\parbox[c][25mm]{30mm}{
\insertfig{2.5}{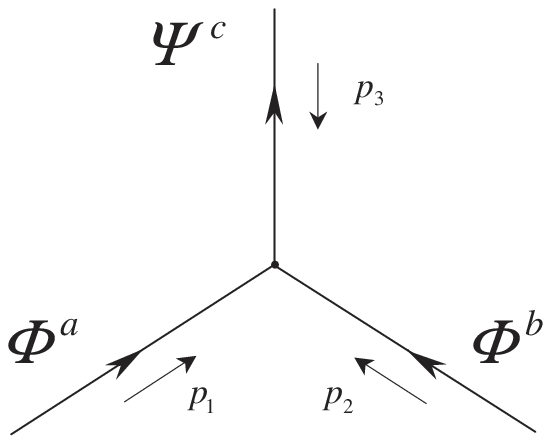} } \!\!\!\!\!\!\!&=& - 2i g \sigma_\mathcal{N} f^{abc}
(p_1, p_2)^\ast
\\
\parbox[c][25mm]{30mm}{\insertfig{2.5}{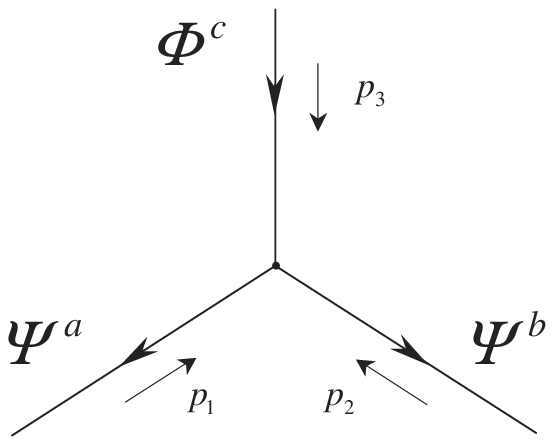} }
\!\!\!\!\!\!\!&=& - 2 i g \sigma_\mathcal{N} f^{abc} \frac{(p_1,
p_2)[p_1,p_2]}{(p_{1+} p_{2+})^2} {(p_{3+})^{2-\mathcal{N}}}
\\
\parbox[c][25mm]{30mm}{\insertfig{2.5}{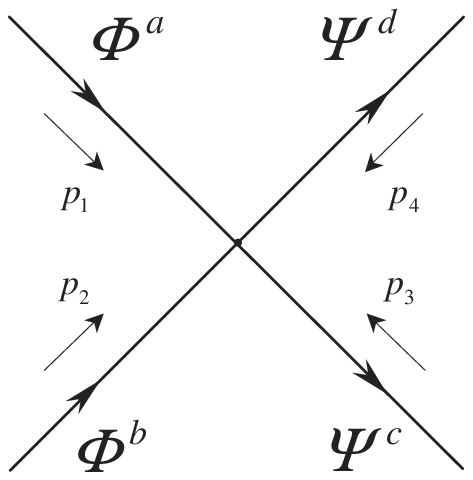} }
\!\!\!\!\!\!\!&=& - 2 i g^2 \sigma_\mathcal{N}\Bigg\{ f^{ace} f^{bde} \left(
\frac{[p_4, p_2] p_{1+} (p_{2+})^{2-\mathcal{N}}}{p_{4+} (p_{1} + p_{3})_+^2} +
\frac{ [p_1, p_3] (p_{1+})^{2-\mathcal{N}}p_{2+}}{p_{3+} (p_{2} + p_{4})_+^2}
\right)
\nonumber\\[-6mm]
&&\!\!\!\!\!\!\!\quad + f^{ade} f^{bce} \left( \frac{[p_2, p_3]p_{1+}
(p_{2+})^{2-\mathcal{N}}}{p_{3+} (p_{1} + p_{4})_+^2} + \frac{[p_4,
p_1](p_{1+})^{2-\mathcal{N}}p_{2+} }{p_{4+} (p_{2} + p_{3})_+^2} \right) \Bigg\}
\end{eqnarray*}

\noindent \underline{\bf\large Feynman rules for ${\cal N} = 4$:}

\vspace*{-5mm}
\begin{eqnarray*}
\parbox[c][25mm]{30mm}{
\insertfig{2.5}{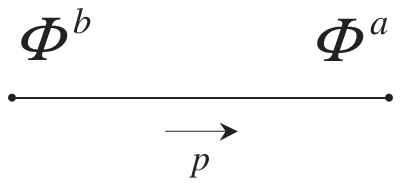} } \!\!\!\!\!\!\!&=& \frac{i \delta^{ab}}{p^2 + i 0}
\\[-6mm]
\parbox[c][25mm]{30mm}{
\insertfig{2.5}{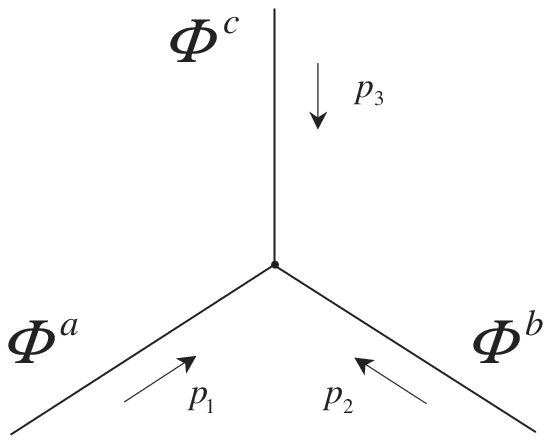} } \!\!\!\!\!\!\!&=& - 2 i g f^{abc} \bigg\{ (p_1, p_2)^\ast
+ \frac{(p_1, p_2) [p_1, p_2]}{( p_{1+} p_{2+} p_{3+} )^2} \bigg\}
\\
\label{N4gluon4}
\parbox[c][25mm]{30mm}{
\insertfig{2.5}{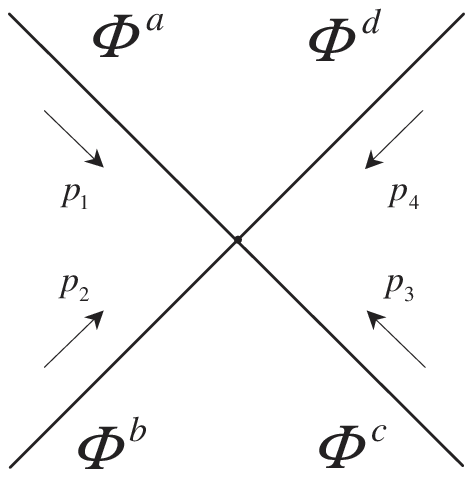} } \!\!\!\!\!\!\!&=&  - i \frac{g^2}{2}
\\[-6mm]
\!\!\!\!\!\!\!\!\!&\times&\!\!\!\! \Bigg\{ \! \left( f^{eab} f^{ecd} \frac{(p_1 -
p_2)_+ (p_3 - p_4)_+}{(p_1 + p_2)_+ (p_3 + p_4)_+} + f^{ead} f^{ebc} - f^{eac}
f^{edb} \right) \left( \frac{[p_1, p_2]}{(p_{1+} p_{2+})^2} + \frac{[p_3,
p_4]}{(p_{3+} p_{4+})^2} \right)
\nonumber\\
&&\!\!\!\!\!\!\!\!\! + \left( f^{eac} f^{edb} \frac{(p_1 - p_3)_+ (p_4 -
p_2)_+}{(p_1 + p_3)_+ (p_2 + p_4)_+} + f^{eab} f^{ecd} - f^{ead} f^{ebc} \right)
\left( \frac{[p_2, p_4]}{(p_{2+} p_{4+})^2} + \frac{[p_1, p_3]}{(p_{1+}
p_{3+})^2} \right)
\nonumber\\
&&\!\!\!\!\!\!\!\!\! + \left( f^{ead} f^{ebc} \frac{(p_1 - p_4)_+ (p_2 -
p_3)_+}{(p_1 + p_4)_+ (p_2 + p_3)_+} + f^{eac} f^{edb} - f^{eab} f^{ecd} \right)
\left( \frac{[p_1, p_4]}{(p_{1+} p_{4+})^2} + \frac{[p_2, p_3]}{(p_{2+}
p_{3+})^2} \right) \! \Bigg\}  \nonumber
\end{eqnarray*}
Here $f^{abc}$ are the $SU(N_c)$ structure constants. The signature factor
$\sigma_\mathcal{N}$ was defined in \re{sigma} ($\sigma_0=1$ and
$\sigma_1=\sigma_2=-1$). The brackets $(p_1,p_2)$ and $[p_1,p_2]$ were introduced
in \re{brackets}. Notice that $[p_1,p_2]=1$ for $\mathcal{N}=0$.

\section{Computation of one-loop dilatation operator}

To calculate one-loop correction to the dilatation operator in the light-cone
formalism, we employ the dimensional regularization with $D=4-2\varepsilon$ and
extract divergent $\sim 1/\varepsilon$ part of the corresponding Feynman
diagrams. The essential steps in the calculation of the Feynman diagrams are:

$\bullet$~Simplify the color factors using the $SU(N_c)$ identities
\begin{equation}
f^{abc} f^{abc'}= N_c \delta^{cc'},\qquad t^{a'} t^{b'} f^{a'b'c} f^{abc} =
\frac{N_c}{2} [ t^a t^b - t^b t^a ],\qquad t^{a'} t^{b'} f^{a'ac}f^{b'bc} =
\frac{N_c}2 \, t^a t^b + \frac14 \delta^{ab}\,.
\label{color-factors}
\end{equation}
In the multi-color limit, we shall drop the term $\sim \delta^{ab}$ in the last
relation.

$\bullet$~ Simplify the momentum integral using the identities
\begin{eqnarray}
&& (p_1,k)(p_2,k)^* = p_{1+}p_{2+} k \bar k +\ldots = \frac12 p_{1+}p_{2+}
\bit{k}_\perp^2+\ldots
\label{round-bracket}
\\
&& {} [p_j,k] = \sigma_\mathcal{N} \,\left(p_{j+}\right)^\mathcal{N}
\delta^{(\mathcal{N})}\left(\pi_k-\pi_j\frac{k_+}{p_{j+}}\right)\,,\qquad
(j=1,2).
\label{square-bracket}
\end{eqnarray}
Hereafter ellipses denote terms which do not produce divergent contribution.

$\bullet$~Get rid of transverse components of the loop momentum
\begin{equation}
{\bit{k}_\perp^2 } = - k^2\frac{(p - k)_+}{p_{+}} - (p - k)^2\frac{k_+}{p_+}+
\ldots\, ,
\label{k-perp}
\end{equation}
with $p^\mu$ being either $p_1^\mu$ or $-p_2^\mu$, and perform integration over
the $k-$momentum by making use of the following relation~\cite{Capper84}
\begin{equation}
\label{PPmomentumIntegral}
\int \frac{d^D k}{(2 \pi)^D} \frac{{\rm e}^{- i k_+ z}}{k^2 (p -
k)^2}\frac{p_+^n}{[k_+]^n_{\scriptscriptstyle \rm ML}} = \frac{i}{(4 \pi)^2
\varepsilon} \int_0^1 \frac{d \alpha}{\alpha^n} \left( {\rm e}^{- i \alpha p_+
z} - \sum_{l = 0}^{n - 1} \left( - i \alpha p_+ z \right)^l \right)
+\mathcal{O}(\varepsilon^0) \, ,
\end{equation}
where the $\alpha-$variable has the meaning of the momentum fraction
$k_+=\alpha p_+$. Here the pole at $k_+=0$ is integrated using the
Mandelstam-Leibbrandt prescription \re{ML-def}.

\subsection*{D1.\quad Self-energy}

Self-energy corrections renormalize the superfields. In the light-cone formalism,
in the light-like axial gauge the renormalization constants for the SYM coupling
$g$ and the superfields are equal to each other
\begin{equation}
{\Phi}^{(0)} = \sqrt{{\cal Z}} {\Phi} \, , \qquad {\Psi}^{(0)} =
\sqrt{{\cal Z}} {\Psi} \, , \qquad g^{(0)} = \frac{\mu^{\varepsilon}
g}{\sqrt{{\cal Z}}} \, ,
\end{equation}
where the superscript $(0)$ stands for the bare field/coupling. Because of
this, the beta-function and the anomalous dimensions of the superfields
coincide
\begin{equation}
\frac{\beta (g)}{g} = \gamma (g) = \frac{g^2}{16 \pi^2}\beta_0 +
\mathcal{O}(g^4)\, .
\label{beta0}
\end{equation}
In the $\mathcal{N}=2$ SYM theory the exact beta-function is given by the
one-loop result while in the $\mathcal{N}=4$ SYM theory it equals zero to all
loops.

\begin{figure}[t]
\begin{center}
\mbox{
\begin{picture}(0,40)(200,0)
\put(0,0){\insertfig{14}{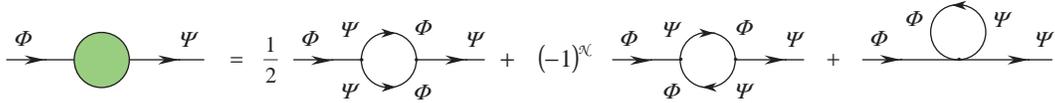}}
\end{picture}
}
\end{center}
\caption{\label{OneLoopSelfEnergy} Self-energies with included combinatoric
factors.}
\end{figure}

The renormalization factor and the anomalous dimension are computed from the
self-energy insertions
\begin{equation}
\gamma (g) \equiv  \frac{d\ln {\cal Z}}{d \ln \mu^2}  \,,\qquad {\cal Z} = 1 + (-
1)^{{\cal N} ({\cal N} + 1)/2} \frac{i}{p^2} {\Sigma}^{\rm div} (p^2) \,,
\end{equation}
where ${\Sigma}^{\rm div} (p^2)$ denotes divergent part of the self-energy.
To one-loop level, ${\Sigma}(p^2)$ receives contribution from the Feynman
diagrams shown in Fig.~\ref{OneLoopSelfEnergy}. Their calculation in the
$\mathcal{N}-$extended SYM theory leads to the following result
\begin{eqnarray}
{\Sigma}_{{\cal N} = 0} (p) \!\!\!&=&\!\!\! 2 g^2 N_c \int \frac{d^D k}{(2
\pi)^D} \frac{1}{k^2 (p - k)^2} \left\{ \frac{(k, p)(k, p)^\ast}{(p - k)_+^2}
\left( \frac{p_+^2}{k_+^2} + 2 \frac{k_+^2}{p_+^2} \right) + 2 k^2 \frac{p_+ (p -
k)_+}{k_+^2} \right\}
\nonumber\\
\!\!\!&\stackrel{\rm div}{=}&\!\!\! - \frac{i p^2}{\varepsilon} \frac{11}{3} N_c
\frac{g^2\mu^{2\varepsilon}}{(4 \pi)^2} \, .
\end{eqnarray}
for ${\cal N} = 0$,
\begin{eqnarray}
{\Sigma}_{{\cal N} = 1} (p) \!\!\!&=&\!\!\! - 2 g^2 N_c \int \frac{d^D k
}{(2 \pi)^D}\int d \pi\, \frac{[k, p]}{k^2 (p - k)^2} \left\{ \frac{(k, p)(k,
p)^\ast}{(p - k)_+^2} \left( \frac{p_+}{k_+^2} + 2 \frac{k_+}{p_+^2} \right) + 2
k^2 \frac{(2 p - k)_+}{k_+^2} \right\}
\nonumber\\
\!\!\!&\stackrel{\rm div}{=}&\!\!\! \frac{i p^2}{\varepsilon}  \{ 2 - 1 + 2 \}
N_c \frac{g^2\mu^{2\varepsilon}}{(4 \pi)^2} = \frac{i p^2}{\varepsilon} 3 N_c
\frac{g^2\mu^{2\varepsilon}}{(4 \pi)^2} \, ,
\label{Sigma-N=1}
\end{eqnarray}
for ${\cal N} = 1$,
\begin{eqnarray}
{\Sigma}_{{\cal N} = 2} (p) \!\!\!&=&\!\!\! 2 g^2 N_c \int \frac{d^D k}{(2
\pi)^D} \int d^2 \pi \frac{[k, p]}{k^2 (p - k)^2} \left\{ \frac{(k, p)(k,
p)^\ast}{(p - k)_+^2} \left( \frac{1}{k_+^2} + 2 \frac{1}{p_+^2} \right) + 2 k^2
\frac{1}{k_+^2} \right\}
\nonumber\\
\!\!\!&\stackrel{\rm div}{=}&\!\!\! \frac{i p^2}{\varepsilon}  \{ 0 + 2 + 0 \}
N_c \frac{g^2\mu^{2\varepsilon}}{(4 \pi)^2} = \frac{i p^2}{\varepsilon} 2 N_c
\frac{g^2\mu^{2\varepsilon}}{(4 \pi)^2} \, ,
\label{Sigma-N=2}
\end{eqnarray}
for ${\cal N} = 2$. For ${\cal N} = 4$, ${\Sigma}_{{\cal N} = 4} (p)$
remains finite for $\varepsilon\to 0$. In Eqs.~\re{Sigma-N=1} and \re{Sigma-N=2},
the terms/numbers in the curly brackets correspond to diagrams in Fig.\
\ref{OneLoopSelfEnergy}. We recover from this calculation the well-known results
for the one-loop beta-function,  $\beta_0= - \frac{11}{3} N_c, - 3 N_c , - 2 N_c$
and $0$ for $\mathcal{N}=0,1,2$ and $4$, respectively.

\subsection*{D2.\quad Dilatation operator in the ${\Psi}{\Psi}-$sector}

The one-loop dilatation operator in this sector receives nonvanishing
contribution from Feynman diagrams shown in Fig.~\ref{HolomorphicKernel}(a), (b)
and (d). Using the Feynman rules for $\mathcal{N}\le 2$, their sum can be written
as
\begin{eqnarray}
\vev{\mathbb{O}_{{\scriptscriptstyle\Psi\Psi}}^{(1)}(Z_1,Z_2,...)} &=&\!\!\!
-i g^2 {N_c}\sigma_{\scriptscriptstyle\mathcal{N}}\, \mu^{4-D}\!\! \int\!
\frac{d^D k}{(2 \pi)^D} \frac{ {\rm e}^{-i (p_1 - k) \cdot z_1 - i (p_2 + k)
\cdot z_2} }{ (p_1 - k)^2 (p_2 + k)^2 }\int d^\mathcal{N} \pi\,{\rm
e}^{-(\pi_1-\pi)_A\theta_1^A-(\pi_2+\pi)_A\theta_2^A} \nonumber
\\
&\times&\!\!\! \bigg\{ \frac{2 (k, p_1) (k, p_2)^\ast}{k^2} \frac{(p_1 -
k)_+^{2-\mathcal{N}}}{(k_+ p_{1+})^2} [p_1,k] + \frac{2 (k, p_1)^* (k, p_2)}{k^2}
\frac{(p_2 + k)_+^{2-\mathcal{N}}}{(k_+ p_{2+})^2}  [p_2,k]
\nonumber\\
&&  + \frac{(p_{1} - k)_+^{2-\mathcal{N}} (p_{2} + k)_+}{p_{1+} (k_+)^2}[p_1,k]+
\frac{(p_1 - k)_+ (p_2 + k)_+^{2-\mathcal{N}}}{p_{2+} (k_+)^2}[p_2,k] \bigg\} \,
.
\label{Psi-Psi-Feynman}
\end{eqnarray}
Here the first two terms correspond to the diagrams in
Fig.~\ref{HolomorphicKernel}(a) and (b) containing triple vertices and the last
two terms correspond to the diagram in Fig.~\ref{HolomorphicKernel}(d) with
quartic vertex. The color factor accompanying these diagrams can be easily
calculated in the multi-color limit with a help of the identities
\re{color-factors}. For example, the color factor for the diagram in
Fig.~\ref{HolomorphicKernel}(a) and (b) is
\begin{equation}
t^a t^b f^{aa'c} f^{bb'c} \widetilde {\Psi}_1^{a'} \widetilde
{\Psi}_2^{b'} =\frac{N_c}2\widetilde {\Psi}_1\widetilde {\Psi}_2
\end{equation}
with $\widetilde {\Psi}_k\equiv \widetilde {\Psi}^a(p_k,\pi_k) t^a$.
One examines the coefficient in front of $[p_1,k]$ in \re{Psi-Psi-Feynman} and
simplifies it with a help of \re{round-bracket} and \re{k-perp} as
\begin{equation}
\frac{2(k,p_1)(k,p_2)^*}{p_{1+} k^2}+(p_2+k)_+ = \frac{(p_2+k)^2}{k^2}k_+\,.
\label{rule-1}
\end{equation}
Similar relation holds for the coefficient in front of $[p_2,k]$
\begin{equation}
\frac{2(k,p_1)^*(k,p_2)}{p_{2+} k^2}+(p_1-k)_+ = -\frac{(p_1-k)^2}{k^2}k_+\,.
\label{rule-2}
\end{equation}
Replacing $[p_1,k]$ and $[p_2,k]$ by their expressions in terms of
delta-functions, Eq.~\re{square-bracket}, one arrives at Eq.~\re{Psi-Psi-int}.
The calculation of the momentum integral in \re{Psi-Psi-int} is straightforward
by making use of the identity (\ref{PPmomentumIntegral}). In this way, one gets
\re{Psi-Psi-div} and casts it into the operator form \re{Psi-Psi-vev}.

\subsection*{D3.\quad Dilatation operator in the ${\Phi}{\Phi}-$sector for $\mathcal{N}\le 2$}

The Feynman diagrams defining one-loop dilatation operator in the
${\Phi}{\Phi}-$sector for $\mathcal{N}\le 2$ are similar to those in
the ${\Psi}{\Psi}-$sector, Fig.~\ref{HolomorphicKernel}. The only
difference is that the direction of all lines has to be flipped. The self-energy
diagram is calculated in Sect.~D1 and the annihilation diagram vanishes as
before, Eq.~\re{annihilation}. Applying the Feynman rules given in Appendix~B,
one finds for the sum of the remaining three diagrams
\begin{eqnarray}
\vev{\mathbb{O}_{{\scriptscriptstyle\Phi\Phi}}^{(1)}(Z_1,Z_2,...)}
\!\!\!&=&\!\!\!
-i g^2 {N_c}\sigma_{\scriptscriptstyle\mathcal{N}}\, \mu^{4-D}\!\! \int \frac{d^D
k}{(2 \pi)^D} \frac{ {\rm e}^{- i (p_1 - k) \cdot z_1 - i (p_2 + k) \cdot z_2} }{
(p_1 - k)^2 (p_2 + k)^2 }\int d^\mathcal{N} \pi\,{\rm
e}^{-(\pi_1-\pi)_A\theta_1^A-(\pi_2+\pi)_A\theta_2^A} \nonumber
\\
&\times&\!\!\!\! \bigg\{  \frac{2 (k, p_1)^* (k, p_2)}{k^2}
\frac{(p_{1+})^{2-\mathcal{N}}}{(k_+ (p_1-k)_{+})^2}  [p_1,k]+\frac{2 (k, p_1)
(k, p_2)^\ast}{k^2} \frac{(p_{2+})^{2-\mathcal{N}}}{(k_+ (p_2+k)_+)^2} [p_2,k]
\!\!\!\!\nonumber\\
&&  + \frac{ (p_{1+})^{2-\mathcal{N}}p_{2+}}{(p_1-k)_+(k_+)^2}[p_1,k]+ \frac{
p_{1+}(p_{2+})^{2-\mathcal{N}}}{(p_2+k)_+(k_+)^2}[p_2,k] \bigg\} \, .
\label{Phi-Phi-Feynman}
\end{eqnarray}
One repeats the same steps as in the previous case, applies the identities
Eqs.~\re{rule-1}, \re{rule-2} and arrives at \re{Phi-Phi-int}.

The expression inside square brackets in \re{Phi-Phi-int} can be rewritten as
\re{aux1}. To see this, one examines the difference between
$[\cdots]_\mathcal{N}$ and the first two terms in the right-hand side\ of \re{aux1}
\begin{equation}
-\frac1{k_+ k^2} \left[\frac{p_{1+}}{(p_1-k)^2}+\frac{p_{2+}}{(p_2+k)^2}\right]
\left[ \frac{p_{1+}^2}{(p_1-k)_+^2}\delta^{(\mathcal{N})}\! \left(
\pi-\pi_1\ft{k_+}{p_{1+}} \right)
-\frac{p_{2+}^2}{(p_2+k)_+^2}\delta^{(\mathcal{N})}\!
\left(\pi-\pi_2\ft{k_+}{p_{2+}}\right) \right] \,.
\label{difference}
\end{equation}
The delta-functions in this relation can expressed in terms of
$\delta^{(\mathcal{N})}\!\left(\pi-\Pi\right)$ with the momentum $\varpi_A$
introduced in Eq.~\re{Pi-def}. To this end, one rewrites the momentum $\varpi_A$
as
\begin{equation}
\varpi_A =\pi_{1A}\frac{k_+}{p_{1+}}
+p_{2+}\frac{(p_1-k)_+}{(p_1+p_2)_+}\left(\frac{\pi_{1A}}{p_{1+}}-
\frac{\pi_{2A}}{p_{2+}}\right) =\pi_{2A}\frac{k_+}{p_{2+}}+
p_{1+}\frac{(p_2+k)_+}{(p_1+p_2)_+}\left(\frac{\pi_{1A}}{p_{1+}}-
\frac{\pi_{2A}}{p_{2+}}\right)
\label{Pi-app}
\end{equation}
and simplifies the expression for
$\delta^{(\mathcal{N})}\!\left(\pi-\varpi\right)-
\delta^{(\mathcal{N})}\!\left(\pi-\pi_j{k_+}/{p_{j+}}\right)$ by taking into
account that the odd $\delta-$function is polynomial in its argument,
Eq.~\re{delta-function},
\begin{equation}
\delta^{(\mathcal{N})}(\pi + \pi') - \delta^{(\mathcal{N})}(\pi) =
\ft1{(\mathcal{N}-1)!}\epsilon^{A_1\ldots A_{\mathcal{N}-1}A_\mathcal{N}}
\pi_{A_1}\ldots \pi_{A_{\mathcal{N}-1}}\pi'_{A_\mathcal{N}}+ \ldots +
\delta^{(\mathcal{N})}(\pi')\,.
\label{Delta-app}
\end{equation}
Here the expansion goes in powers of $\pi'$. Substituting $\pi \to \pi-\varpi$
and $\pi'\to \varpi-\pi_j{k_+}/{p_{j+}}$ into \re{Delta-app} and applying the
resulting identities in \re{difference}, one recovers after some algebra the last
term in the right-hand side\ of \re{aux1}. The resulting expression for the one-loop
matrix element \re{Phi-Phi-Feynman} involves the Feynman integral which is
defined below in \re{Big-integral}. As described in Appendix~D4, it calculation
leads to desired expression for the dilatation operator in the $\Phi\Phi-$sector,
Eq.~\re{H-Phi-Phi-final}.

\subsection*{D4.\quad Dilatation operator in the ${\Phi}{\Phi}-$sector for $\mathcal{N}= 4$}

For $\mathcal{N}=4$ the one-loop dilatation operator receives contribution from
the Feynman diagrams similar to those shown in
Fig.~\ref{HolomorphicKernel}(b)--(e). The only difference is that the lines
should not have arrows. For $\mathcal{N}=4$ both self-energy and annihilation
diagrams vanish. One applies the $\mathcal{N}=4$ Feynman rules listed in
Appendix~C and obtains for the sum of the remaining diagrams the expression given
in \re{Phi-Phi-N=4}.

To begin with, one eliminates $\bit{k}_\perp^2$ from \re{Phi-Phi-N=4} with a help
of \re{k-perp} and rewrites the second term in the curly bracket in
\re{Phi-Phi-N=4} as
\begin{eqnarray}
& & -4\frac{(p_1+k)^2 }{k^2}\frac{p_{2+}}{k_+}
\frac{[p_2,k]}{(p_{2+}(p_2-k)_+)^2}+4\frac{(p_2-k)^2}{k^2 }\frac{p_{1+}}{k_+}
\frac{[p_1,k]}{(p_{1+}(p_1+k)_+)^2}
\nonumber \\
& &\qquad +2\frac{(p_1+p_2)_+}{k_+}\left(\frac{[p_2,k]}{(p_{2+}(p_2-k)_+)^2}
-\frac{[p_1,k]}{(p_{1+}(p_1+k)_+)^2}\right)
\label{N=4-aux}
\end{eqnarray}
After substitution into \re{Phi-Phi-N=4} the first two terms match \re{aux1} for
$\mathcal{N}=4$ and can be easily integrated with a help of the identity
\re{PPmomentumIntegral}. According to \re{square-bracket} the last term in
\re{N=4-aux} contains the same combination of delta-functions as \re{difference}.
As before, one applies \re{Pi-app} and \re{Delta-app} for $\mathcal{N}=4$ and
expands the delta-functions in powers of $\pi_{A}-\varpi_A$. The same set of
transformations is applied to the remaining two terms inside the curly brackets
in \re{Phi-Phi-N=4}. Namely, one rewrites them in terms of delta-functions by
taking into account Eqs.~\re{square-bracket} and \re{pp-bracket} together with
the identities
\begin{eqnarray}
&& [p_1+k,p_2-k] = (p_1+p_2)_+^4\,\delta^{(4)}\left(\pi-\varpi\right) \,,\qquad
\label{PP-identity}
\\[2mm]
{}&& [p_2,p_1+k] =p_{2+}^4\,
\delta^{(4)}\left(\pi+\pi_1-\pi_2\ft{(p_1+k)_+}{p_{2+}}\right)
\nonumber \\
{}&& [p_1,p_2-k] =p_{1+}^4\,
\delta^{(4)}\left(\pi-\pi_2+\pi_1\ft{(p_2-k)_+}{p_{1+}}\right)
\nonumber
\end{eqnarray} and, then, expands the delta-functions in powers of
$\pi_{A}-\varpi_A$ with a help of \re{Delta-app}. For instance the explicit form
of the expansion for $[k, p_1]$ is
\begin{eqnarray}
[k, p_1] \!\!\!&=&\!\!\! \left( \frac{p_{1+}}{(p_1 + p_2)_+}\right)^4 [k - p_1, k
+ p_2] + \left( \frac{(p_1 - k)_+}{(p_1 + p_2)_+}\right)^4 [p_1, p_2]
\nonumber\\
&+&\!\!\! \frac{1}{3!} p_{1+}^4 p_{2+} \frac{(p_1 - k)_+}{(p_1 + p_2)_+}
\varepsilon^{ABCD} (\pi - \varpi)_{ABC} \left( \frac{\pi_1}{p_{1+}} -
\frac{\pi_2}{p_{2+}} \right)_D
\nonumber\\
&+&\!\!\! \frac{1}{3!} p_{1+} \left( p_{1+} p_{2+} \frac{(p_1 - k)_+}{(p_1 +
p_2)_+} \right)^3 \varepsilon^{ABCD} (\pi - \varpi)_A \left( \frac{\pi_1}{p_{1+}}
- \frac{\pi_2}{p_{2+}} \right)_{BCD}
\nonumber\\
&+&\!\!\! \frac{1}{4} p_{1+}^2 \left( p_{1+} p_{2+} \frac{(p_1 - k)_+}{(p_1 +
p_2)_+} \right)^2 \varepsilon^{ABCD} (\pi - \varpi)_{AB} \left(
\frac{\pi_1}{p_{1+}} - \frac{\pi_2}{p_{2+}} \right)_{CD} \, ,
\end{eqnarray}
where the notation was introduced for $(\pi)_{A\cdots C} = \pi_A\ldots \pi_C$.
One substitutes this and analogous expression for $[k, p_2]$ into the momentum
integral \re{Phi-Phi-N=4} and discovers that the terms containing
$\pi_{A}-\varpi_A$ in a power smaller than three cancel against each other. The
remaining terms give rise to the last term in \re{aux1} with
$\mathcal{X}_{\mathcal{N}=4}$ given by \re{X-N=4}.

In this way, the $\mathcal{N}=4$ Feynman integral entering \re{Phi-Phi-N=4} can
be written as
\begin{eqnarray}
\label{N4kernel}
&& i\int \frac{d^D k}{(2 \pi)^D} \int d^\mathcal{N} \pi {\rm e}^{ - i (P_1 + K)
\cdot Z_1 - i (P_2 - K) \cdot Z_2} \Bigg\{ \frac{[k + p_2, k - p_1]}{(p_2 +
k)^2(p_1 - k)^2} \frac{ p_{1+} (p_2 + k)_+ + p_{2+} (p_1 - k)_+ }{ (p_1 - k)_+^2
(k + p_2)_+^2 (p_1 + p_2)_+^2 }
\nonumber\\
&&\quad  +  \frac{1}{(p_2 + k)^2(p_1 - k)^2} \frac{ (p_1 + p_2)_+ }{ (p_1 - k)_+
(k + p_2)_+ }  \frac{\varepsilon^{A_1\cdots
A_{\mathcal{N}-1}A_\mathcal{N}}}{(\mathcal{N}-1)!}(\pi - {\varpi})_{A_1\cdots
A_{\mathcal{N}-1}} (\pi_{1 A_\mathcal{N}} p_{2+} - \pi_{2 A_\mathcal{N}} p_{1+})
\nonumber\\
&&\quad  - \frac{[p_1 , k]}{p_{1+} k_+ (p_1 - k)_+^2 k^2 (p_1 - k)^2} +
\frac{[p_2 , k]}{p_{2+} k_+ (p_2 + k)_+^2 k^2 (p_2 + k)^2} \Bigg\} \equiv \frac{1
}{(4 \pi)^2 \varepsilon} \sum_{i = 1}^4 \mathcal{J}_i \, ,
\label{Big-integral} 
\end{eqnarray}
where $\mathcal{J}_i$ stands for the contribution of $i$th term inside curly
brackets. Here we displayed the $\mathcal{N}-$dependence because calculation of
the one-loop matrix element in the $\Phi\Phi-$sector for $\mathcal{N}\le 2$ (see
\re{Phi-Phi-int} and Appendix~D3) leads to the same Feynman integral
\re{Big-integral}.

The calculation of \re{Big-integral} is straightforward and relies on the
identities \re{square-bracket}, \re{PP-identity} and \re{PPmomentumIntegral}. To
represent the result in a concise manner, one introduces plane waves in the
superspace
\begin{equation}
\phi (Z_1 , Z_2) \equiv {\rm e}^{- i P_1 \cdot Z_1 - i P_2 \cdot Z_2} \, .
\end{equation}
Then the momentum integration in \re{Big-integral} leads to
\begin{eqnarray*}
\mathcal{J}_1 \!\!\!&=&\!\!\! \int_{0}^1 \frac{d \alpha}{\bar\alpha} \bigg\{
\left( 1 - i \alpha \frac{p_{1+}}{(p_1 + p_2)_+} (P_1 + P_2) \cdot Z_{12}
\right) \left[ \phi (Z_2, Z_2) - \phi (\bar\alpha Z_1 + \alpha Z_2,
\bar\alpha Z_1 + \alpha Z_2) \right]
\nonumber\\
&&\qquad \ \, + \left( 1 - i \alpha \frac{p_{2+}}{(p_1 + p_2)_+} (P_1 + P_2)
\cdot Z_{21} \right) \left[ \phi (Z_1, Z_1) - \phi (\alpha Z_1 + \bar\alpha
Z_2, \alpha Z_1 + \bar\alpha Z_2) \right] \bigg\}
\, , \nonumber\\
\mathcal{J}_3 \!\!\!&=&\!\!\! \int_0^1 \frac{d \alpha}{\alpha \bar\alpha^2}
\Big\{ \phi ( \bar\alpha Z_1 + \alpha Z_2 , Z_2 ) - \left( 1 - \bar\alpha^2 - i
\alpha
\bar\alpha P_1 \cdot Z_{12} \right) \phi ( Z_2 , Z_2 ) -
\bar\alpha^2 \phi ( Z_1 , Z_2 )
\Big\} \, ,
\end{eqnarray*}
where $\bar\alpha=1-\alpha$. Note that $\mathcal{J}_4 = \mathcal{J}_3|_{P_1
\leftrightarrow P_2, Z_1 \leftrightarrow Z_2}$. To calculate the integral
$\mathcal{J}_2$, one applies the identity
\begin{eqnarray}
\int d^\mathcal{N} \pi \, {\rm e}^{- \pi \theta} (\pi - \varpi)_{1\cdots
\mathcal{N}-1} = -(-1)^{\mathcal{N}(\mathcal{N} - 1)/2} {\rm e}^{- \varpi \theta}
\theta^\mathcal{N} \, ,
\end{eqnarray}
changes the integration variable $k'=k+p_1$ and takes into account
\re{PPmomentumIntegral}. This leads to
\begin{eqnarray*}
\mathcal{J}_2 \!\!\!&=&\!\!\! \frac{ \theta_{12}^A \, (\pi_{1A} p_{2+} -
\pi_{2A} p_{1+})}{(p_1 + p_2)_+}\int_{0}^1 \frac{d \alpha}{\bar\alpha}\Big\{
\phi (Z_1 , Z_1) + \phi (Z_2 , Z_2)
\\
& & \qqqquad - \phi (\bar\alpha Z_1 + \alpha Z_2, \bar\alpha Z_1 + \alpha Z_2)
- \phi (\bar\alpha Z_2 + \alpha Z_1, \bar\alpha Z_2 + \alpha Z_1) \Big\} \, .
\nonumber
\end{eqnarray*}
The integrals $\mathcal{J}_1,\ldots,\mathcal{J}_4$ assume the same form for the
SYM theories with ${\cal N}=0,1,2,4$. Their sum reduces to the desired form
\re{H-Phi-Phi-final} and \re{N=4-kernel}
\be
{\rm Eq.\,\re{Big-integral}}=\frac{1}{(4 \pi)^2 \varepsilon} \left\{
\mathbb{V}^{(-1/2,-1/2)} (1 - \Pi_{\Phi\Phi}) \phi (Z_1, Z_2) +
\Delta_{\Phi\Phi}\phi (Z_1, Z_2) \right\}\,,
\ee
where after some algebra the remnant $\Delta_{\Phi\Phi}$ can be cast into the
form
\begin{eqnarray}
\Delta_{\Phi\Phi}\phi (Z_1, Z_2)  \!\!\!&=&\!\!\! - \frac{i}{2} \frac{(p_1 -
p_2)_+}{(p_1 + p_2)_+} (P_1 + P_2) \cdot Z_{12}
\\
&\times&\!\!\! \int_0^1 d\alpha \, \Big\{ \phi (Z_1, Z_1) + \phi (Z_2, Z_2) - 2
\phi (\bar\alpha Z_1 + \alpha Z_2, \bar\alpha Z_1 + \alpha Z_2) \Big\}
\nonumber\\
&=&\!\!\! \frac{(p_{1} - p_{2})_+}{(p_{1} + p_{2})_+} \left( 1 - \ft12 Z_{21}
\partial_{Z_2} \right) \phi (Z_2, Z_2) + \frac{(p_{2} - p_{1})_+}{(p_{1} + p_{2})}
\left( 1 - \ft12 Z_{12} \partial_{Z_1} \right) \phi (Z_1, Z_1) \, , \nonumber
\end{eqnarray}
with $ Z_{jk} \partial_{Z_j} \equiv z_{jk}  { \partial_{z_j}} + \theta^A_{jk}
 {\partial_{\theta^A_j}} \, . $

\subsection*{D5.\quad Dilatation operator in the ${\Phi}{\Psi}-$sector}

The one-loop dilatation operator in the ${\Phi}{\Psi}-$sector is given
by Feynman diagrams shown in Fig.~\ref{MixedcKernel}. Since the superfields are
different, one encounters a new diagram, Fig.~\ref{MixedcKernel}(c), in which two
superfields are interchanged on the light-cone. As compared with the diagonal
${\Phi}{\Phi}-$ and ${\Psi}{\Psi}-$sectors, the one-loop
dilatation operator is given by the sum of two terms corresponding to the
transitions ${\Phi}{\Psi}\to{\Phi}{\Psi}$ and ${\Phi}
{\Psi}\to{\Psi}{\Phi}$. For $\mathcal{N}=0$, they are defined in
Eqs.~\re{Phi-Psi-a} and \re{Phi-Psi-b}, respectively.

Let us consider these two contributions separately. Applying the $\mathcal{N}=0$
Feynman rules one finds for the sum of Feynman diagrams in
Figs.~\ref{MixedcKernel}(a), (b) and (e)
\begin{eqnarray}
\vev{\mathbb{O}_{{\scriptscriptstyle\Phi\Psi}}^{(1)}}
\!\!\!&\stackrel{{\Phi}{\Psi}\to {\Phi}{\Psi}}{=}&\!\!\! - i
g^2 N_c\mu^{4-D}\! \int \frac{d^D k}{(2 \pi)^D} \frac{ {\rm e}^{- i (p_1 - k)
\cdot z_1 - i (p_2 + k) \cdot z_2} }{(p_1 - k)^2 (p_2 + k)^2}
\frac{p_{1+}^2}{(p_1 - k)_+^2}
\nonumber\\
&&\!\!\! \times\Bigg\{ \frac{\bit{k}_\perp^2}{k^2 (k_+)^2} \left( p_{1+} p_{2+} +
\frac{(p_1 - k)_+^2 (p_2+ k)_+^2}{p_{1+} p_{2+}} \right)
\nonumber\\
&&\!\!\! -\frac{(p_1 - k)_+(p_2 + k)_+}{p_{1+} p_{2+}} \frac{p_{1+} (p_1 - k)_+ +
p_{2+} (p_2 + k)_+}{(p_{1} + p_{2})_+^2}
\nonumber\\
&&\!\!\! +\frac{(p_1 - k)_+ (p_2 + k)_+}{k_+^2} \left( 1 + \frac{(p_1 - k)_+ (p_2
+ k)_+}{p_{1+} p_{2+}} \right) \Bigg\} \, ,
\end{eqnarray}
and for the sum of Feynman diagrams in Figs.~\ref{MixedcKernel}(c) and (e)
\begin{eqnarray}
\vev{\mathbb{O}_{{\scriptscriptstyle\Phi\Psi}}^{(1)}}
\!\!\!&\stackrel{{\Phi}{\Psi}\to {\Psi}{\Phi}}{=}&\!\!\! - i
g^2N_c\mu^{4-D}\!\int \frac{d^D k}{(2 \pi)^D} \frac{ {\rm e}^{- i (p_1 - k) \cdot
z_2 - i (p_2 + k) \cdot z_1} }{(p_1 - k)^2 (p_2 + k)^2}  \frac{p_{1+}^2}{(p_2 +
k)_+^2}
\nonumber\\
&&\!\!\! \times\Bigg\{ \frac{\bit{k}_\perp^2 (k_+)^2}{k^2 p_{1+} p_{2+}} +
\frac{(p_1 - k)_+(p_2 + k)_+}{p_{1+} p_{2+}} \frac{p_{1+} (p_2 + k)_+ + p_{2+}
(p_1 - k)_+}{(p_1 + p_2)_+^2} \Bigg\} \, .
\end{eqnarray}
One applies the identity \re{k-perp} to get rid of $\bit{k}_\perp^2$ and performs
integration with a help of \re{PPmomentumIntegral}.  The calculation gives
\begin{eqnarray}
\vev{\mathbb{O}_{{\scriptscriptstyle\Phi\Psi}}^{(1)}}
\!\!\!&\stackrel{{\Phi}{\Psi}\to
{\Phi}{\Psi}}{=}&\!\!\!\frac{g^2 N_c}{(4 \pi)^2 \varepsilon} \Big\{
\int_0^1 \frac{d \alpha}{(1-\alpha) \alpha^2}\bigg[ \alpha^4
\phi_1(z_1,(1-\alpha) z_1 + \alpha z_2)
\\
&&\qquad\qquad\qquad + \phi_1(\alpha z_1 + (1-\alpha) z_2,z_2) - 2 \alpha^2
\phi_1(z_1,z_2) \bigg] - \phi_3(z_1,z_2) \Big\} \, , \nonumber
\\
\vev{\mathbb{O}_{{\scriptscriptstyle\Phi\Psi}}^{(1)}}
\!\!\!&\stackrel{{\Phi}{\Psi}\to {\Psi}{\Phi}}{=}&\!\!\!
\frac{g^2 N_c}{(4 \pi)^2 \varepsilon} \left\{ \int_0^1 d \alpha
\frac{(1-\alpha)^3}{\alpha^2} \, \phi_2(\alpha z_1 + (1-\alpha) z_2,z_2) +
\phi_3(z_1,z_2) \right\} \, .
\end{eqnarray}
Here the notation was introduced for the functions
\begin{eqnarray}
\phi_1(z_1,z_2)&=&\left( 1 - {\Pi}_{\Phi\Psi} \right) {\rm e}^{- i p_{1+} z_1 -
i p_{2+} z_2}\,, \nonumber
\\[2mm]
\phi_2(z_1,z_2)&=&\left( 1 - {\Pi}_{\Psi\Phi} \right) {\rm e}^{- i p_{2+} z_1 -
i p_{1+} z_2}\,, \nonumber
\\[2mm]
\phi_3(z_1,z_2)&=&- \frac{p_{1+}}{(p_1 + p_2)_+} \left( 2 - z_{12}
\partial_{z_2} \right) {\rm e}^{- i (p_1 + p_2)_+ z_2}\,,
\end{eqnarray}
and the projectors ${\Pi}_{\Phi\Psi}$ and ${\Pi}_{\Psi\Phi}$ were defined in
\re{projector-phi-psi} and \re{projector-psi-phi}, respectively. Identifying the
plane waves as matrix elements of the light-cone operator, Eq.~\re{Born}, one
arrives at Eqs.~\re{Phi-Psi-a} and \re{Phi-Psi-b}.



\end{document}